\documentclass[review]{elsarticle}
\usepackage{lineno,hyperref}
\usepackage{graphicx}
\usepackage[latin1]{inputenc}
\usepackage{amsmath}
\usepackage{amsfonts,dsfont,ulem}
\usepackage{amssymb}
\usepackage{algorithm}
\usepackage{algorithmic}
\usepackage{bm}
\usepackage{subfigure}
\usepackage{epstopdf}
\usepackage[french, english]{babel}
\usepackage{multirow}
\usepackage[latin1]{inputenc}
\usepackage[T1]{fontenc}
\usepackage{multicol}
\usepackage{amsmath}
\usepackage{amsfonts}
\usepackage{amssymb}
\usepackage{graphicx}
\usepackage{amssymb}
\usepackage{algorithm}
\usepackage{algorithmic}
\usepackage{bm}
\usepackage{subfigure}
\usepackage{epstopdf}
\usepackage{multirow}
\usepackage[latin1]{inputenc}
\usepackage[T1]{fontenc}
\usepackage{multicol}
\usepackage{color}
\usepackage{mathtools}
\usepackage[bottom]{footmisc}

    \makeatletter
    \def\ps@pprintTitle{%
      \let\@oddhead\@empty
      \let\@evenhead\@empty
      \let\@oddfoot\@empty
      \let\@evenfoot\@oddfoot
    }
    \makeatother
\newcommand{\ubar}[1]{\text{\b{\text{$#1$}}}}   

\addto\captionsenglish{%
}
\modulolinenumbers[5]

\begin{document}

\begin{frontmatter}

\title{AC OPF in Radial Distribution Networks - Part I:\\
On the Limits of the Branch Flow Convexification
and the Alternating Direction Method of Multipliers}


\author[mymainaddress,mysecondaryaddress]{Konstantina Christakou}
\ead{konstantina.christakou@epfl.ch}

\author[mymainaddress2]{Dan-Cristian Tomozei}
\ead{dtomozei@cisco.com}

\author[mymainaddress]{Jean-Yves Le Boudec}
\ead{jean-yves.leboudec@epfl.ch}

\author[mysecondaryaddress]{Mario Paolone\corref{mycorrespondingauthor}}
\cortext[mycorrespondingauthor]{Corresponding author. Phone number: $+41$ $21$ $69$ $32662$, Postal address: EPFL STI IEL DESL
ELL 136 (B\^{a}timent ELL), Station 11, CH-1015 Lausanne}
\ead{mario.paolone@epfl.ch}

\address[mymainaddress]{Laboratory for Communications and Applications 2, Ecole Polytechnique F\'{e}d\'{e}rale de Lausanne,
CH-1015 Lausanne, Switzerland}
\address[mysecondaryaddress]{Distributed Electrical Systems Laboratory, Ecole Polytechnique F\'{e}d\'{e}rale de Lausanne, CH-1015
Lausanne, Switzerland}
\address[mymainaddress2]{Cisco Systems, Inc, EPFL Innovation Park Building E 1015 Lausanne, Switzerland}

\begin{abstract}
The optimal power-flow problem (OPF) has always played a key role in the planning and operation of power systems. Due to the non-linear nature of the AC power-flow equations, the OPF problem is known to be non-convex, therefore hard to solve. {\color{black}{During the last few years several methods for solving the OPF have been proposed. The majority of them rely on approximations, often applied to the network model, aiming at making OPF convex and yielding inexact solutions. Others, kept the non-convex nature of the OPF with consequent increase of the computational complexity, inadequateness for real time control applications and sub-optimality of the identified solution.}} Recently, Farivar and Low proposed a method that is claimed to be exact for the case of radial distribution systems under specific assumptions, despite no apparent approximations. In our work, we show that it is, in fact, not exact. {\color{black}{On one hand, there is a misinterpretation of the physical network model related to the ampacity constraint of the lines' current flows. On the other hand, the proof of the exactness of the proposed relaxation requires unrealistic assumptions and, in particular, (i) full controllability of loads and generation in the network and (ii) no upper-bound on the controllable loads.}}  We also show that the extension of this approach to account for exact line models might provide physically infeasible solutions. 
{\color{black}{In addition to the aforementioned convexification method, recently several contributions have proposed OPF algorithms that rely on the use of the alternating-direction method of multipliers (ADMM).}} However, as we show in this work, there are cases for which the ADMM-based solution of the non-relaxed OPF problem fails to converge. To overcome the aforementioned limitations, we propose a specific algorithm for the solution of a non-approximated, non-convex OPF problem in radial distribution systems. 
In view of the complexity of the contribution, this work is divided in two parts. In this first part, we specifically discuss the limitations of both BFM and ADMM to solve the OPF problem.
\end{abstract}



\end{frontmatter}


\section{Introduction}
\label{intro}
The category of optimal power-flow problems (OPFs) represents the main set of problems for the optimal operation of power systems. The first formulation of an OPF problem appeared in the early 1960s and has been well-defined ever since \cite{carpentier1962contribution}. It consists in determining the operating point of controllable resources in an electric network in order to satisfy a specific network objective subject to a wide range of constraints. Typical controllable resources considered in the literature are generators, storage systems, on-load tap changers (OLTC), flexible AC transmission systems (FACTS) and loads (e.g.,~\cite{6170586,6313956,1178804,997946,6716076}). The network objective is usually the minimization of losses or generation costs, and typical constraints include power-flow equations, capability curves of the controllable resources, as well as operational limits on the line power-flows and node voltages (e.g.,~\cite{4548149}).

The OPF problem is known to be non-convex, thus difficult to solve efficiently (e.g.,~\cite{lesieutre2005convexity,hiskens2001exploring,4470566}). Since the problem was first formulated, several techniques have been used for its solution. Among others, non-linear and quadratic programming techniques, Newton-based methods, interior point methods in the earlier years, as well as heuristic approaches based on genetic algorithms, evolutionary programming, and particle-swarm optimization in recent years (e.g.,~\cite{frank2012optimal,frank2012optimal2,qiu2009literature,7038504}). {\color{black}{These techniques, even though they have been shown to sucessfully solve instances of the non-convex OPF problem, seek to find a local optimal solution of the OPF. They, generally, utilize powerful general purpose solvers or in-house developed software but they cannot guarantee the identification of the global optimal solution. In general, they are characterized by high computational complexity. The first category of approaches make use of gradient-based optimization algorithms or even require the use of hessian matrices related to the problem. Therefore, such techniques require several assumptions on the OPF problem formulation such as analytic and smooth objective functions. Heuristics have been applied widely in the literature as a solution technique, for instance in cases where the OPF problem is non-smooth, non-differentiable and highly non-linear.}}

Recently, the OPF problem is becoming more compelling due to the increasing penetration of embedded generation in
distribution networks, essentially composed by renewable resources\footnote{{\color{black}{It is worth noting that transmission and distribution systems are different with respect to (i) topology, (ii) electrical line parameters, (iii) power flow values, (iv) nature and number of controllable devices. Therefore, these systems require dedicated OPF algorithms that account for their specific characteristics. The focus of this work is on OPF algorithms specifically designed for the case of distribution networks.}}}. {\color{black}{The distributed nature of such resources, as well as their large number and potential stochasticity increase significantly the complexity and the size of the OPF problem and bring about the need for distributed solutions. In this direction, several algorithms have been proposed in the literature to handle large-scale OPF problems (e.g.,~\cite{6510541,Capitanescu20111731,918290}).}} Additionally, several contributions have proposed specific distributed algorithms for the solution of the OPF problem.  In~\cite{bolognani2013distributed,6687943} the authors design a dual-ascent algorithm for optimal reactive power-flow with power and voltage constraints. In~\cite{dall2013distributed,lam2012optimal} dual decomposition is used as the basis for the distributed solution of the OPF problem. Finally, a significant number of contributions propose distributed formulations of the OPF problem that are based on the alternating direction method of multipliers (ADMM) (e.g.,~\cite{kraning2013dynamic,dall2013distributed,sun2013fully,vsulc2013optimal,peng2014distributed,erseghe2014distributed}). 

{\color{black}{Recently a lot of emphasis is put on the convexification of the OPF problem. The reason behind this emerging trend is that convex problems provide convergence guarantees to an optimal solution and therefore such methods can be deployed within the context of control applications for power systems and specifically distribution networks.}} However, most of the proposed convexification schemes either do not guarantee to yield an optimal solution or they are based on approximations that convexify the problem in order to guarantee convergence. These approximations, often, either lead to (i) misinterpretation of the system model~\cite{bakirtzis2003decentralized} or (ii) solutions that, even though mathematically sound, might be far away from the real optimal solution, thus having little meaning for the grid operation~\cite{6120344}.

Recently, Farivar and Low proposed in~\cite{farivar2013branch,6507352} a convexification of the problem that is claimed to be exact for radial networks. In Part I of this paper, we show that this claim is not exact, as the convexification of the problem leads to an inexact system model. We also show that the method of ADMM-based decomposition, which comes together with the convexification, does not work for a correct system model. {\color{black}{In this first part of the paper we focus on the Farivar-Low convexification and ADMM algorithms since they are considered as the most prominent ones by the recent literature on the subject.}}  As an alternative, we propose in Part II an algorithm for the solution of the correct AC OPF problem in radial networks. Like ADMM, it uses an augmented Lagrangian, but unlike ADMM, it uses primal decomposition~\cite{palomar2006tutorial} and does not require that the problem be convex. We consider a direct-sequence representation of the electric distribution grid and we present both a centralized and a decentralized asynchronous version of the algorithm.

The structure of this first part is as follows. In
Section~\ref{sec:opf} we present the generic formulation of the OPF
problem in radial distribution systems and we classify several OPF
algorithms based on the approximations and assumptions on which they
rely. In Section~\ref{sec:farivar} we discuss the limitations and
applicability of the Farivar-Low formulation of the OPF problem
proposed in~\cite{farivar2013branch,6507352}. We provide, in
Section~\ref{sec:admm}, the ADMM-based solution of the original
non-approximated OPF problem. In the same section, we highlight
specific cases where the ADMM-based algorithm fails to
converge. Finally, we provide the main observations and concluding
remarks for this part in Section~\ref{sec:conclusion}.

\section{Generic Formulation of the OPF Problem}
\label{sec:opf}
\subsection{Notation and Network Representation}
\begin{figure}
\begin{centering}
\includegraphics[width=1\linewidth, clip=true, trim=150 230 100 150]{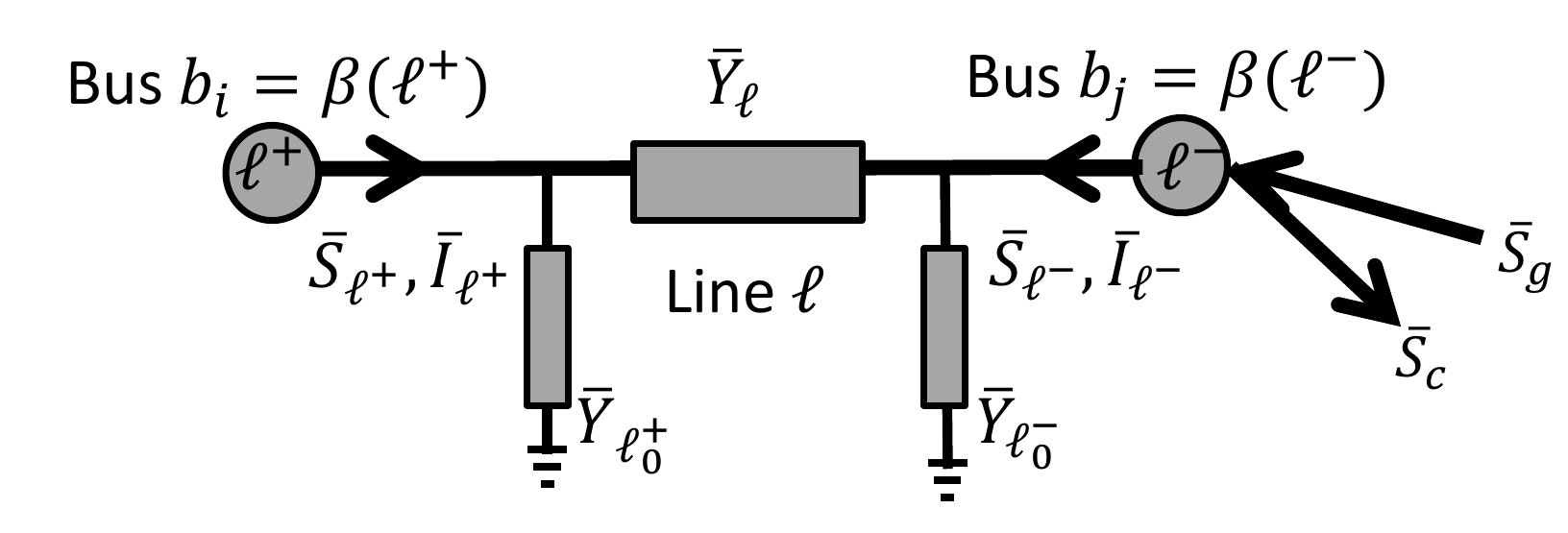}
\vspace{-25pt}
\caption{{\color{black}{Notation used in this paper for the OPF formulation.}}}
\label{fig:notation}
\end{centering}
\end{figure}

In the rest of the paper, we consider a balanced radial network composed of buses ($\mathcal{B}$), lines ($\mathcal{L}$), generators ($\mathcal{G}$) and loads ($\mathcal{C}$). The network admittance matrix is denoted by $Y$. Several generators/loads can be connected to a bus $b{\in}\mathcal{B}$. We denote that a generator $g{\in}\mathcal{G}$ or a load $c{\in}\mathcal{C}$ is connected to a bus by ``$g{\in}b$'' and ``$c{\in}b$''. {\color{black}{ We assume that the nodal-power injections are voltage-independent.}} A line $\ell{\in}\mathcal{L}$ is represented using its exact $\pi$-equivalent model and it has a receiving and a sending end denoted by $\ell^+$ and $\ell^-$. Each line is connected to two adjacent buses: $\beta(\ell^+)$ and $\beta(\ell^-)$, respectively. $\bar{Y}_\ell$ denotes the longitudinal admittance of a line, $\bar{Y}_{\ell^+_0} $ ($\bar{Y}_{\ell^-_0} $) is the shunt capacitance at the receiving (sending) end of the line\footnote{In the rest of the paper, complex numbers are denoted with a bar above (e.g., $\bar{V}$) and complex conjugates with a bar below (e.g.,$\ubar{V}$).}. The notation adopted is shown in detail in Fig.~\ref{fig:notation} where the network branch connecting the generic network nodes $i$ and $j$ is represented.

\subsection{Generic OPF Formulation}
\label{sec:classic_OPF}
The traditional formulation of the OPF problem consists in minimizing a specific network objective:
\begin{align}
\label{eq:OPF-class_obj}
\min_{\substack{\bar{S}_g,\bar{S}_c,\bar{S}_\ell^+,\bar{S}_\ell^-,\bar{I}_\ell^+,\bar{I}_\ell^-,\bar{V}_b}} \sum_{g \in \mathcal{G}} C_g(\bar{S}_{g})+\sum_{c \in \mathcal{C}} C_c(\bar{S}_{c})
\end{align}
The first term of the network objective ($C_g$) in (\ref{eq:OPF-class_obj}) is typically a non-decreasing convex function accounting for the minimization of the generation costs or the network real power losses. The second term ($C_c$) is included in the objective when the cost of non-supplied load is taken into account.

The following set of constraints is considered\footnote{{\color{black}{Note that the proposed formulation can be extended without loss of generality to the case of multi-phase unbalanced grids by adopting the so-called compound network admittance matrix, (i.e., the 3-phase representation of the grid model which takes into account the various couplings between the network phases) instead of the single-phase equivalents. In this cases, each of the constraints in (\ref{eq:Sbal})-(\ref{eq:cap_curve}) needs to be formulated separately for each network phase.}}}:

\begin{align}
&\sum_{g \in b} \bar{S}_g - \sum_{c \in b} \bar{S}_c + \sum_{\beta(\ell^+)=b} \bar{S}_{\ell^+} + \sum_{\beta(\ell^-)=b} \bar{S}_{\ell^-}=0, \quad \forall b \in \mathcal{B}  \label{eq:Sbal}\\
&  \bar{S}_{\ell^+}  =  \bar{V}_{\beta(\ell^+)}\ubar{I}_{\ell^+}, \quad \bar{S}_{\ell^-}  =  \bar{V}_{\beta(\ell^-)}\ubar{I}_{\ell^-}, \quad \forall \ell \in \mathcal{L} \label{eq:Sl_ac}\\
&\bar{I}_{\ell^+}  =  \bar{Y}_\ell (\bar{V}_{\beta(\ell^+)} -\bar{V}_{\beta(\ell^-)} )+ \bar{Y}_{\ell^+_0} \bar{V}_{\beta(\ell^+)},\quad \forall \ell \in \mathcal{L} \label{eq:Ilp}\\
&\bar{I}_{\ell^-}  = \bar{Y}_\ell (\bar{V}_{\beta(\ell^-)} -\bar{V}_{\beta(\ell^+)} )+ \bar{Y}_{\ell^-_0} \bar{V}_{\beta(\ell^-)}, \quad \forall \ell \in \mathcal{L} \label{eq:Ilm}\\
& V_{min} \leq |\bar{V}_b| \leq V_{max}, \quad \forall b \in \mathcal{B} \label{eq:volt_limits}\\
& |\bar{S}_{\ell^+}| \leq S_{\ell_{max}}, \quad \text{or} \quad |\bar{I}_{\ell^+}| \leq I_{\ell_{max}}, \quad \forall \ell \in \mathcal{L} \label{eq:OPF_lineflows}\\
& |\bar{S}_{\ell^-}| \leq S_{\ell_{max}}, \quad \text{or} \quad  |\bar{I}_{\ell^-}| \leq I_{\ell_{max}}, \quad \forall \ell \in \mathcal{L}\\
& \bar{S}_g \in \mathcal{H}_g, \quad \forall g \in \mathcal{G} \quad \text{and} \quad \bar{S}_c \in \mathcal{H}_c, \quad \forall c \in \mathcal{C}\label{eq:cap_curve}
\end{align}
where, $\bar{S}$ denotes the complex power\footnote{We use the convention that positive values represent power injection and negative power consumption.}, $\bar{V}_b$ is the direct sequence phase-to-ground voltage of node $b$, $\bar{I}_{\ell^+}$ $(\bar{I}_{\ell^-})$ is the current flow in the receiving (sending) end of line $\ell$, and $\mathcal{H}_g, \mathcal{H}_c$ are the capability curve of the generator $g$ and the limits of the load $c$ respectively\footnote{Note that different types of controllable generators or loads can be accounted for via their corresponding capability curves/limits.}. If a generator (load) is non-controllable then the set $\mathcal{H}_g$ ($\mathcal{H}_c$) is limited to a single point.

The first constraint (\ref{eq:Sbal}) corresponds to the power balance constraint at each network bus, whereas (\ref{eq:Sl_ac}) is an alternative way to define the AC power flow equations. Constraints (\ref{eq:volt_limits}) and (\ref{eq:OPF_lineflows}) are so-called node voltage and lines ampacity contraints, i.e., limits on node voltages and line power/current flows. The last constraints (\ref{eq:cap_curve}) represent the capability limits that each of the controllable devices should respect.

The equality constraints (\ref{eq:Sl_ac}) render the OPF problem non-convex and, therefore, difficult to solve efficiently. The majority of the proposed algorithms in the literature rely on several approximations and/or convex relaxations and seek a solution to a modified OPF problem. In what follows, we describe and discuss the most common approximations.

\subsection{Approximations of the OPF Problem}
In general, the approximations used in the formulation of an OPF problem can be categorized in two large groups: approximations of the physical network models and methods that relax the space of the solutions and/or control variables.

In the first case, we can find OPF formulations that rely mainly on linearizations of the AC power flow equations. Such attempts typically (i) consider the DC power flow, (ii) use the decoupled AC power flow or (iii) neglect the network losses and/or the transverse parameters of the lines. Specifically, the concepts of the DC and the decoupled OPF have been extensively used in the literature (e.g.,~\cite{sun2007dc,biskas2005decentralized,ferreira2014flexible,1216149}), as they approximate the OPF problem with linear programming problems and, therefore, enable its fast resolution. {\color{black}{Such techniques have been extensively used for the OPF solution in the case of transmission networks where the DC approximation might be reasonable since the resistance to inductance ratio of the transmission lines is negligible and the reactive power flow is supplied locally. However, the DC OPF is not applicable in distribution networks due to the large resistance over reactance ratio of the lines. Furthermore, the DC-based OPF algorithms have several shortcomings like, for instance, the inability to optimize the reactive power dispatch and the fact that they always provide a solution even when it is physically infeasible.}} In the same direction, the authors in~\cite{vsulc2013optimal} use the so-called Dist-Flow equations (\cite{19266}) to linearize the power flows and propose an ADMM-based OPF algorithm that neglects the real and reactive losses.  Finally, several contributions rely on simplified network line-models that neglect the transverse parameters, resulting in inaccuracies of the physical system model (e.g.,~\cite{bose2012quadratically,lowpscc,jabr2006radial}).

In the second case, we can find OPF formulations where, typically, the constraints are relaxed in order to convexify the problem. In particular, a large number of contributions recently proposed a SDP formulation of the OPF problem, where the rank-one constraint of a matrix is relaxed and the algorithm is claimed to yield zero-duality gap for radial distribution networks (e.g.,~\cite{LavaeiLow,dall2013distributed,lam2012optimal}). Another relaxation is proposed in~\cite{jabr2006radial} where the OPF problem is cast as a second order cone programming. A similar technique is used in~\cite{farivar2011inverter}, where the equality constraints of the branch flows are relaxed.

In both the aforementioned categories of approximations, the modified OPF formulations guarantee convergence of the proposed algorithms. The reached solutions, however, even though mathematically sound, are not always meaningful for the grid operation. The DC and the decoupled OPF work sufficiently well for transmission systems, nevertheless they can introduce large errors when used for solving the OPF in the case of distribution systems (e.g.,~\cite{stott2009dcrev}). As far as the semidefinite relaxation is concerned, its limitations have been recently investigated. The authors in~\cite{6120344} show through practical examples, that in the case of negative locational marginal prices or strict line-flow constraints it can lead to solutions that are not valid, namely for which the duality gap is not zero. Furthermore, in~\cite{6581918} the authors show the existence of multiple local optima of the OPF problem due to the feasible region being disconnected and due to the nonlinearities of the constraints; they show that the SDP formulation of the OPF problem fails to find the global optimum in cases where there are multiple local optima. In the same direction, a recent review (\cite{lowpscc}) summarizes the semidefinite relaxations applied to the OPF problem and discusses their limitations.

Recently, another formulation of the OPF problem has been proposed (\cite{farivar2012branch,li2012exact,farivar2013branch,6507352,gan2012branch}). This formulation also belongs to the category of the semidefinite relaxations and uses the so-called branch-flow model (BFM) for describing the network. The BFM essentially describes the network flows by using as variables the currents and the powers of the various network branches, instead of the nodal injections. In~\cite{farivar2013branch,6507352} Farivar and Low propose an OPF formulation that relies on the BFM representation of the network and they present a two-step relaxation procedure that turns the problem into a second-order cone program (SOCP). The authors prove that under specific assumptions both relaxation steps are exact for the case of radial networks, hence a globally optimal OPF solution can be retrieved by solving the relaxed convex problem.

In what follows, we first briefly recall the formulation of the OPF problem in~\cite{farivar2013branch,6507352} and then we investigate the applicability of the branch flow model to the OPF formulation. We show, on one hand, that this model misinterprets the physical network representation by imposing the ampacity constraint on a fictitious line-current that neglects the contribution of the shunt components of the line. We also show that, on the other hand, the proof of the exactness of the proposed relaxation requires several unrealistic assumptions. {\color{black}{In particular, the OPF formulation in~\cite{farivar2013branch,6507352} assumes full controllability of both loads and generators connected in the network buses. As a matter of fact, this is a strong assumption, as in a real setting the DNO has very few specific control points available in the network with controllable resources' capability curves that are typically complex. In addition to this, the controllable loads are required to have infinite upper bounds in order to prove the exactness of the proposed SOCP relaxation in~\cite{farivar2013branch,6507352}. In a realistic scenario, such an assumption implies that in cases where excessive production of the generators causes violations of the voltage or line-flows limits, local demand is invoked to compensate for the increased generation even beyond the possible power consumption from installed loads.}}

\section{On the Limits of the Branch-Flow Convexification for the Solution of the OPF Problem}
\label{sec:farivar}

\subsection{The BFM-based Formulation of the OPF Problem}
\begin{figure}[h!]
\begin{centering}
\includegraphics[width=1\linewidth, clip=true, trim=50 450 50 180]{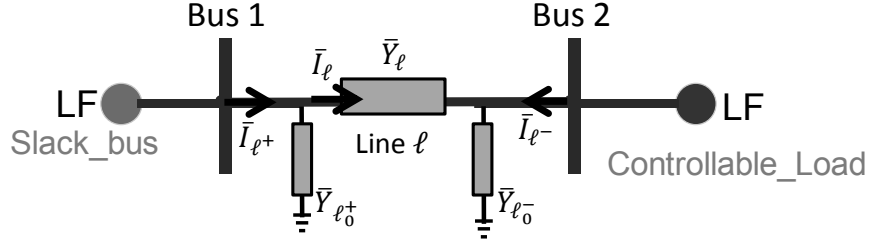}
\vspace{-25pt}
\caption{{\color{black}{Notation used in this paper for the BFM convexification of the OPF problem.}}}
\label{fig:notation_bfm}
\end{centering}
\end{figure}

We assume the same objective function as in Eq.~\ref{eq:OPF-class_obj} and again consider that the network lines are represented using a $\pi$-model. Contrary to the formulation in (\ref{eq:Sbal})-(\ref{eq:cap_curve}), we reformulate the constraints of the OPF problem by using the branch power and current flows as variables, similarly to~\cite{farivar2013branch}. To this end, we denote by $\bar{S}_\ell$ and $\bar{I}_\ell$ the power and the current that flow across the longitudinal elements of a network line $\ell$ from the receiving toward the sending end, for which it holds that
\begin{align}
&  \bar{I}_{\ell}  =  \bar{Y}_\ell (\bar{V}_{\beta(\ell^+)} -\bar{V}_{\beta(\ell^-)}), \quad \forall \ell \in \mathcal{L} \label{eq:Il}\\
&  \bar{S}_{\ell}  =  \bar{V}_{\beta(\ell^+)}\ubar{I}_{\ell}, \quad \forall \ell \in \mathcal{L} \label{eq:Sl}
\end{align}
The power and current flows along the shunt elements of the lines are taken into account in the bus power balance constraints as nodal injections. In this direction, we denote by $\bar{Y}_{b_0}$ the sum of all the shunt elements of the lines that are adjacent to bus $b$. In particular, the notation used for the BFM convexification of the OPF problem is shown in Fig.~\ref{fig:notation_bfm}.

Using this nomenclature, the constraints of the OPF problem are reformulated as follows:

\begin{align}
&\sum_{g \in b} \bar{S}_g - \sum_{c \in b} \bar{S}_c =  \sum_{\mathclap{\beta(\ell^+)=b}} \bar{S}_{\ell} -\sum_{\mathclap{\beta(\ell^-)=b}} (\bar{S}_{\ell}-\bar{Y}^{-1}_\ell |\bar{I}_\ell|^2)-\bar{Y}_{b_0} |\bar{V}_b|^2, \, \forall b \in \mathcal{B} \label{eq:Sbal_BFM} \\
& |\bar{I}_\ell|^2 = \frac{|\bar{S}_\ell|^2}{|\bar{V}_{\beta(\ell^+)}|^2}, \, \forall \ell \in \mathcal{L}\label{eq:quad_eq_BFM} \\
&  |\bar{V}_{\beta(\ell^-)}|^2 = |\bar{V}_{\beta(\ell^+)}|^2 + |\bar{Y}^{-1}_\ell|^2 |\bar{I}_\ell|^2-(\bar{Y}^{-1}_\ell\ubar{S}_\ell+\ubar{Y}^{-1}_\ell\bar{S}_\ell),\, \forall \ell \in \mathcal{L}\label{eq:curr_elim_BFM}\\
& V_{min}^2 \leq |\bar{V}_b|^2 \leq V_{max}^2, \, \forall b \in \mathcal{B} \label{eq:volt_limits_BFM}\\
& |\bar{I}_{\ell}|^2 \leq I_{\ell_{max}}^2, \, \forall \ell \in \mathcal{L} \label{eq:OPF_lineflows_BFM}\\
&Re(\bar{S}_g) \in [P_{g_{min}}, P_{g_{max}}] \, , Im(\bar{S}_g) \in [Q_{g_{min}}, Q_{g_{max}}], \, \forall g \in \mathcal{G} \label{eq:cap_curve_BFM} \\
&Re(\bar{S}_c) \in [P_{c_{min}}, P_{c_{max}}] \, , Im(\bar{S}_c) \in [Q_{c_{min}}, Q_{c_{max}}], \, \forall c \in \mathcal{C} \label{eq:cap_curve_cons_BFM}
\end{align}

Note that in this formulation of the OPF problem, the
capability curves of the controllable loads and generators, i.e.,
constraints (\ref{eq:cap_curve_BFM},\ref{eq:cap_curve_cons_BFM}) on the nodal power $\bar S$
are limited to rectangular regions. This is essential for the conic relaxation
proposed in~\cite{farivar2013branch,6507352}.

Starting from this formulation, the equality constraints in (\ref{eq:quad_eq_BFM}) are relaxed to inequalities and the aforementioned problem is casted as a second-order cone program. They also prove that for radial networks a global solution of the original OPF problem can be recovered from the solution of the relaxed problem if there are no upper bounds on the loads. In other words, the OPF problem is solved (\ref{eq:Sbal_BFM})-(\ref{eq:cap_curve_cons_BFM}) by setting $P_{c_{max}}{=}\infty$ and $Q_{c_{max}}{=}\infty$ in constraint (\ref{eq:cap_curve_cons_BFM}).

We show, in what follows, that this formulation is not equivalent to (\ref{eq:OPF-class_obj}-\ref{eq:cap_curve}). In particular, constraint (\ref{eq:OPF_lineflows_BFM}) (constraint (9) in~\cite{farivar2013branch}) is only an approximation of the ampacity constraints and, moreover, the assumptions on the controllability and bounds of the energy resources in the network are unrealistic.

\subsection{Misinterpretation of the Physical Network Model in the BFM-based OPF Formulation}

The branch-flow model has been often used in load-flow studies (e.g.,~\cite{107303,52734}) and constitutes an accurate representation of the network model. The first problem with the formulation in (\ref{eq:Sbal_BFM})-(\ref{eq:cap_curve_cons_BFM}) is that it misinterprets the physical network model when constraining the line flows in the network. Even though the power-flow equations in (\ref{eq:Sbal_BFM})-(\ref{eq:curr_elim_BFM}) are exact when the shunt capacitances are considered as nodal injections, the constraint (\ref{eq:OPF_lineflows_BFM}) is imposed on a fictitious current flow across the longitudinal component of the lines, thus \textit{does not} account for the current flow toward the shunt elements. Therefore, the optimum of problem (\ref{eq:Sbal_BFM})-(\ref{eq:cap_curve_cons_BFM}) can be such that the line ampacity constraint is violated.

\begin{figure}[b!]
\begin{centering}
\includegraphics[width=1\linewidth, clip=true, trim=0 500 0 180]{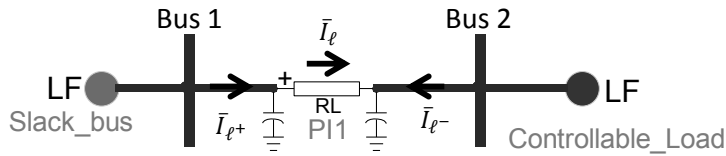}
\caption{{\color{black}{The test network used for the numerical comparison of the current flows at the sending/receiving end of the lines and the current flow along the longitudinal line impedance.}}}
\label{fig:network_bfm_EMTP}
\end{centering}
\end{figure}

\begin{figure}
\begin{centering}
\includegraphics[width=1\linewidth, clip=true, trim=0 0 0 70]{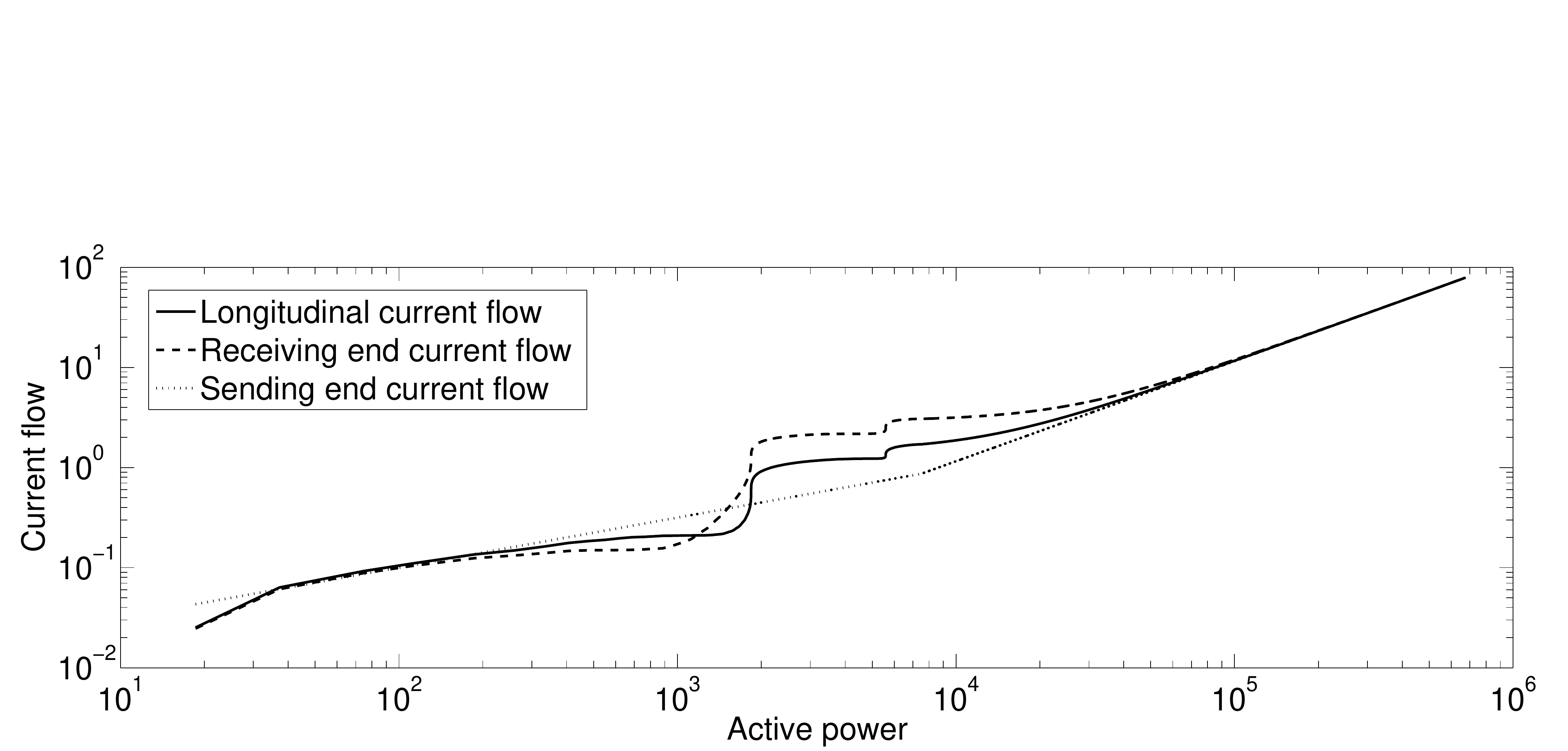}
\vspace{-25pt}
\caption{{\color{black}{Current flows at the sending/receiving end of the line and along the longitudinal line impedance (log-log scale).}}}
\label{fig:bfm_current_mismatch}
\end{centering}
\end{figure}

To better clarify why this occurs, we use a single-branch toy network, as shown in Fig.~\ref{fig:network_bfm_EMTP}.
The line parameters, as well as the base values of the system are given in Table~\ref{bfm_emtp_param}. A purely resistive load is connected to bus $2$ that we vary linearly in the range of $[100-10000]$Ohms in order to numerically quantify the mismatch between those quantities. We measure the current flows at the two ends of the line, as well as the flow along the longitudinal impedance of the line. Fig.~\ref{fig:bfm_current_mismatch} shows the measured quantities as a function of the load. It can be observed that the current flowing across the longitudinal impedance of the line under-estimates the actual current flow in the receiving end of the line.

\begin{table}
\begin{center}
\caption{Parameters of the test network in Fig.\ref{fig:network_bfm_EMTP}}
\label{bfm_emtp_param}
\begin{tabular}{|c|c|}
  \hline
  \textbf{Parameter} &  \textbf{Value} \\ \hline
  Network rated voltage, $V$(kV) & 15  \\ \hline
  Line parameters, $R$(Ohms), $L$(H), $C$(uF) & (1,0.003,0.54)\\ \hline
\end{tabular}
\end{center}
\end{table}

\begin{figure}[t!]
\begin{centering}
\includegraphics[width=1\linewidth, clip=true, trim=120 240 100 200]{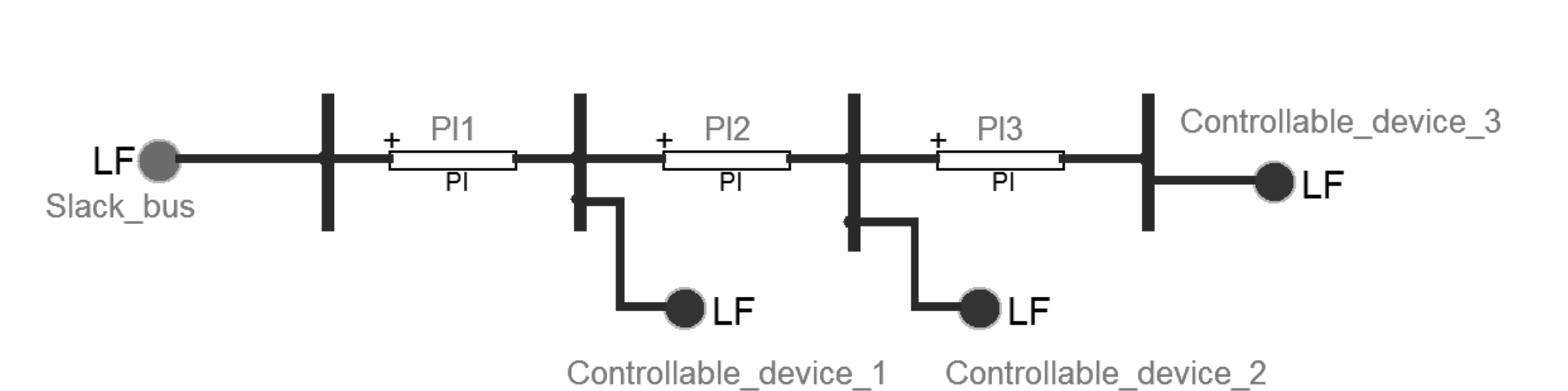}
\vspace{-25pt}
\caption{Network used in the study of the BFM-based OPF formulation.}
\label{fig:network_bfm_EMTP_OPF}
\end{centering}
\end{figure}

As a consequence, in this formulation of the OPF problem setting the limit on the longitudinal current flow below the line ampacity does not guarantee that the actual line current will respect this limit. In order to illustrate such a scenario, we consider yet another simple test network shown in Fig.~\ref{fig:network_bfm_EMTP_OPF}.
All the network lines are built by using the same values of resistance, reactance and capacitance per km, but by assuming different values of their length\footnote{{\color{black}{Typical values of medium-voltage underground cables are considered for the resistance, reactance and shunt capacitances of the lines taken from~\cite{nexans}.}}}. We assume a first test case where the controllable device connected to bus $4$ is a generator, whereas controllable loads are connected to buses $2$ and $3$.  The network characteristics, the base values, the capability limits of the controllable resources\footnote{The upper bounds of the active and reactive power of the loads are considered to be infinite, as required in~\cite{farivar2013branch,6507352}.}, and the voltage and ampacity bounds are provided in Table~\ref{bfm_emtp_OPF_param}. We assume that the controllable generation operates at a unity power factor. The problem in (\ref{eq:Sbal_BFM})-(\ref{eq:cap_curve_cons_BFM}) is formulated and solved in Matlab. The objective function accounts for loss minimization, as well as utility maximization of the controllable generation units:

\begin{align}
\label{eq:OPF-BFM_obj}
\min_{\bar{S}_g,\bar{S}_\ell,|\bar{V}_b|,|\bar{I}_\ell|} -\sum_{g \in \mathcal{G}} Re(\bar{S}_{g}) + \sum_{\ell \in \mathcal{L}} Re(\bar{Y}_\ell)|\bar{I}_\ell|^2
\end{align}

\begin{table}
\begin{center}
\caption{Parameters of the test network in Fig.\ref{fig:network_bfm_EMTP_OPF} used for the investigation of the line ampacity limit violation}
\label{bfm_emtp_OPF_param}
\begin{tabular}{|c|c|}
  \hline
  \textbf{Parameter} &  \textbf{Value} \\ \hline
  Network rated voltage and base power, $V$(kV),$S$(MVA)& 24.9,5  \\ \hline
  Line parameters, $R$(Ohms/km), $L$(mH/km), $C$(uF/km) & (0.193,0.38,0.24)\\ \hline
  {\color{black}{Lines length, (km)}} & {\color{black}{(2.5,3,3.5)}}\\ \hline
  $[P_{g_{min}},P_{g_{max}}]$ (MW)& $[0,2]$\\ \hline
  $P_{c_{min}}$(MW) (bus2, bus3)& $(0.05,0.06)$\\ \hline
 $Q_{c_{min}}$(Mvar) (bus2, bus3)& $(0.03,0.027)$\\ \hline
  $[V_{min},V_{max}]$ (p.u)& $[0.9,1.1]$\\ \hline
  $I_{max}$ (A)& $80$\\ \hline
\end{tabular}
\end{center}
\end{table}

\begin{figure}
\begin{centering}
\includegraphics[width=1\linewidth, clip=true, trim=0 0 0 70]{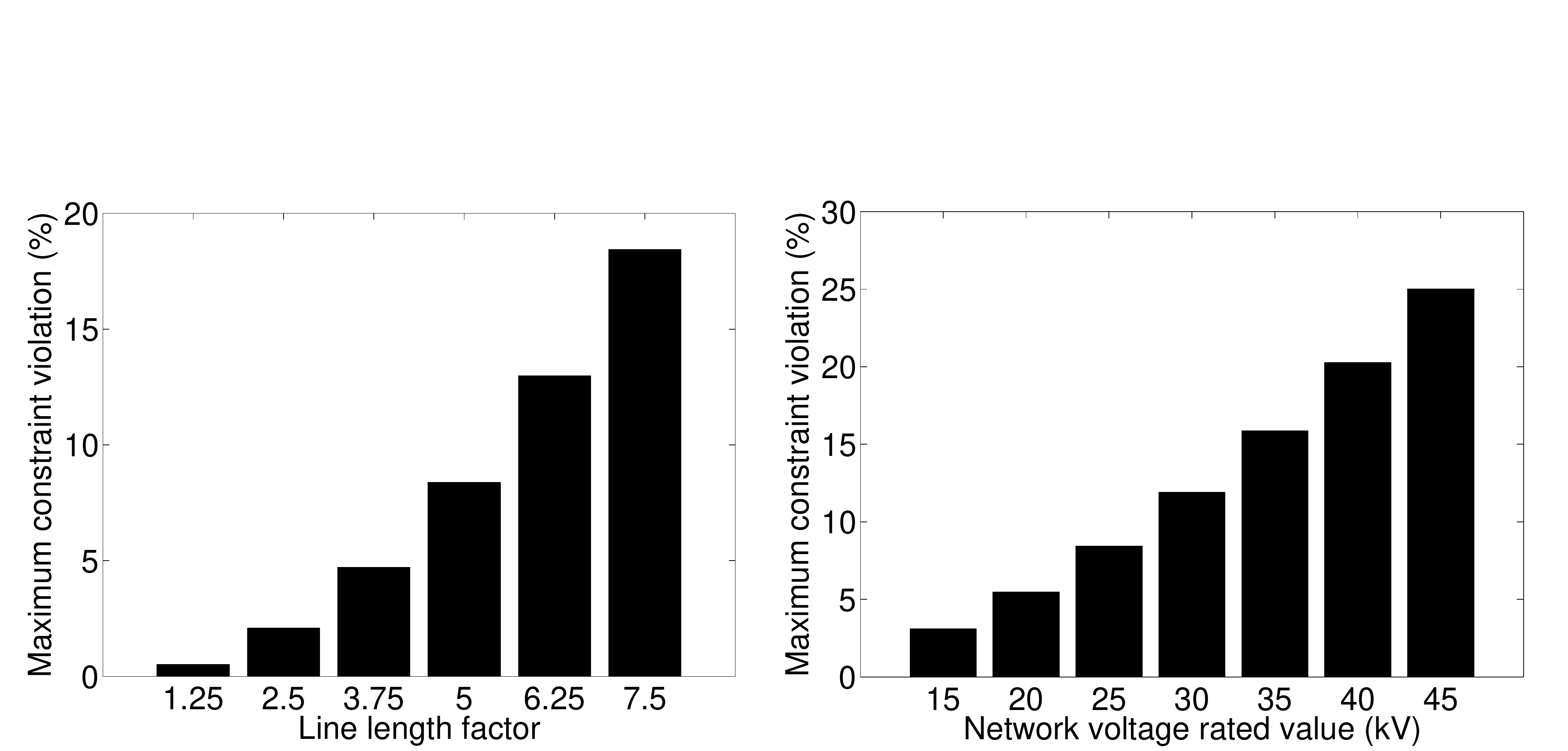}
\vspace{-25pt}
\caption{Maximum ampacity constraint violation as a function of the line lengths and the network voltage rated value.}
\label{fig:current_bfm_EMTP_OPF_op_point}
\end{centering}
\end{figure}

In order to investigate the order of magnitude of the violation of the ampacity constraint, we solve the OPF problem for various line lengths and network voltage-rated values. In particular, we assume that the line lengths are uniformly multiplied by a factor in the range $[1.25-7.5]$ (while keeping the network voltage rated value to its nominal value) and the network voltage rated value varies in the range $[15-45]kV$ (while keeping the line lengths to their nominal values). Once the optimal solution is computed in each case, we calculate the actual current flows in the sending/receiving end of the lines and we compute the maximum constraint violation. The results are shown in Fig.~\ref{fig:current_bfm_EMTP_OPF_op_point}. As the line length increases, the current flowing toward the shunt capacitors increases, thus neglecting its contribution to the line flow leads to significant violations of the ampacity limit. At $7.5$ times the initial line length, the violation reaches a value of $18.4\%$. The effect of the network voltage-rated value is similar, with a maximum constraint violation of $25\%$ when the voltage value is 45kV.

\begin{figure}
\begin{centering}
\includegraphics[width=1\linewidth, clip=true, trim=0 0 10 70]{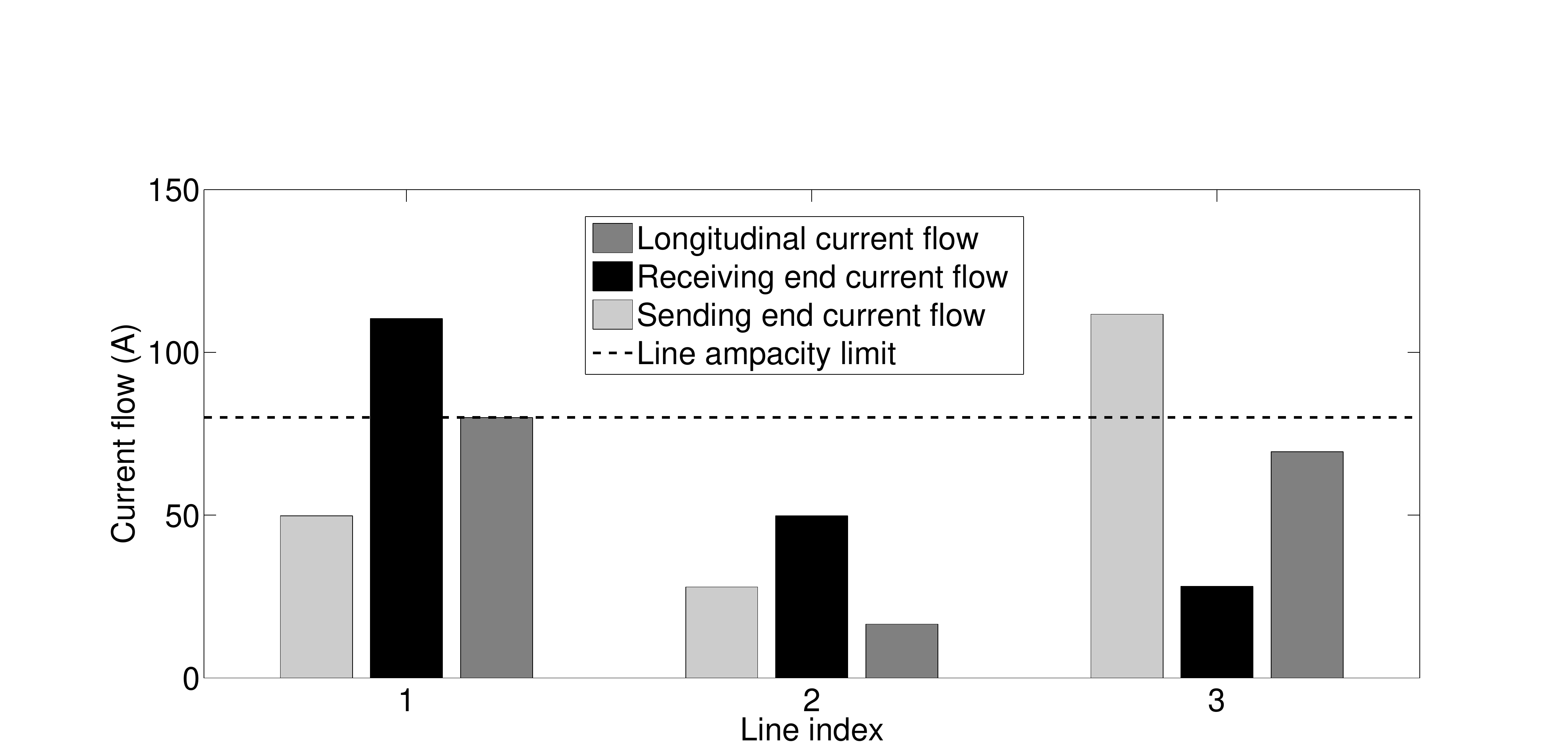}
\vspace{-25pt}
\caption{OPF solution for the current flows at the sending/receiving end of the network lines and across
the longitudinal line impedance under heavy consumption and light generation conditions.}
\label{fig:current_bfm_EMTP_OPF}
\end{centering}
\end{figure}

In addition to the effect of the line lengths and the network voltage-rated value, we study the effect of the network operating point on the ampacity violation. To this end, we consider a second test case where the controllable device connected to bus $4$ is a load and generators are connected to buses $2$ and $3$. The capability limits of the controllable resources are provided in Table~\ref{bfm_emtp_OPF_param_op_state}. For this setting, Fig.~\ref{fig:current_bfm_EMTP_OPF} shows the solution of the BFM-based OPF problem, namely current flows at the receiving/sending end of the network lines, as well as across the longitudinal impedance. We can observe that the maximum violation of the ampacity constraint is in the order of $39.6\%$.

\begin{table}
\begin{center}
\caption{Parameters of the test network in Fig.\ref{fig:network_bfm_EMTP_OPF} used for the investigation of the network operating point on the line ampacity limit violation}
\label{bfm_emtp_OPF_param_op_state}
\begin{tabular}{|c|c|}
  \hline
  \textbf{Parameter} &  \textbf{Value} \\ \hline
  $[P_{g_{min}},P_{g_{max}}]$(MW) (bus 2)& $[0,0.01]$\\ \hline
  $[P_{g_{min}},P_{g_{max}}]$(MW) (bus 3)& $[0,0.012]$\\ \hline
  $(P_{c_{min}},Q_{c_{min}})$(MW,Mvar) (bus 4)& $0.3,0.15$\\ \hline
\end{tabular}
\end{center}
\end{table}
In order to avoid current flows that exceed the lines' ampacity limits, i.e., in order to use the BFM in an accurate way, the aforementioned formulation should either consider the actual current flows in the receiving/sending ends of the lines as optimization variables, or should add the contribution of the current flows toward the shunt elements of the lines to the longitudinal current flow in the inequality constraint (\ref{eq:OPF_lineflows_BFM}). By adopting either of the two approaches, however, (\ref{eq:Sbal_BFM})-(\ref{eq:cap_curve_cons_BFM}) can no longer be solved efficiently as proposed in~\cite{farivar2013branch,6507352}. Therefore, the generic OPF problem cannot be convexified by using the approach in~\cite{farivar2013branch,6507352}.

\subsection{On the Assumptions Required for the Exactness of the SOCP Relaxation}
{\color{black}{In addition to the aforementioned fundamental problem, which is related to the physical network model, the proof of exactness of the proposed SOCP relaxation in~\cite{farivar2013branch,6507352} requires specific assumptions, related to the controllability of the demand in the network. Several of these assumptions might not be realistic. The goal of this section is to discuss these assumptions and their consequences in a realistic setting. In order to do so, we consider realistic case-studies and we show that the solution of the OPF problem can result in unrealistic values for the control variables.}}

\begin{figure}
\begin{centering}
\includegraphics[width=1\linewidth, clip=true, trim=0 5 10 0]{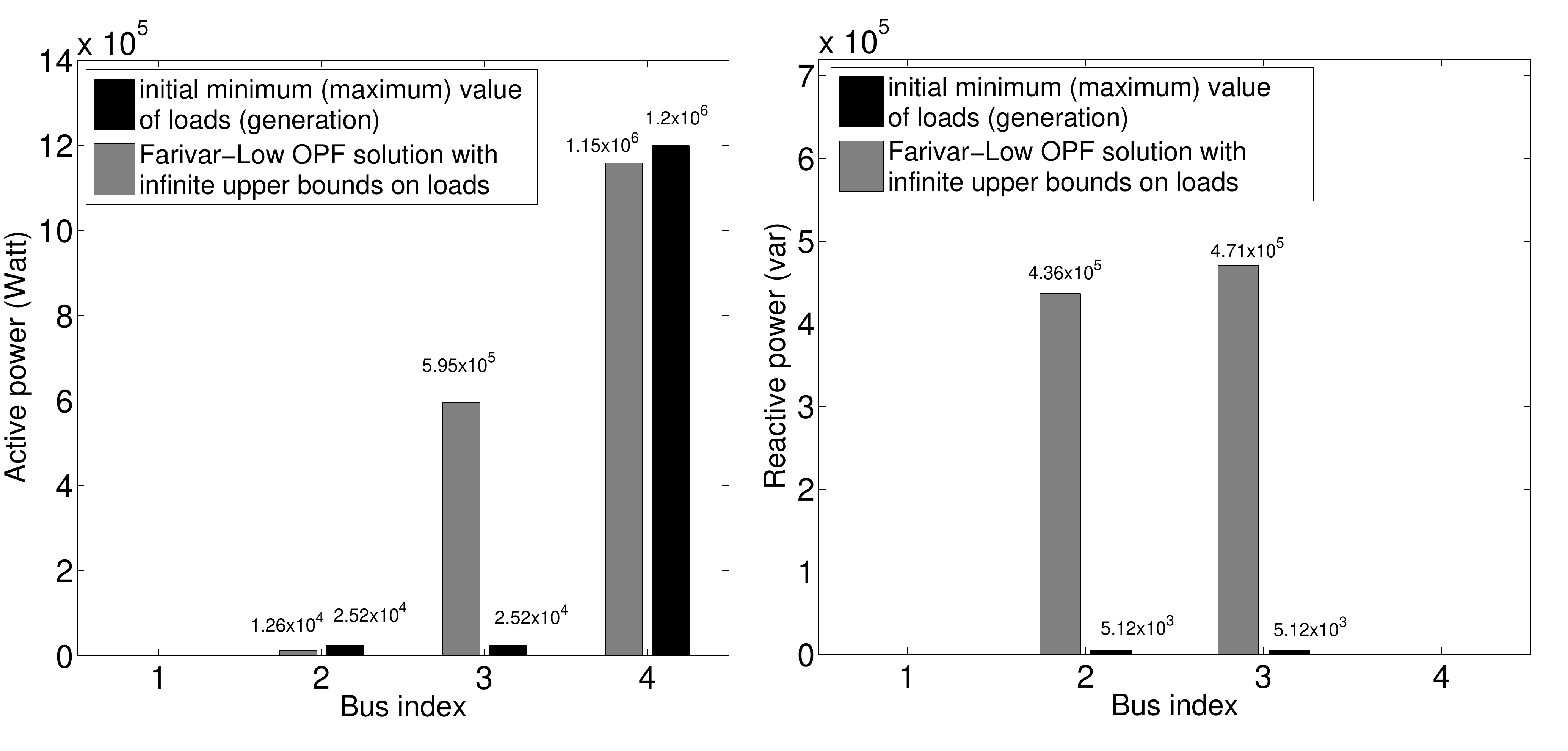}
\vspace{-25pt}
\caption{Optimal solution of the OPF formulation for the active and reactive power set-points when upper bounds on loads are infinite.}
\label{fig:unbounded_loads}
\end{centering}
\end{figure}

To begin with, the OPF formulation in~\cite{farivar2013branch,6507352} assumes controllability of both loads and generators in the network buses and, in particular, assumes rectangular bounds on the powers of loads/generators. This is quite a strong assumption, as usually the DNO has very few specific control points available in the network with capability curves that are typically more complex and that account, among others, for capabilities of power electronics and limitations of machinery. An even more serious limitation is that the model in~\cite{farivar2013branch,6507352} considers no upper bounds on the controllable loads in order to prove the exactness of the proposed relaxation. This implies that in cases where excessive production of the generators causes violations of the voltage or line-flows limits, local demand is invoked to compensate for the increased generation. In order to illustrate such a setting and to show that the result of the OPF problem can result in unrealistic values for demand, we consider the same network in Fig.~\ref{fig:network_bfm_EMTP_OPF} and we assume that there is high penetration of distributed generation and a low demand. The values of loads and generation, as well as the corresponding limits are shown in Table~\ref{bfm_emtp_OPF_param_unb}. Solving the optimization problem and considering infinite upper bounds on the demand results in load values that are significantly increased, compared to the minimum values shown in Table~\ref{bfm_emtp_OPF_param_unb}.
\begin{table}
\begin{center}
\caption{Parameters of the test network in Fig.\ref{fig:network_bfm_EMTP_OPF} used for the investigation of the unboundedness of the consumption}
\label{bfm_emtp_OPF_param_unb}
\begin{tabular}{|c|c|}
  \hline
  \textbf{Parameter} &  \textbf{Value} \\ \hline
  $[P_{g_{min}},P_{g_{max}}]$ (MW)& $[0,1.2]$\\ \hline
  $(P_{c_{min}},Q_{c_{min}})$ (MW,Mvar) (buses 2,3) & $(0.0125,0.0026)$\\ \hline
\end{tabular}
\end{center}
\end{table}
The resulting optimal power points are shown in Fig.~\ref{fig:unbounded_loads}. We show in black the initial values for active and reactive power of loads and generation (corresponding to the values of Table~\ref{bfm_emtp_OPF_param_unb}), and in gray the results of the OPF solution (when not accounting for upper bounds on loads). It is worth observing that the optimal active power consumption of bus 3 is increased $23.6$ times and the reactive power consumption at buses 2 and 3 is increased $85.3$ and $92$ times, respectively. In a realistic setting, even if part of the demand in the network is controllable, the amount of available demand-response is limited and such an increase in the consumption is most likely not possible. Therefore, in such a case, the congestion and voltage problems should be solved by properly controlling the generator within its capability limits. In addition to this, typically, the active and reactive power consumption should be linked via the corresponding power factor. We observe, however, that the OPF solution in this scenario results in very large values for the reactive power consumption and, in particular, the power factor of bus 2 is $0.03$ after the OPF solution, whereas initially its value is $0.98$.
In an attempt to relax this assumption, it is shown in~\cite{gan2012branch} that the infinite upper bound on the loads, when not applicable, can be replaced by equivalent conditions. However, not only are these conditions unrealistic, they are also not applicable in our context as they require no upper bound on the voltage magnitudes. This is in contradiction with the actual problem we target, i.e., voltage rise due to high penetration of renewable energy resources.

\subsection{On the Extension of the SOCP Relaxation to Networks with Lines Modeled as $\pi$-equivalents}
In this paragraph, we discuss two different approaches that can be used to extend the initial formulation in (\ref{eq:Sbal_BFM})-(\ref{eq:cap_curve_cons_BFM}) in order to properly account for the shunt elements of the lines and the line ampacity constraints. {\color{black}{The goal of this paragraph is to show through concrete examples that extending the approach in~\cite{farivar2013branch,6507352} to a system model that is correctly represented results in a convex problem, however it cannot guarantee the exactness of the SOCP relaxation and, thus, the retrieval of a feasible OPF solution.

The first straightforward way to properly account for the line ampacity constraints in (\ref{eq:OPF_lineflows_BFM}) is to include in these inequality constraints the contribution of the current flowing towards the shunt elements of the line. In order to do so, we keep the same branch flow variables of the initial formulation in (\ref{eq:Sbal_BFM})-(\ref{eq:cap_curve_cons_BFM}) and we define a new set of line constraints for the case of $\pi$-model lines. In this case, the total line flowing, for instance, in the receiving end of the line is the sum of the longitudinal current plus the contribution of the shunt, resulting in the following constraint:
\begin{align}
& |\bar{I}_{\ell} + \bar{Y}_{\ell_0^+}\bar{V}_{\beta(\ell^+)}|^2\leq I_{\ell_{max}}^2, \, \forall \ell \in \mathcal{L} \label{eq:OPF_lineflows_wshunt}
\end{align}

Similarly for the sending end of the lines. Now, after expanding the square, the line constraints (\ref{eq:OPF_lineflows_BFM}) are reformulated as follows\footnote{{\color{black}{Note that with the inclusion of the shunt elements of the lines two inequality constraints are required per line in order to properly account for the ampacity limits. The reason is that in this case, the currents at the two ends of the line are no longer equal and both need to be constrained below the line ampacity limit.}}}:
\begin{align}
& |\bar{I}_{\ell}|^2 + |\bar{Y}_{\ell_0^+}|^2|\bar{V}_{\beta(\ell^+)}|^2+2Re(\bar{Y}_{\ell_0^+}\bar{S}_{\ell})\leq I_{\ell_{max}}^2, \, \forall \ell \in \mathcal{L} \label{eq:OPF_lineflows_BFM_wshunt}\\
& |\bar{I}_{\ell}|^2 + |\bar{Y}_{\ell_0^-}|^2|\bar{V}_{\beta(\ell^-)}|^2+2Re(\bar{Y}_{\ell_0^-}(\bar{Y}^{-1}_\ell|\bar{I}_{\ell}|^2 -\bar{S}_{\ell}))\leq I_{\ell_{max}}^2, \, \forall \ell \in \mathcal{L} \label{eq:OPF_lineflows_BFMminus_wshunt}
\end{align}

It is worth noting that these new constraints, that account also for the line flows towards the shunt elements, are still convex in the branch-flow variables, i.e., ($|\bar{V}_{\beta(\ell^+)}|^2,|\bar{I}_{\ell}|^2,Re(\bar{S}_{\ell}),Im (\bar{S}_{\ell})$). Therefore, these constraints can be added to the initial formulation in~\cite{farivar2013branch,6507352} without losing convexity. The main question now is whether the SOCP relaxation continues to be exact with the reformulation of the lines ampacity constraints as above, i.e., whether the optimal solution of the relaxed SOCP problem is guaranteed to be a physically feasible one. As we show below, there are cases for which the attained solutions can be physically infeasible. In other words, even with a correct system model, the proposed SOCP relaxation in ~\cite{farivar2013branch,6507352}  cannot guarantee a solution that is meaningful for the grid operation.}}

Let us consider the following simple example. We use once again the simple test network shown in Fig.~\ref{fig:network_bfm_EMTP_OPF}. We assume a test case where the controllable device connected to bus $4$ is a load, whereas controllable generators are connected to buses $2$ and $3$.  The network characteristics, the base values, the capability limits of the controllable resources, and the voltage and ampacity bounds are provided in Table~\ref{bfm_emtp_OPF_param2}. We assume that the controllable generators operate at a unity power factor. Note that the upper bounds for the loads are considered infinite as required in the original formulation in~\cite{farivar2013branch,6507352}.

\begin{table}[b!]
\begin{center}
\caption{Parameters of the test network in Fig.\ref{fig:network_bfm_EMTP_OPF}}
\label{bfm_emtp_OPF_param2}
\begin{tabular}{|c|c|}
  \hline
  \textbf{Parameter} &  \textbf{Value} \\ \hline
  Network rated voltage and base power, $V$(kV),$S$(MVA)& 24.9,5  \\ \hline
  Line parameters, $R$(Ohms/km), $L$(mH/km), $C$(uF/km) & (0.193,0.38,0.24)\\ \hline
  $P_{g_{max}}$ (bus2, bus3)(MW)& $(1,1.2)$\\ \hline
  $P_{c_{min}}$(MW) (bus4)& $0.1$\\ \hline
 $Q_{c_{min}}$(Mvar) (bus4)& $0.05$\\ \hline
  $[V_{min},V_{max}]$ (p.u)& $[0.9,1.1]$\\ \hline
  $I_{max}$ (A)& $80$\\ \hline
\end{tabular}
\vspace{-0.2cm}
\end{center}
\end{table}
We solve the problem in Matlab, using the interior-point algorithm provided by the \textit{fmincon} solver, and the resulting values for the SOCP inequalities for each network line are shown in Table~\ref{table:SOCP_case2}. In this case, after the solution of the OPF problem not all the inequalities in (\ref{eq:quad_eq_BFM}) are satisfied with equality, namely the SOCP relaxation is inexact, therefore, the obtained solution has no physical meaning and a physically feasible solution cannot be recovered.

\begin{table}
\begin{center}
\caption{SOCP inequalities in (\ref{eq:quad_eq_BFM})}
\label{table:SOCP_case2}
\begin{tabular}{|c|c|}
  \hline
  \textbf{Line} &  \textbf{Value} \\ \hline
  $1-2$& $1.7E-16$\\ \hline
  $2-3$& $-0.0461$\\ \hline
  $3-4$& $ 1.3E-16$\\ \hline
\end{tabular}
\vspace{-0.6cm}
\end{center}
\end{table}

A second way to extend the approach in~\cite{farivar2013branch} to networks with lines represented as $\pi$-equivalents is to reconstruct the BFM in order to include the shunt elements of the lines. To this end, we consider an undirected graph $G=(\mathcal{N},\mathcal{E})$. $\mathcal{N}$ and $\mathcal{E}$ represent the set of nodes and lines respectively. 
{\color{black}{The power flow equations that define the branch-flow model with the inclusion of shunt elements are in this case:
\begin{align}
&\label{pbalbfm2}\sum_{g \in b} \bar{S}_g - \sum_{c \in b} \bar{S}_c - \sum_{\mathclap{\beta(\ell^+)=b}} \bar{S}_{\ell^+} -\sum_{\mathclap{\beta(\ell^-)=b}} \bar{S}_{\ell^-}=0, \, \forall b \in \mathcal{B} \\
&\label{scopbfm2}\bar{S}_{\ell^+}=\bar{V}_{\beta(\ell^+)}\ubar{I}_{\ell^+}, \quad \bar{S}_{\ell^-}=\bar{V}_{\beta(\ell^-)}\ubar{I}_{\ell^-}, \,\quad  \forall \ell \in \mathcal{L}  \\
&\bar{I}_{\ell^+}  =  \bar{Y}_\ell (\bar{V}_{\beta(\ell^+)} -\bar{V}_{\beta(\ell^-)} )+ \bar{Y}_{\ell^+_0} \bar{V}_{\beta(\ell^+)},\quad \forall \ell \in \mathcal{L} \label{eq:Ilpbfm2}\\
&\bar{I}_{\ell^-}  = \bar{Y}_\ell (\bar{V}_{\beta(\ell^-)} -\bar{V}_{\beta(\ell^+)} )+ \bar{Y}_{\ell^-_0} \bar{V}_{\beta(\ell^-)}, \quad \forall \ell \in \mathcal{L} \label{eq:Ilmbfm2}
\end{align}

Note that, contrary to the formulation in(\ref{eq:Il})-(\ref{eq:Sl}), the power flows and current flows variables are now defined for both ends of the lines.

Similarly to the procedure described in Section III.A in~\cite{farivar2013branch}, we substitute (\ref{eq:Ilpbfm2}) and (\ref{eq:Ilmbfm2}) in (\ref{scopbfm2}) resulting in the following set of equations per line:
\begin{align}
&(\bar{Y}_\ell + \bar{Y}_{\ell^+_0})\bar{V}^2_{\beta(\ell^+)}=\bar{S}_{\ell^+}+\bar{Y}_\ell\bar{V}_{\beta(\ell^+)}\bar{V}_{\beta(\ell^-)}\label{eq:Ilpbfmrel}\\
&(\bar{Y}_\ell + \bar{Y}_{\ell^-_0})\bar{V}^2_{\beta(\ell^-)}=\bar{S}_{\ell^-}+\bar{Y}_\ell\bar{V}_{\beta(\ell^+)}\bar{V}_{\beta(\ell^-)}\label{eq:Ilmbfmrel}
\end{align}

Then taking the magnitude of the resulting equations squared, we can eliminate the angles from (\ref{eq:Ilpbfmrel})-(\ref{eq:Ilmbfmrel}).  Doing so, the OPF problem in (\ref{eq:Sbal_BFM})-(\ref{eq:cap_curve_cons_BFM}) is reformulated as:}}
\begin{align}
\label{obj}
&\min_{\substack{\bar{S}_g,\bar{S}_c,\bar{S}_{\ell^+},|\bar{I}_{\ell^+}|^2,|\bar{V}_{\beta(\ell^+)}|^2,\\\bar{S}_{\ell^-},|\bar{I}_{\ell^-}|^2,|\bar{V}_{\beta(\ell^-)}|^2}} \sum_{g \in \mathcal{G}} C_g(\bar{S}_g)+\sum_{c \in \mathcal{C}} C_c(\bar{S}_c)\\
&\label{pbal}\text{subject to: }\sum_{g \in b} \bar{S}_g - \sum_{c \in b} \bar{S}_c - \sum_{\mathclap{\beta(\ell^+)=b}} \bar{S}_{\ell^+} -\sum_{\mathclap{\beta(\ell^-)=b}} \bar{S}_{\ell^-}=0, \, \forall b \in \mathcal{B} \\
&\label{scop}|\bar{S}_{\ell^+}|^2 -|\bar{V}_{\beta(\ell^+)}|^2|\bar{I}_{\ell^+}|^2\leq 0, \quad |\bar{S}_{\ell^-}|^2 -|\bar{V}_{\beta(\ell^-)}|^2|\bar{I}_{\ell^-}|^2\leq 0, \,\quad  \forall \ell \in \mathcal{L}  \\
&\label{vplus}|\alpha_{\ell^+}|^2|\bar{V}_{\beta(\ell^+)}|^2-|\bar{V}_{\beta(\ell^-)}|^2 = 2Re(\alpha_{\ell^+}\ubar{Y}^{-1}_\ell \bar{S}_{\ell^+})-|\bar{Y}^{-1}_\ell||\bar{I}_{\ell^+}|^2, \,\quad  \forall \ell \in \mathcal{L} \\
&\label{vminus}|\alpha_{\ell^-}|^2|\bar{V}_{\beta(\ell^-)}|^2-|\bar{V}_{\beta(\ell^+)}|^2 = 2Re(\alpha_{\ell^-}\ubar{Y}^{-1}_\ell \bar{S}_{\ell^-})-|\bar{Y}^{-1}_\ell||\bar{I}_{\ell^-}|^2, \,\quad  \forall \ell \in \mathcal{L}
\end{align}

where $\alpha_{\ell^+} := 1+\bar{Y}^{-1}_\ell \bar{Y}_{\ell_0^+} $, $\alpha_{\ell^-} :=  1+\bar{Y}^{-1}_\ell \bar{Y}_{\ell_0^-}$.

At this point it is important to note that, in order to recover a solution of the original ACOPF problem, we need to recover the line angle from the solution of the above problem in a way similar to~\cite{farivar2013branch}. In the original paper, it is shown that for radial distribution networks the angle relaxation step is always exact. On the contrary, in the formulation above, both angles $\beta_{\ell^+}=\angle(\ubar{\alpha}_{\ell^+}|\bar{V}_{\beta(\ell^+)}|^2-\ubar{Y}^{-1}_\ell\bar{S}_{\ell^+})$ and $\beta_{\ell^-}=\angle(\ubar{\alpha}_{\ell^-}|\bar{V}_{\beta(\ell^-)}|^2-\ubar{Y}^{-1}_\ell\bar{S}_{\ell^-})$ are defined. In order for a solution to be physically meaningful, namely the angle relaxation step to be exact, these line angles should satisfy $\beta_{\ell^+}+\beta_{\ell^-}=0$. However, there is no guarantee that this will occur in the obtained solution. In fact, as we show in the example that follows the angle relaxation is not exact when using this formulation even in the case of radial networks.

To support the above claim, we consider the same simple test network shown in Fig.~\ref{fig:network_bfm_EMTP_OPF}. For the sake of simplicity, we assume a test case where the controllable device connected to bus $4$ is a generator, whereas controllable loads are connected to buses $2$ and $3$.  The network characteristics, the base values, the capability limits of the controllable resources, and the voltage and ampacity bounds are provided in Table~\ref{bfm_emtp_OPF_param}. We assume that the controllable generation operates at a unity power factor. Note that the upper bounds for the loads are considered infinite as in the original paper (\cite{farivar2013branch}).\\

\begin{table}
\begin{center}
\caption{Parameters of the test network in Fig.\ref{fig:network_bfm_EMTP_OPF}}
\label{bfm_emtp_OPF_param}
\begin{tabular}{|c|c|}
  \hline
  \textbf{Parameter} &  \textbf{Value} \\ \hline
  Network rated voltage and base power, $V$(kV),$S$(MVA)& 24.9,5  \\ \hline
  Line parameters, $R$(Ohms/km), $L$(mH/km), $C$(uF/km) & (0.193,0.38,0.24)\\ \hline
  $[P_{g_{min}},P_{g_{max}}]$ (MW)& $[0,2]$\\ \hline
  $P_{c_{min}}$(MW) (bus2, bus3)& $(0.05,0.06)$\\ \hline
 $Q_{c_{min}}$(Mvar) (bus2, bus3)& $(0.03,0.027)$\\ \hline
  $[V_{min},V_{max}]$ (p.u)& $[0.9,1.1]$\\ \hline
  $I_{max}$ (A)& $80$\\ \hline
\end{tabular}
\vspace{-0.2cm}
\end{center}
\end{table}

The objective function has two terms, namely loss minimization and utility maximization of the controllable generation unit:
\begin{align}
&\min_{\substack{\bar{S}_g,\bar{S}_c,\bar{S}_{\ell^+}, |\bar{I}_{\ell^+}|^2, |\bar{V}_{\ell^+}|^2,\\ \bar{S}_{\ell^-}, |\bar{I}_{\ell^-}|^2,|\bar{V}_{\ell^-}|^2}} \sum_{\ell \in \mathcal{L}}Re(\bar{Y}^{-1}_\ell)( |\bar{I}_{\ell^+}|^2+ |\bar{V}_{\ell^+}|^2{\bar{Y}_{\ell_0^+}}^2-Im(\bar{S}_{\ell^+}))+\\
&Re(\bar{Y}^{-1}_\ell) (|\bar{I}_{\ell^-}|^2+ |\bar{V}_{\ell^-}|^2{\bar{Y}_{\ell_0^-}}^2-Im(\bar{S}_{\ell^-}))-\sum_{g \in \mathcal{G}}Re(\bar{S}_g)
\end{align}

It is worth mentioning that in view of the new formulation, the current used for the real losses computation is no longer represented by the variable $|\bar{I}_{\ell}|^2$, but needs to be computed as the difference between the current flowing from one end of the line to the other and the current flowing towards the shunt elements. This has two implications. First, the current across the series admittance can be computed twice using $|\bar{I}_{\ell^+}|^2$ or $|\bar{I}_{\ell^-}|^2$.  Neglecting one of the two currents in the objective function results in a non-exact SOCP relaxation. Second, computing the longitudinal component of the current results in the objective function not being independent of the power flow variables, which is one of the assumptions used in~\cite{farivar2013branch,6507352} to prove exactness of the proposed relaxation.
\begin{figure}[b!]
\begin{centering}
\includegraphics[width=1\linewidth, clip=true, trim=0 0 0 0]{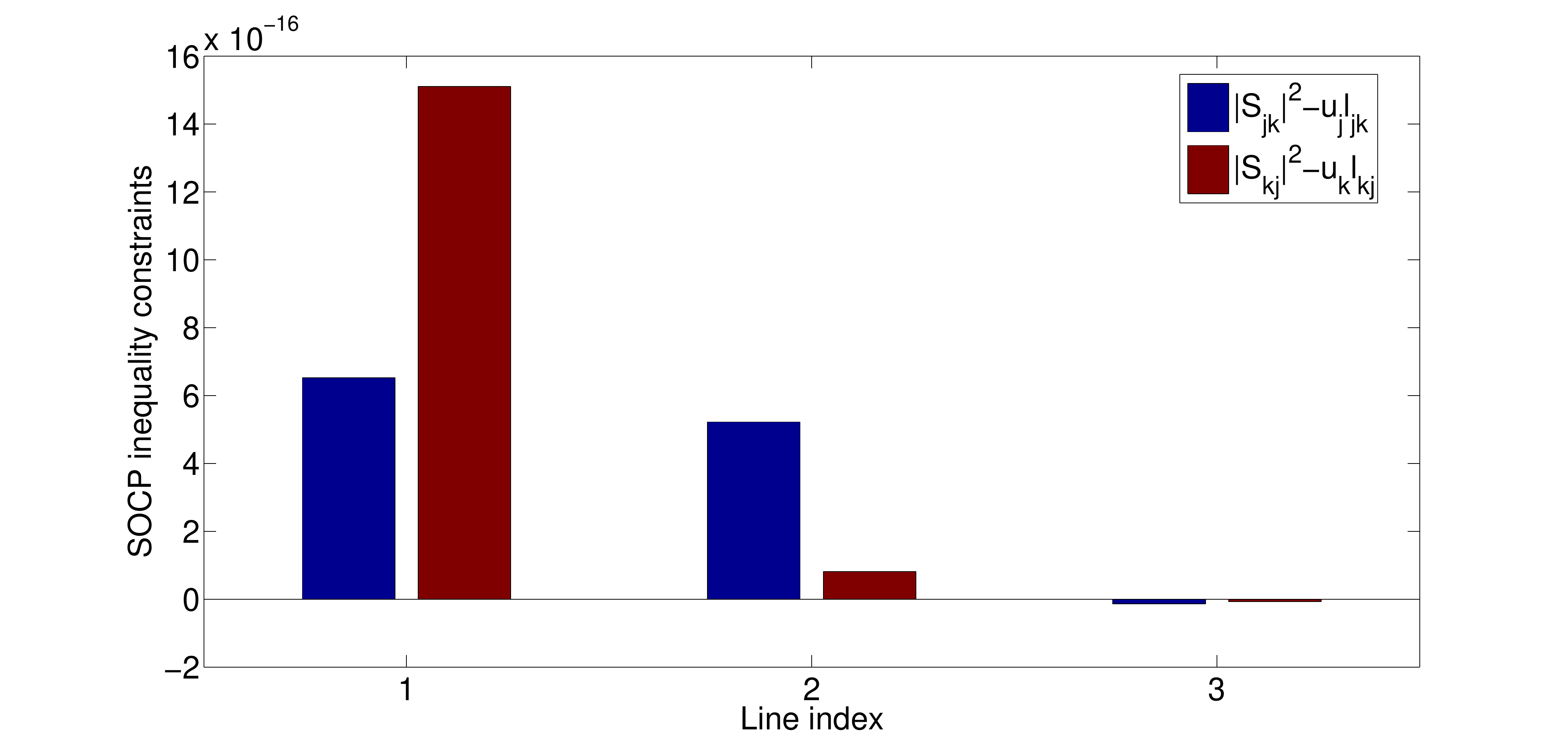}
\vspace{-25pt}
\caption{The relaxed inequalities values for the problem (\ref{obj})-(\ref{vminus}).}
\label{fig:SOCP_case1}
\end{centering}
\end{figure}

\begin{figure}[t!]
\begin{centering}
\includegraphics[width=1\linewidth, clip=true, trim=0 0 0 0]{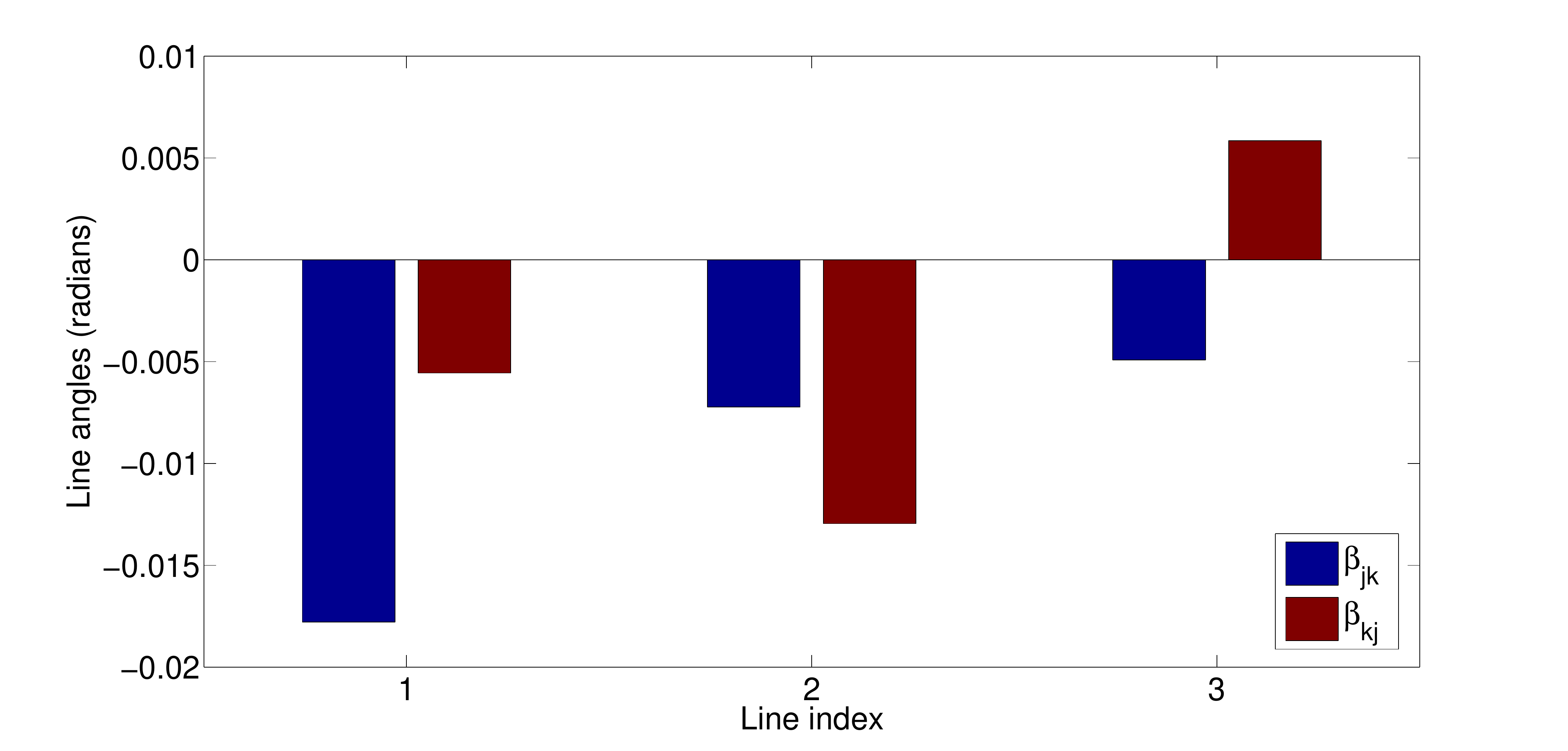}
\vspace{-25pt}
\caption{The line angles obtained after the OPF solution of problem (\ref{obj})-(\ref{vminus}).}
\label{fig:angles_case1}
\end{centering}
\end{figure}

In this case, after the solution of the OPF problem the inequalities in (\ref{scop}) are satisfied with equality, namely the SOCP relaxation is exact. The values for the two inequalities for each network line are shown in Fig.~\ref{fig:SOCP_case1}. For the same test case, we obtain the line angles
$\beta_{\ell^+}$ and $\beta_{\ell^-}$. The results are shown in Fig.~\ref{fig:angles_case1}. One can clearly observe that the obtained line angles do not satisfy $\beta_{\ell^+}+\beta_{\ell^-}=0$ and, therefore, the obtained solution has, again, no physical meaning.\\

The aforementioned examples indicate that the proposed SOCP relaxation in~\cite{farivar2013branch,6507352} cannot be trivially extended to lines represented as $\pi$-model equivalents even with convex constraints applied to the line exact $\pi$-model.

{\color{black}{
\subsection{Discussion}
\label{sec:discussion_F&L}
In the previous sections, we have investigated the BFM-based OPF formulation that allows a SOCP convexification of the OPF problem, therefore an efficient resolution of the problem with guaranteed convergence. This particular convexification proposed originally in~\cite{farivar2013branch,6507352} is valid in specific cases. In particular, when the DNO knows a-priori that the network lines are operated far away from their ampacity limit, therefore constraints (\ref{eq:OPF_lineflows_BFM}) are irrelevant. Finally, when the network lines are electrically short, namely when their shunt elements, combined with the line operating voltage, drain a negligible amount of capacitive reactive power. Practical examples are composed by short overhead lines operating at nominal voltages less than 20kV. However, as we have shown in the previous sections there are several case-studies involving realistic network topologies and operating points for which the original formulation presented in~\cite{farivar2013branch,6507352} results in violations of the line ampacity limits. Additionally, the assumptions required for the exactness of the proposed relaxation can result in solutions that are unrealistic for the grid operation, when for instance increased connection of distributed generation units are connected to the grid. Finally, we have extended the original problem formulation to include the shunt capacitances in an effort to properly account for their contribution in the line ampacity constraints of the OPF problem. As we have shown, such an extension preserves convexity, however cannot guarantee exactness of the SOCP relaxation. In this work, we are interested in the generic formulation of the OPF problem without any restriction on the grid topology and its operating point. As a consequence, there is a need to design algorithms that target the original non-approximated OPF problem that remains inherently non-convex. Recent trends are in favor of using ADMM for the solution of the OPF problem. Even though ADMM requires the underlying problem to be convex in order to guarantee convergence, it has been applied also to the case of non-convex AC OPF problems with promising convergence performance (e.g.,~\cite{sun2013fully,erseghe2014distributed}). In what follows we first present the ADMM solution of the problem in (\ref{eq:OPF-class_obj})-(\ref{eq:cap_curve}) and then we highlight specific scenarios for which ADMM fails to converge when applied to the non-approximated OPF problem.}}

\section{On the Application of ADMM for the Solution of the OPF Problem}
\label{sec:admm}

\subsection{ADMM-based Solution of the OPF Problem}
The ADMM-based solution of the OPF problem requires that the control variables are split into two separate groups and that the objective function is separable across this splitting~\cite{boyd2011distributed}. To this end, we introduce additional slack variables, $\bar{z}$, for the devices' and loads' power injections and for the line power flows and we reformulate the OPF problem as follows\footnote{In what follows we assume that demand is non-controllable. Also, as in~\cite{kraning2013dynamic} the constraints (\ref{eq:Sl_ac}),(\ref{eq:cap_curve}) are considered internal constraints of the lines and devices respectively and $\bar{I}_\ell^+,\bar{I}_\ell^-$ are internal variables of the lines.
}:

\begin{align}
\label{reform_optim}
&\min_{\substack{\bar{S}_g, \bar{z}_g,\bar{S}_c, \bar{z}_c,\bar{S}_{\ell^+}, \bar{z}_{\ell^+},\bar{S}_{\ell^-}\\ \bar{z}_{\ell^-},\bar{E}_{\ell^+}, \bar{E}_{\ell^-},\bar{I}_{\ell^+}, \bar{I}_{\ell^-} \bar{V}_{b}}}-\sum_{g}U_g(Re(\bar{S}_{g}))+\sum_{b}J_V(|\bar{V}_b|)+\\
&\sum_{\ell}J_I(|\bar{I}_\ell^+|,|\bar{I}_\ell^-|)+\sum_{b}\phi(\sum_{g \in b} \bar{z}_g - \sum_{c \in b} \bar{z}_c + \sum_{\mathclap{\beta(\ell^+)=b}} \bar{z}_{\ell^+} + \sum_{\mathclap{\beta(\ell^-)=b}} \bar{z}_{\ell^-}) \nonumber\\
&\text{subject to:   }\bar{S}_g=\bar{z}_g\, ,\forall \, g \in \mathcal{G} , \quad \text{and} \quad \bar{S}_c=\bar{z}_c \, ,\forall \, c \in \mathcal{C} \label{eq:admm_Sd} \\
&\bar{S}_{\ell^+}=\bar{z}_{\ell^+}, \quad \text{and} \quad  \bar{S}_{\ell^-}=\bar{z}_{\ell^-}, \,\forall \, \ell \in \mathcal{L} \label{eq:admm_Slp} \\
& \bar{E}_{\ell^+}=\bar{V}_{\beta(\ell^+)} , \quad \text{and} \quad \, \bar{E}_{\ell^-}=\bar{V}_{\beta(\ell^-)}, \,\forall \, \ell \in \mathcal{L} \label{eq:admm_Elp}
\end{align}
where $\phi$ is the characteristic function of the set $\{\bar{x} \in \mathbb{C}:\bar{x}=0\}$, $J_V$ is a penalty function with value $0$ if $V_{min} \leq |\bar{V}_b| \leq V_{max}$ and $\infty$ otherwise and $J_I$ is a penalty function with value $0$ if $\max(|\bar{I}_\ell^+|,|\bar{I}_\ell^-|) \leq I_{\ell_{max}}$ and $\infty$ otherwise.

The augmented Lagrangian for this problem is as follows:
\begin{align}
\label{augm_langr}
&L_\omega(\bar{S}_g,\bar{S}_c,\bar{S}_{\ell^+},\bar{S}_{\ell^-},\bar{E}_{\ell^+},\bar{E}_{\ell^-},\bar{I}_{\ell^+}, \bar{I}_{\ell^-},\bar{z}_g,\bar{z}_c,\bar{z}_{\ell^+},\bar{z}_{\ell^-},
\bar{V}_b,\bar{\mu},\bar{\nu},\bar{\lambda})\nonumber \\
&=-\sum_{g}U_g(Re(\bar{S}_{g}))+\sum_{b}J_V(|\bar{V}_b|)
+\sum_{\ell}J_I(|\bar{I}_\ell^+|,|\bar{I}_\ell^-|) \nonumber\\
&+\sum_{b}\phi(\sum_{g \in b} \bar{z}_g - \sum_{c \in b} \bar{z}_c + \sum_{\mathclap{\beta(\ell^+)=b}} \bar{z}_{\ell^+} \sum_{\mathclap{\beta(\ell^-)=b}} \bar{z}_{\ell^-}) \nonumber \\
&+\frac{\omega}{2}\{\sum_{\ell}|\bar{E}_{\ell^+}-\bar{V}_{\beta(\ell^+)}+\bar{\mu}_\ell|^2 +\sum_{\ell}|\bar{E}_{\ell^-}-\bar{V}_{\beta(\ell^-)}+\bar{\nu}_\ell|^2 \nonumber \\
&+\sum_{g}|\bar{S}_{g}-\bar{z}_g+\bar{\lambda}_g|^2+\sum_{c}|\bar{S}_{c}-\bar{z}_c+\bar{\lambda}_c|^2 \nonumber\\
&+\sum_{\ell}|\bar{S}_{\ell^+}-\bar{z}_{\ell^+}+\bar{\lambda}_{\ell^+}|^2 +\sum_{\ell}|\bar{S}_{\ell^+}-\bar{z}_{\ell^-}+\bar{\lambda}_{\ell^-}|^2\}
\end{align}
where $\bar{\mu},\bar{\nu},\bar{\lambda}$ are the lagrange multipliers associated with the equality constraints (\ref{eq:admm_Sd})-(\ref{eq:admm_Elp}).

The ADMM algorithm at the $k{-}th$ iteration consists of the following steps:
\begin{enumerate}
\item {First, all the devices, loads and lines update in parallel the primary variables, and their internal variables, i.e., ($\bar{S}_g,\bar{S}_c,\bar{S}_{\ell^+},\bar{S}_{\ell^-},\bar{E}_{\ell^+},\bar{E}_{\ell^-},\bar{I}_{\ell^+},\bar{I}_{\ell^-}$) with the secondary variables, and the dual variables fixed~\footnote{Note that demand is not controllable, hence the loads do not require the solution of an optimization problem to update their power consumption.}:
\begin{align}
&\text{For each network line $\ell$:}\nonumber \\
&(\bar{S}_{\ell^+}^{k+1},\bar{S}_{\ell^-}^{k+1},\bar{E}_{\ell^+}^{k+1},\bar{E}_{\ell^-}^{k+1},\bar{I}_{\ell^+}^{k+1},\bar{I}_{\ell^-}^{k+1})= \nonumber\\
&\underset{\bar{S}_{\ell^+},\bar{S}_{\ell^-},\bar{E}_{\ell^+},\bar{E}_{\ell^-},\bar{I}_{\ell^+},\bar{I}_{\ell^-}}{\operatorname{argmin}}
J_I(|\bar{I}_\ell^+|,|\bar{I}_\ell^-|)+ \nonumber \\
&\frac{\omega}{2}(|\bar{E}_{{\ell}^+}-\bar{V}_{\beta({\ell}^+)}^k+\bar{\mu}_{\ell}^k|^2+|\bar{E}_{\ell^-}-\bar{V}_{\beta(\ell^-)}^k+\bar{\nu}_{\ell}^k|^2 \nonumber \\
&+|\bar{S}_{{\ell^+}}-\bar{z}_{\ell^+}^k+\bar{\lambda}_{\ell^+}^k|^2+|\bar{S}_{{\ell^-}}-\bar{z}_{\ell^-}^k+\bar{\lambda}_{\ell^-}^k|^2) \label{admm_line}
\end{align}
 \begin{align}
&\text{ subject to:  } \bar{S}_{\ell^+}  =  \bar{E}_{\ell^+}\ubar{I}_{\ell^+} \quad \text{and} \quad  \bar{S}_{\ell^-}  =  \bar{E}_{\ell^-}\ubar{I}_{\ell^-} \label{eq_admm_Slp} \\
&\bar{I}_{\ell^+} = \bar{Y}_\ell (\bar{E}_{\ell^+} -\bar{E}_{\ell^-} )+ \bar{Y}_{\ell^+_0} \bar{E}_{\ell^+}  \label{eq_admm_Ilp}\\
&\bar{I}_{\ell^-} = \bar{Y}_\ell (\bar{E}_{\ell^-} -\bar{E}_{\ell^+} )+ \bar{Y}_{\ell^-_0} \bar{E}_{\ell^-} \label{eq_admm_Ilm}
\end{align}
 \begin{align}
&\text{For each device $g$: } \label{admm_device} \\ \nonumber
&\bar{S}_{g}^{k+1}=\underset{\bar{S}_{g}}{\operatorname{argmin}}-U_g(Re(\bar{S}_{g}))
+\frac{\omega}{2}(|\bar{S}_{g}-\bar{z}_{g}^k+\bar{\lambda}_g^k|^2) \\
&\text{subject to:  } \bar{S}_{g}\in \mathcal{H}_g \nonumber\\
&\text{For each load $c$: }   \bar{S}_{c}^{k+1}=\bar{S}_{c} \label{admm_load}
\end{align}}
\item {Then, by using the updated primary variables, the secondary variables are updated, i.e.,$(\bar{z},\bar{V}_b)$,
 on a bus level. We denote by $\bar{z}_b$ the vector of complex powers of all the devices, loads and lines that are connected
 to bus $b$, i.e., $\bar{z}_b\triangleq (\bar{z}_{g:g \in b},\bar{z}_{c:c \in b},\bar{z}_{\ell^+:  \beta(\ell^+)=b},\bar{z}_{\ell^-:  \beta(\ell^-)=b})$:
\begin{align}\label{admm_z_b}
&\bar{z}_b^{k+1}=\underset{\bar{z}_b}{\operatorname{argmin}}
(\phi(\sum_{g \in b} \bar{z}_g - \sum_{c \in b} \bar{z}_c + \sum_{\mathclap{\beta(\ell^+)=b}} \bar{z}_{\ell^+} \sum_{\mathclap{\beta(\ell^-)=b}} \bar{z}_{\ell^-}) \\ \nonumber
&+\frac{\omega}{2}\{\sum_{\mathclap{g \in b}}|\bar{S}_{g}^{k+1}-\bar{z}_g+\bar{\lambda}_g^{k}|^2+\sum_{c \in b}|\bar{S}_{c}^{k+1}-\bar{z}_c+\bar{\lambda}_c^{k}|^2 \\ \nonumber
&+\sum_{\mathclap{\beta(\ell^+)=b}}|\bar{S}_{\ell^+}^{k+1}-\bar{z}_{\ell^+}+\bar{\lambda}_{\ell^+}^{k}|^2+\sum_{\mathclap{\beta(\ell^-)=b}}|\bar{S}_{\ell^-}^{k+1}-\bar{z}_{\ell^-}+\bar{\lambda}_{\ell^-}^{k}|^2\})
\end{align}
\begin{align}\label{admm_V_b}
\bar{V}_b^{k+1}&=\underset{\bar{V}_b}{\operatorname{argmin}}
(J(\bar{V}_b)+\frac{\omega}{2}\{\sum_{{\mathclap{\beta(\ell^+)=b}}}|\bar{E}_{\ell^+}^{k+1}-\bar{V}_{b}+\bar{\mu}_{\ell}^k|^2\nonumber \\
&+\sum_{{\mathclap{\beta(\ell^-)=b}}}|\bar{E}_{\ell^-}^{k+1}-\bar{V}_{b}+\bar{\nu}_{\ell}^k|^2\})
\end{align}
}
\item {Finally, dual variables, i.e., $\bar{\mu},\bar{\nu},\bar{\lambda}$ are updated:
\begin{align}
\label{admm_mu}
&\bar{\mu}_\ell^{k+1}=\bar{\mu}_\ell^{k}+(\bar{E}_{\ell^+}^{k+1}-\bar{V}_{\beta(\ell^+)}^{k+1}) \\
\label{admm_nu}
&\bar{\nu}_\ell^{k+1}=\bar{\nu}_\ell^{k}+(\bar{E}_{\ell^-}^{k+1}-\bar{V}_{\beta(\ell^-)}^{k+1})\\
\label{admm_lam_d}
&\bar{\lambda}_g^{k+1}=\bar{\lambda}_g^{k}+(\bar{S}_{g}^{k+1}-\bar{z}_{g}^{k+1})\\
\label{admm_la_c}
&\bar{\lambda}_c^{k+1}=\bar{\lambda}_c^{k}+(\bar{S}_{c}^{k+1}-\bar{z}_{c}^{k+1})\\
\label{admm_la_p}
&\bar{\lambda}_{\ell^+}^{k+1}=\bar{\lambda}_{\ell^+}^{k}+(\bar{S}_{{\ell^+}}^{k+1}-\bar{z}_{{\ell^+}}^{k+1})\\
\label{admm_la_m}
&\bar{\lambda}_{\ell^-}^{k+1}=\bar{\lambda}_{\ell^-}^{k}+(\bar{S}_{{\ell^-}}^{k+1}-\bar{z}_{{\ell^-}}^{k+1})
\end{align}

}
\end{enumerate}

The stopping criterion for this algorithm is that the primal and dual residuals (defined as in~\cite{boyd2011distributed}) are less than a small predefined tolerance or that a maximum number of iterations has been reached.

In what follows, we show specific scenarios where the ADMM algorithm fails to converge to a solution.

\subsection{Investigation of the Convergence of the ADMM-based Solution of the OPF Problem}
We consider the same network in Fig.~\ref{fig:network_bfm_EMTP_OPF}. Each network bus, apart from the slack bus, has a load and a generator connected to it. The demand in the network is assumed to be non-controllable, whereas the generators are assumed to be distributed solar panels with typical PV-type capability constraints. For this scenario, the capability limits and the values of loads and generation are given in Table~\ref{bfm_emtp_OPF_param_admm}.
\begin{table}[b!]
\begin{center}
\caption{Parameters of the test network in Fig.\ref{fig:network_bfm_EMTP_OPF} used for the ADMM-based solution of the OPF problem}
\label{bfm_emtp_OPF_param_admm}
\begin{tabular}{|c|c|}
  \hline
  \textbf{Parameter} &  \textbf{Value} \\ \hline
  Generators' power, $|\bar{S}_{i_{g_{max}}}|, i=2,3,4$ (MVA)& $0.40,0.39,0.46$\\ \hline
  Generators' power factor, $cos\phi_{i_{g}}, i=2,3,4$ & $0.9$ \\ \hline
  Loads' active power, $P_{i_{c}}, i=2,3,4$ (MW)& $2.76,2.16,2.46$\\ \hline
  Loads' reactive power, $Q_{i_{c}}, i=2,3,4$ (MW)& $1.38,1.08,1.23$\\ \hline
  Shunt capacitor (bus 2), case I and II (uF) & $(239, 859)$ \\ \hline
  Penalty term gain, $\omega$ & $1$ \\ \hline
  Tolerance and maximum number of iterations & $10^{-4}, 10^4$\\ \hline
\end{tabular}
\end{center}
\end{table}
In addition to the loads and generation, we consider that a shunt capacitor is connected to bus 2. In order to model this shunt capacitor, we consider that it is part of the first network line. In particular, we consider that the shunt capacitance on the sending end of the $\pi$-model of the line connecting buses 1 and 2 is modified accordingly to account for the shunt capacitor.{\color{black}{ It is worth noting that switched capacitor banks, if present in the network, can be taken into account in the problem formulation in a similar way provided they are not included in the OPF control variables. The control of these discrete elements is possible provided that we assume they are continuous control variables which are rounded to the nearest integer upon solution of the OPF problem. In particular, switched capacitor banks, can be readily taken into account as control variables in the OPF formulation and in particular as nodal injections with controllable reactive power in a range defined by their capacity limits.}}

\begin{figure}[h!]
\begin{centering}
\includegraphics[width=1\linewidth, clip=true, trim=0 5 5 130]{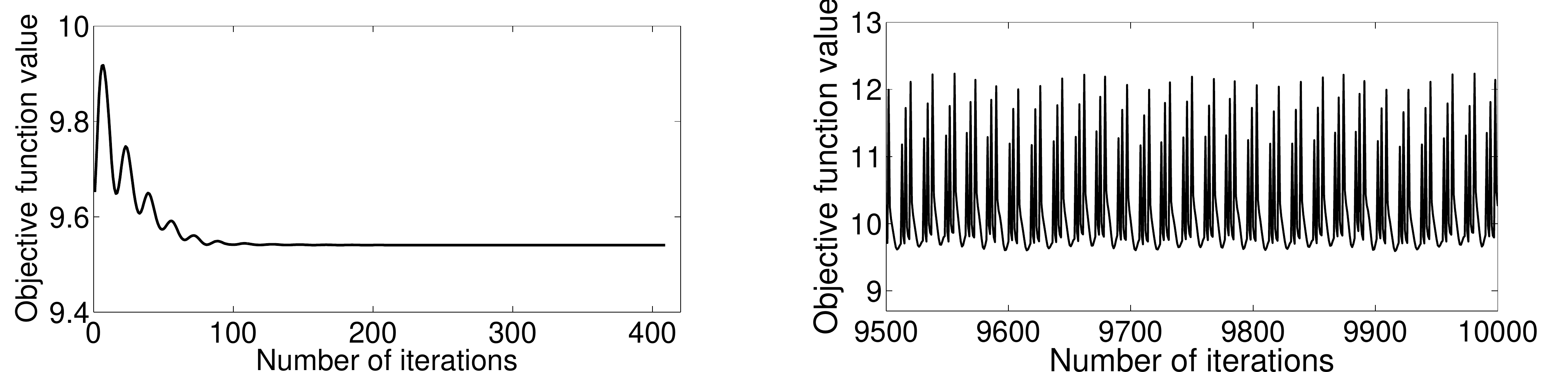}
\vspace{-25pt}
\caption{Objective function value for case I and II (last 500 iterations).}
\label{fig:obj_all}
\end{centering}
\end{figure}

\begin{figure}[h!]
\begin{centering}
\includegraphics[width=1\linewidth, clip=true, trim=0 5 0 110]{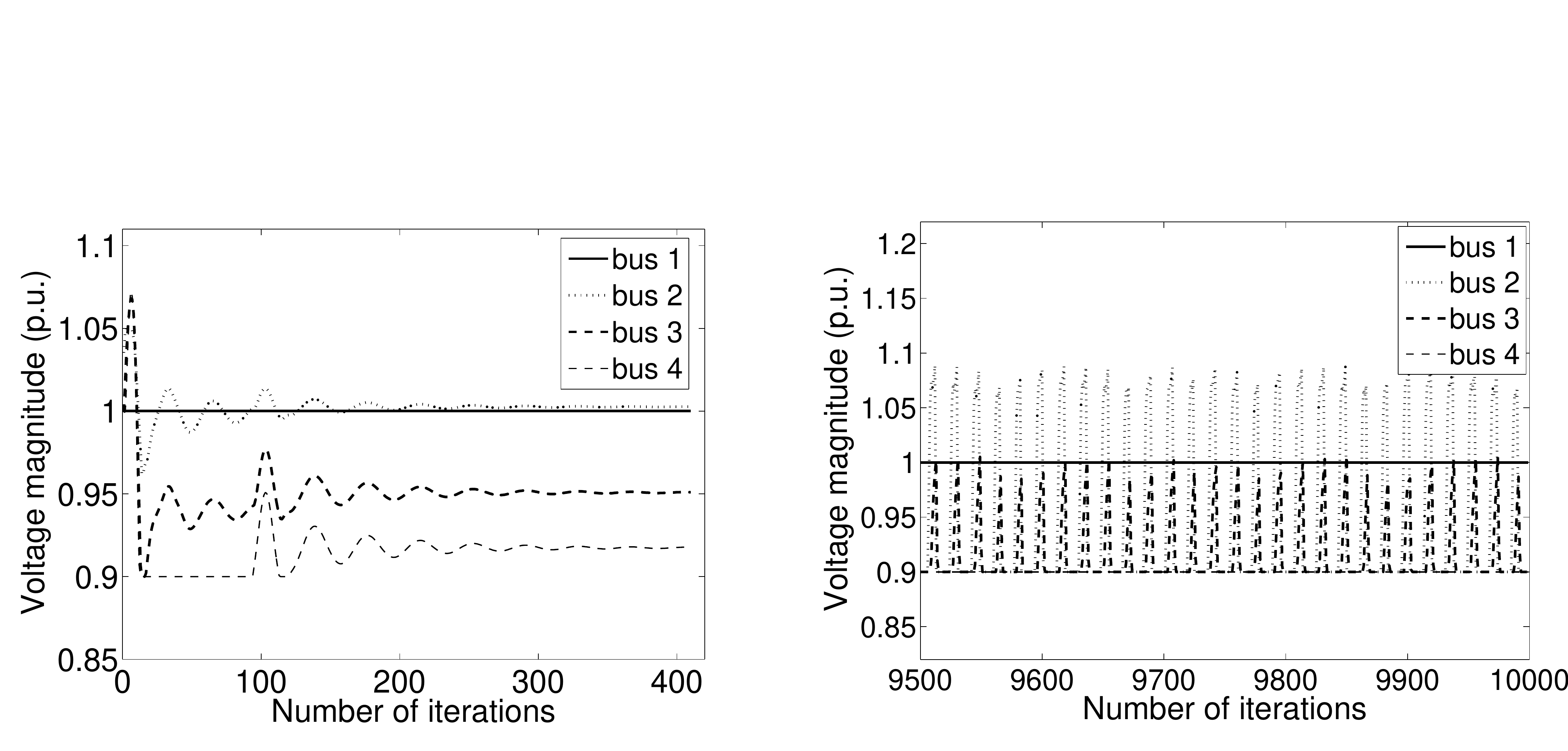}
\vspace{-25pt}
\caption{Voltage magnitude evolution for cases I and II (last 500 iterations).}
\label{fig:volt_all}
\end{centering}
\end{figure}

\begin{figure}
\begin{centering}
\includegraphics[scale = 0.31, clip=true, trim=20 5 5 170]{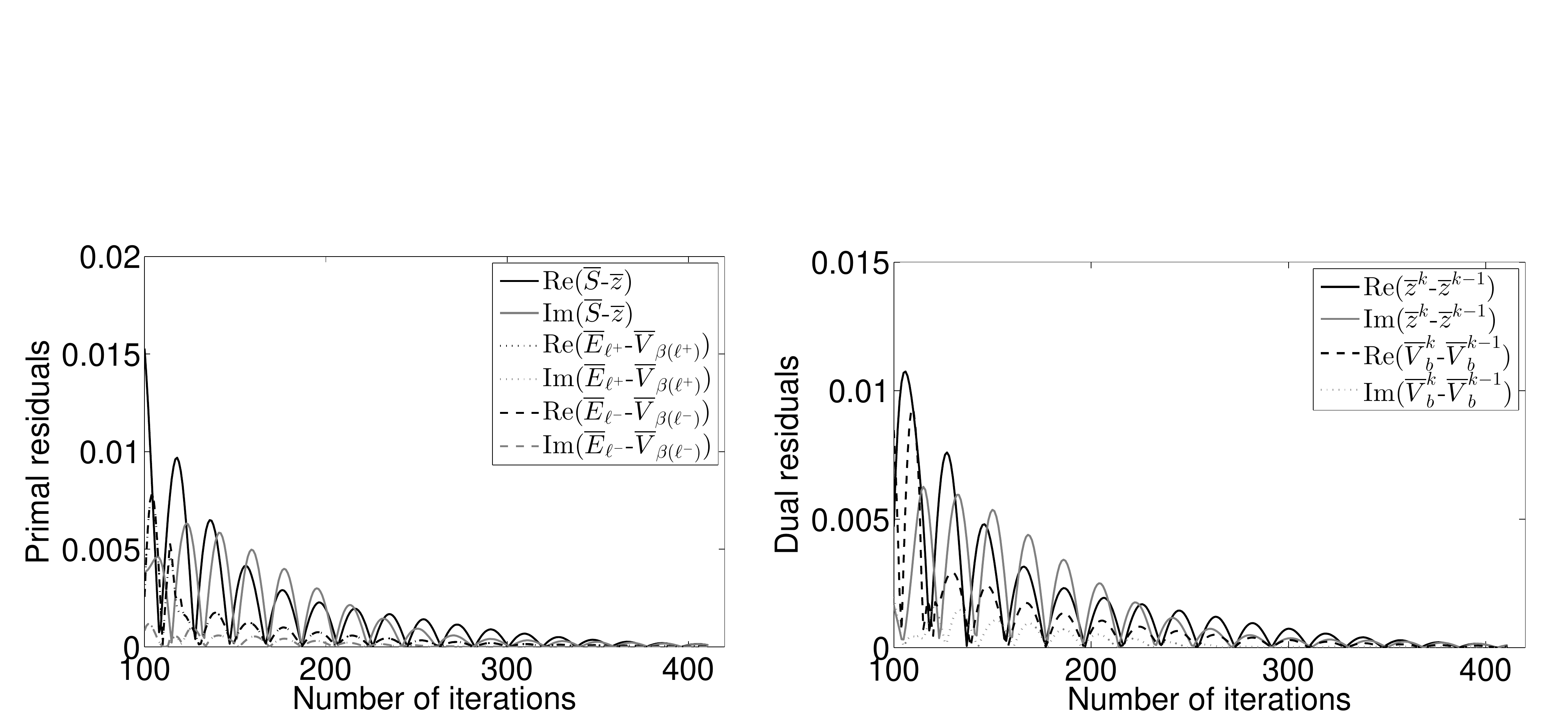}
\vspace{-25pt}
\caption{Norm of the primal/dual residuals for case I (last 311 iterations).}
\label{fig:res_caseI}
\end{centering}
\end{figure}

\begin{figure}[t!]
\begin{centering}
\includegraphics[width=1\linewidth, clip=true, trim=0 5 5 140]{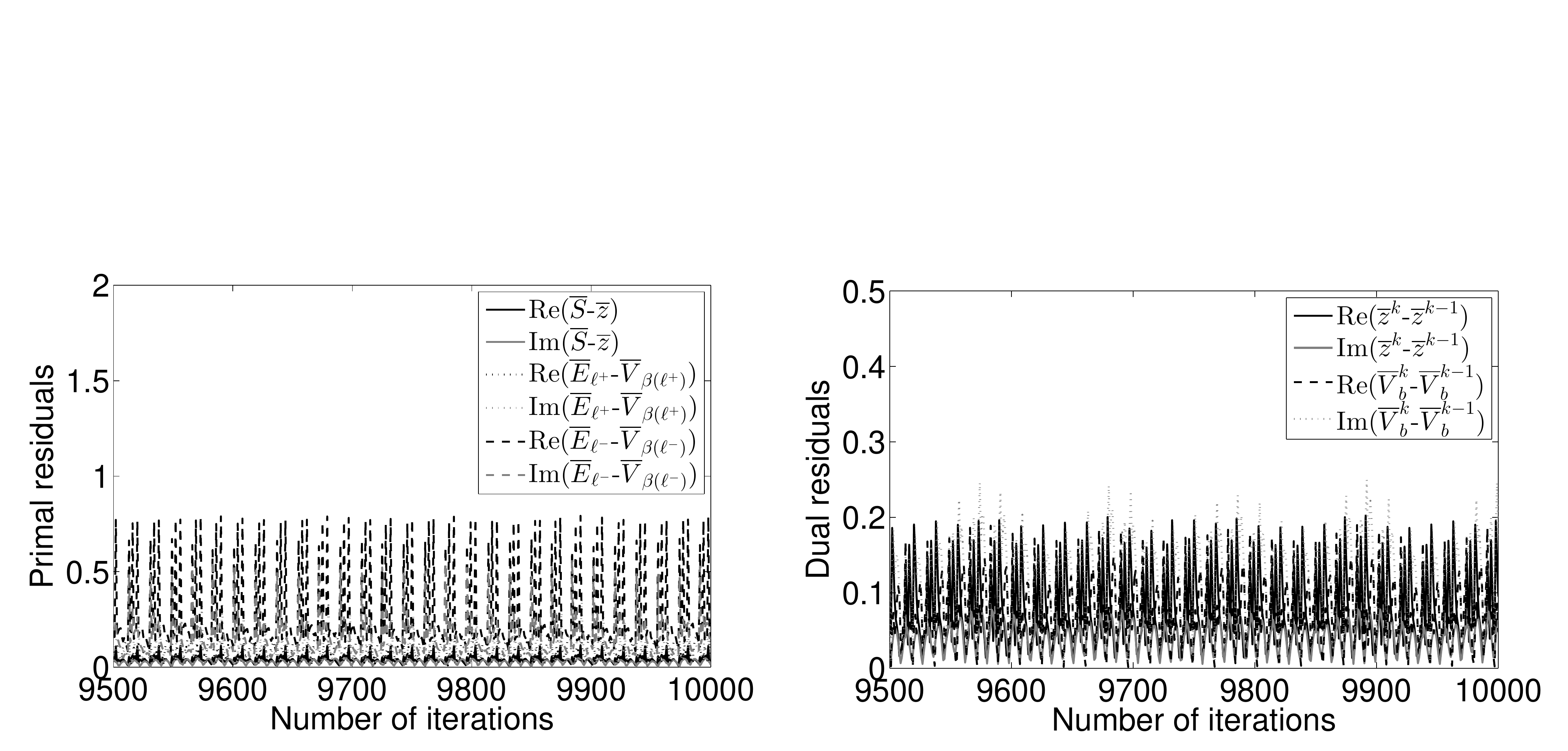}
\vspace{-25pt}
\caption{Norm of the primal/dual residuals for case II (last 500 iterations).}
\label{fig:res_caseII}
\end{centering}
\end{figure}

We implement and solve the ADMM algorithm in Matlab for two different cases that correspond to two different values of the size of the shunt capacitor (see Table~\ref{bfm_emtp_OPF_param_admm}). In Case I, even though the OPF problem solved is the non-approximated non-convex one, ADMM converges, within the predefined tolerance, in $411$ iterations. The left figure in Fig.~\ref{fig:obj_all} shows the objective function value as a function of the number of iterations of ADMM. The left figure in Fig.~\ref{fig:volt_all} shows the convergence of the buses' voltage magnitudes and Fig.~\ref{fig:res_caseI} shows how the primal and dual residuals evolve with the iterations. On the contrary, in Case II, ADMM fails to converge to a solution and reaches the maximum number of iterations.  This is shown in Fig.~\ref{fig:obj_all} (right), \ref{fig:volt_all} (right) and~\ref{fig:res_caseII} where the objective function, as well as the residuals and bus voltages are plotted for the last five hundred iterations until the maximum number of iterations is reached; we can observe that they exhibit oscillations.

In what follows we analyze why the ADMM algorithm converges in Case I but fails in Case II.
To begin with, the first network line has the peculiarity that the voltage at its receiving end $\bar{E}_{\ell^+}$ (i.e., the slack bus voltage) is fixed.\footnote{This holds for all the lines that are connected to the slack bus.} As a consequence, the first equality constraint in (\ref{eq_admm_Slp}) becomes linear in the real and imaginary part of the voltage $\bar{E}_{\ell^-}$, whereas the second equality constraint in (\ref{eq_admm_Slp}) becomes quadratic on the real and imaginary part of the voltage $\bar{E}_{\ell^-}$. In fact, the coefficients of the quadratic terms in the latter constraint are $Re(\bar{Y}_\ell)$ and $-Im(\bar{Y}_\ell)-Im(\bar{Y}_{\ell^-_0})$ for the real and imaginary parts, respectively. Due to the physics of the network, $Re(\bar{Y}_\ell)$ and $Im(\bar{Y}_{\ell^-_0})$ are positive for a network line and $Im(\bar{Y}_\ell)$ is negative. Furthermore, typically, the longitudinal reactance $Im(\bar{Y}_\ell)$ is much larger than the shunt capacitance $Im(\bar{Y}_{\ell^-_0})$ of a network line. Therefore, typically the coefficients of both quadratic terms are positive, and the line problem in (\ref{admm_line}) is convex for the lines that are connected to the slack bus. This is the case for the Case I. However, in Case II the size of the shunt capacitor, connected to bus 2, is such that $Im(\bar{Y}_{\ell^-_0})>-Im(\bar{Y}_\ell)$, thus the coefficient of the aforementioned quadratic term in (\ref{eq_admm_Slp}) is no longer positive and the corresponding line problem becomes non-convex.

{\color{black}{Apart from the aforementioned case of the shunt capacitor, we also discuss the case of OLTCs, which is another discrete control typically used to optimize the grid operation. The ADMM algorithm also fails to converge to a solution when on-load tap changers (OLTCs) are included in the OPF formulation as control variables\footnote{For the sake of brevity we do not include the simulation results for this specific scenario.}. To better understand why this occurs, let us consider a transformer with OLTC capabilities between buses 1 and 2 in the network and let us denote the ideal transformer admittance by $\bar{Y}_t$ and the OLTC ratio by $\alpha \in \mathbb{R}$. Then based on the OLTC model in~\cite{stevenson1994power}, the longitudinal admittance of the first network line equals $\alpha \bar{Y}_t$ and the shunt elements of the receiving and sending ends of the same line are $\alpha (\alpha -1 )\bar{Y}_t$ and $(1-\alpha)\bar{Y}_t$ respectively. If the OLTCs are not included in the set of control variables, then the ratio $\alpha$ has a fixed value and the inclusion of the OLTCs in the OPF formulation does not affect the solution. However, when the OLTCs are considered control variables\footnote{As with the case of switched capacitor banks, when considered control variables, the OLTCs are assumed to be continuous variables rounded to the nearest integer upon solution of the OPF problem.}, their effect is similar to that of the shunt capacitors, in the sense that the line problem in (\ref{admm_line}) becomes once again non-convex for those lines that are connected to regulating transformers.  The reason is that $\alpha$ is now an additional control variable, namely the OLTC ratio appears in the equality constraints (\ref{eq_admm_Slp}) of the first network line problem, and both these constraints become quadratic in $\bar{E}_{\ell^-}$ and $\alpha$ and non-convex.}}

\label{sec:conclusion}
In this first part of the paper we have focused on investigating the limits of the branch flow convexification proposed by Farivar-Low in~\cite{farivar2013branch,6507352} and of the ADMM-based solution of the OPF problem. In particular, we have discussed the misinterpretation of the physical model in the Farival-Low formulation of the OPF problem and the unrealistic assumptions therein. Finally, we have provided the ADMM-based decomposition of the OPF problem and we have shown, through specific examples, cases for which the ADMM-based solution of the non-relaxed OPF problem fails to converge.

\end{document}


\begin{frontmatter}

\title{AC OPF in Radial Distribution Networks - Part II:\\
An Augmented Lagrangian-based OPF Algorithm, Distributable via Primal Decomposition}


\author[mymainaddress,mysecondaryaddress]{Konstantina Christakou}
\ead{konstantina.christakou@epfl.ch}

\author[mymainaddress2]{Dan-Cristian Tomozei}
\ead{dtomozei@cisco.com}

\author[mymainaddress]{Jean-Yves Le Boudec}
\ead{jean-yves.leboudec@epfl.ch}

\author[mysecondaryaddress]{Mario Paolone\corref{mycorrespondingauthor}}
\cortext[mycorrespondingauthor]{Corresponding author. Phone number: $+41$ $21$ $69$ $32662$, Postal address: EPFL STI IEL DESL
ELL 136 (B\^{a}timent ELL), Station 11, CH-1015 Lausanne}
\ead{mario.paolone@epfl.ch}

\address[mymainaddress]{Laboratory for Communications and Applications 2, Ecole Polytechnique F\'{e}d\'{e}rale de Lausanne,
CH-1015 Lausanne, Switzerland}
\address[mysecondaryaddress]{Distributed Electrical Systems Laboratory, Ecole Polytechnique F\'{e}d\'{e}rale de Lausanne, CH-1015
Lausanne, Switzerland}
\address[mymainaddress2]{Cisco Systems, Inc, EPFL Innovation Park Building E 1015 Lausanne, Switzerland}

\begin{abstract}
In the first part of this two-part paper we show that the branch-flow convexification of the OPF problem is not exact and that the ADMM-based decomposition of the OPF fails to converge in specific scenarios. Therefore, there is a need to develop algorithms for the solution of the non-approximated OPF problem that remains inherently non-convex. To overcome the limitations of recent approaches for the solution of the OPF problem, we propose in this paper, a specific algorithm for the solution of a non-approximated, non-convex AC OPF problem in radial distribution systems. It is based on the method of multipliers, as well as on a primal decomposition of the OPF problem. We provide a centralized version, as well as a distributed asynchronous version of the algorithm. We show that the centralized OPF algorithm converges to a local minimum of the global OPF problem and that the distributed version of the algorithm converges to the same solution as the centralized one. Here, in this second part of the two-part paper, we provide the formulation of the proposed algorithm and we evaluate its performance by using both small-scale electrical networks, as well as a modified IEEE 13-node test feeder.
\end{abstract}

\begin{keyword}
OPF, ADMM, decomposition methods, method of multipliers, convex relaxation, active distribution networks, distributed algorithms, asynchronous algorithms.
\end{keyword}

\end{frontmatter}


\section{Introduction}
\label{intro}
In Part I of this two-part paper we present the generic formulation of the non-convex OPF problem and we briefly review several OPF algorithms that are based on approximations and assumptions in order to guarantee convergence. Furthermore, we focus on the branch-flow convexification of the OPF problem that has been recently proposed by Farivar and Low in~\cite{farivar2013branch,6507352} and is claimed to be exact for the case of radial distribution systems under specific assumptions, despite the absence of apparent approximations. We show that this claim, in fact, does not hold, as it leads to an incorrect system model and therefore, there is a need to develop algorithms for the solution of the non-approximated OPF problem that remains inherently non-convex. In detail, we show through practical examples that in~\cite{farivar2013branch,6507352}, on one hand, there is a misinterpretation of the physical network model related to the ampacity constraint of the lines' current flows and, on the other hand, the proof of the exactness of the proposed relaxation requires unrealistic assumptions related to the unboundedness of specific control variables. Furthermore, we investigate the application of ADMM for the solution of the original non-approximated OPF problem. Even though ADMM requires the underlying problem to be convex in order to guarantee convergence, it was applied also to the case of non-convex AC OPF problems with promising convergence performance (e.g.,~\cite{sun2013fully,erseghe2014distributed}). However, we show, through practical examples, cases for which the ADMM-based decomposition of the non-relaxed OPF problem fails to converge.

To overcome the aforementioned limitations, here in this second part, we propose an algorithm for the solution of the non-approximated non-convex AC OPF problem in radial networks. {\color{black}{A large number of algorithms can be found in the literature that tackle the non-approximated non-convex OPF problem ranging from non-linear and quadratic programming techniques, Newton-based methods, interior point methods to heuristic approaches based on genetic algorithms, evolutionary programming, and particle-swarm optimization (e.g.,~\cite{frank2012optimal,frank2012optimal2,qiu2009literature,7038504}). These methods, typically utilize powerful general purpose solvers or in-house developed software but they do not provide, in general, convergence guarantees. Our proposed solution belongs to the family of augmented Lagrangian methods for the solution of the OPF problem. More specifically, our proposed method uses an augmented Lagrangian approach, relies on the method of multipliers (\cite{bertsekas99,powell1978algorithms,hestenes1969multiplier}) and provides convergence guarantees. In particular, we design a centralized OPF algorithm that is proven to converge to a local minimum of the original non-approximated OPF problem.}}

With respect to the case of controlling multiple dispersed energy resources, it is of interest to also define a distributed solution method that is formally equivalent to the centralized formulation. {\color{black}{ Distributed solutions are of interest in several practical cases. Among others, when the problem size is large due to a very large number of small controllable resources like PV panels, when the communication requirements are such that the amount of exchanged information flow between agents needs to be limited, or when an asynchronous solution of the problem is more appealing.}} In fact, several distributed OPF algorithms are proposed in the literature. In~\cite{bolognani2013distributed,6687943} the authors design a dual-ascent algorithm for optimal reactive power flow with power and voltage constraints. In~\cite{dall2013distributed,lam2012optimal} dual decomposition is used as the basis for the distributed solution of the OPF problem. {\color{black}{A further category of distributed OPF has been proposed to solve  multi-area objectives (\cite{nogales2003decomposition,conejo2002decomposition}). However, their applicability to generically decomposable OPF has not been discussed.}} Finally, a significant number of contributions propose distributed formulations of the OPF problem, based on the alternating direction method of multipliers (ADMM) (e.g.,~\cite{kraning2013dynamic,dall2013distributed,sun2013fully,vsulc2013optimal,peng2014distributed,erseghe2014distributed}).

In this direction, we present, here in this second part, a distributed version of the proposed algorithm that, unlike ADMM, is based on a primal decomposition~\cite{palomar2006tutorial} and does not require that the problem be convex. In this decentralized version of the algorithm, at each iteration, local agents, assigned to network buses and network lines, exchange messages with their neighbors using only local information. We prove that the distributed algorithm converges to the same solution as the centralized version. Finally, we present an asynchronous implementation of the distributed algorithm where the messages of the neighboring agents need not be synchronized.

The structure of this second part is the following. In Section II we describe the proposed algorithm for the OPF solution. We present both a centralized, as well as a decentralized asynchronous version of the proposed algorithm. In Section III we investigate the convergence of the proposed algorithm in the cases where the BFM convexification leads to an incorrect solution and ADMM fails to converge to a solution. In Section IV we evaluate the performance of the proposed algorithm using a modified IEEE 13-node test feeder. Finally, in Section V we provide the main observations and concluding remarks for this Part II.

\section{AC OPF in Radial Distribution Systems}
We first write the AC OPF problem presented in Part I in an equivalent form, and then we provide a centralized, as well as a distributed algorithm for its resolution.

We make the following assumptions about the grid model:
\begin{enumerate}
\item[A1.]{We consider a direct sequence representation of the grid\footnote{{\color{black}{Note that the proposed formulation can be extended without loss of generality to the case of multi-phase unbalanced grids by adopting the so-called compound network admittance matrix, (i.e., the 3-phase representation of the grid model which takes into account the various couplings between the network phases) instead of the single-phase equivalents. In this cases, each of the constraints in the OPF formulation needs to be formulated separately for each network phase.}}};}
\item[A2.]{Any two-port component (e.g., lines, transformers etc.) is represented as a $\pi$-equivalent;}
\item[A3.]{We assume a perfect knowledge of the system parameters, i.e., the network admittance matrix is known;}
\item[A4.]{The nodal-power injections are voltage-independent;}
\item[A5.]{The control variables are composed by the nodal power injections/absorptions.}
\end{enumerate}

\subsection{The Proposed Centralized OPF Algorithm}
We are interested in maximizing the social welfare of the economic agents that use the grid, while maintaining an acceptable network voltage profile and respecting the line ampacity limits. Specifically, we tune the line ampacities and the network voltage profiles by controlling the $(P,Q)$-injections of distributed controllable devices ${\cal G}$ ({\it e.g.}, renewable generators) in a ``fair'' way: Each controllable device $g\in {\cal G}$ has a certain utility function $U_g(\cdot)$, and the sum of these utility functions is maximized subject to the satisfaction of the network operation constraints (voltage and ampacity). The resulting set-point is thus Pareto-optimal, {\it i.e.}, no single device can increase its utility without hurting the utility of some other device, and locally-``fair'', {\it i.e.}, the resulting set-point is a local maximizer of the sum of the device utilities lying on the Pareto boundary of feasible set-points. 

By convention, each line $\ell\in{\cal L}$ has a ``receiving'' and a
``sending'' end, which we denote by $\ell^+$ and $\ell^-$,
respectively. These are chosen arbitrarily. A line is connected to two
adjacent buses to which we refer by $\beta(\ell^+)$ and
$\beta(\ell^-)$, respectively. For each line, we introduce two
auxiliary variables $\bElp$ and $\bElm$ representing the complex
voltage at the two ends of the line. Assumptions A1-A3 allow us to
express the corresponding injected currents and powers at the two ends
of line $\ell$:
\begin{align}
  \bIlp = \bIlp(\bElp, \bElm) &= (\bar{Y}_\ell+\bar{Y}_{\ell^+_0})\bElp-\bar{Y}_\ell \bElm \label{eq:Ilp}\\
  \bIlm = \bIlm(\bElp, \bElm) &= (\bar{Y}_\ell+\bar{Y}_{\ell^-_0})\bElm-\bar{Y}_\ell \bElp \label{eq:Ilm}\\
  \bSlp = \bSlp(\bElp, \bElm) &= \bElp \uIlp\label{eq:Slp}\\
  \bSlm = \bSlm(\bElp, \bElm) &= \bElm \uIlm\label{eq:Slm}
\end{align}
In the remainder of this paper, unless otherwise stated, the complex
line currents and powers expressed above are always computed according
to equations \eqref{eq:Ilp}-\eqref{eq:Slm}. They are thus all
functions of $\bElp$ and $\bElm$ {\it exclusively}, although the
arguments are often omitted for the sake of brevity. All quantities
are expressed in ``per-unit'', unless otherwise specified.

For readability, we denote the vector formed by the real and imaginary
parts of variables $(\bElp, \bElm)_\ell$ by $y\in\mathds R^{4L}$,
where $L = |{\cal L}|$ is the number of lines. Note that for a given
value of $y$, the corresponding currents and powers do not necessarily
satisfy Kirchhoff's law.

We call $y$ feasible if it satisfies voltage consistency and per-bus
power-balance. Voltage consistency means that the voltages of all the
lines incident to a specific bus $b\in{\cal B}$ are identical, {\it
  i.e.}, have the same amplitude $V_b$ and the same argument
$\varphi_b$:
\begin{align}
  |\bElp| = V_{\beta(\ell^+)}, \quad &|\bElm| = V_{\beta(\ell^-)} \label{eq:vc-ampl}\\
  \arg(\bElp) = \varphi_{\beta(\ell^+)}, \quad &\arg(\bElm) = \varphi_{\beta(\ell^-)}, &\forall \ell\in{\cal L}. \label{eq:vc-arg}
\end{align}
At each bus $b\in{\cal B}$, power-balance is satisfied if and only if
\begin{align}
  \sum_{\beta(\ell^+)=b} \bSlp {+} \sum_{\beta(\ell^-)=b} \bSlm =-\sum_{g \in b} \bar{S}_g {-} \bar{S}(b), \quad \forall b \in \mathcal{B},  \label{eq:OPF-Sbal}
\end{align}
where $S_g$ is the controlled generated power of device $g$ found at
bus $b$, $\bar{S}(b)$ denotes the non-controllable power injection at
bus $b$, and $\bSlp$, $\bSlm$ are obtained
via~\eqref{eq:Slp}-\eqref{eq:Slm}.

If $y$ is feasible, it is important to note that equations
(\ref{eq:Ilp})-(\ref{eq:Slm}) describe the exact AC power-flow
equations. Hence, we use a non-approximated model of the grid.

We write the OPF formulation (Part I) equivalently:\footnote{Unlike in Part I, we consider wlog that there are two
  types of connected devices: they either have controllable power
  injection $\bar{S}_g$ or impose an overall fixed power injection
  $\bar{S}(b)$ in bus $b$.} \small
\begin{align}
\label{eq:OPF-global2}
&\max_{\substack{\bar{S}_g,V_b,\varphi_b\\\bElp,\bElm}} \sum_{g \in \mathcal{G}} W_g(\bar{S}_{g}) \qquad \mbox{subject to:} \\
&\mbox{Feasibility constraints~\eqref{eq:vc-ampl},~\eqref{eq:vc-arg},~\eqref{eq:OPF-Sbal}}\nonumber\\
&|\bIlp| \leq I_{\ell,{\max}} \mbox{ and } |\bIlm| \leq I_{\ell,{\max}}, & \forall \ell \in \mathcal{L}  \label{eq:amp_lim} \\
&V_{\min} \leq V_b \leq V_{\max}, & \forall b \in \mathcal{B} \label{eq:OPF-Fb}\\
& \bar{S}_g \in \mathcal{H}_g, & \forall g \in
\mathcal{G}\label{eq:cap_curve_gen}
\end{align}
\normalsize As previously stated, the objective function is the sum of
the welfare of the controllable devices $W_g$. In the above
formulation, we denote by ${\cal G}$ the set of controllable devices
and by $\bar{S}_g = P_g + j Q_q$ the controllable injected power by
device $g$, subject to the capability
constraint~\eqref{eq:cap_curve_gen}. The set ${\cal G}$ can contain
both generators and consumers. However, for the sake of presentation
clarity, we consider that ${\cal G}$ contains uniquely PV
generators. This is not a limiting assumption, as our results apply to
any device with controllable power injections (including controllable
loads). 
Non-controllable loads do not appear in the objective function, that
expresses the utility of PV generators (a concave increasing function
$U(\cdot)$ of active power injection) and the losses of the power
converter:
\begin{align}
W_g(\bar S_g) = U_g(P_{g}) - \eta(P_g^2+Q_g^2), \ \forall g\in{\cal G}.
\end{align}
We consider typical capability curves of PV power inverters:
\begin{align}
  {\cal H}_g = \{\bar{S}_g : |\bar{S}_g| \leq S_{g,{\max}}, \
  |arg({\bar{S}_g})| \leq \phi_{g,{\max}}\}. \label{eq:OPF-fd}
\end{align}


In order to solve the problem \eqref{eq:OPF-global2}-\eqref{eq:cap_curve_gen}, we convert the inequality constraints (\ref{eq:amp_lim}) to equality constraints by introducing slack variables $i_{{\ell}^+}$ and $i_{{\ell}^-}$ as follows:
\begin{align}
&|\bIlp| + i_{{\ell}^+} = I_{\ell,{\max}} \mbox{ and } |\bIlm| + i_{{\ell}^-} = I_{\ell,{\max}}, \, \forall \ell \in \mathcal{L} \label{eq:amp-lim-eq}\\
& i_{{\ell}^+} , i_{{\ell}^-} \geq 0, \, \quad  \forall \ell \in \mathcal{L} \label{eq:amp-lim-slack-var}
\end{align}

We denote by $x$ the real vector of variables formed by the artificial
control variables $(V_b, \varphi_b)_{b\in{\cal B}}$, $(i_{\ell^+},
i_{\ell^-})_{\ell\in{\cal L}}$, and the device controllable injected
power $(P_g,Q_g)_{g\in{\cal G}}$.

Notice that all the equality constraints above, \eqref{eq:vc-ampl},
\eqref{eq:vc-arg}, \eqref{eq:OPF-Sbal}, and \eqref{eq:amp-lim-eq} can
be summarized as $g(y) + Ax + b = 0$, where $g(\cdot)$ is a smooth
non-convex function that can be derived from equations
\eqref{eq:Ilp}-\eqref{eq:vc-arg}, and $A$ is a positive definite
matrix. Similarly, the inequality constraints, \eqref{eq:OPF-Fb},
\eqref{eq:cap_curve_gen}, and \eqref{eq:amp-lim-slack-var}, can be
expressed as $h(x)\ge0$, where $h(x)$ is a convex function that can be
derived from equations~\eqref{eq:OPF-Fb},
\eqref{eq:amp-lim-slack-var}, and \eqref{eq:OPF-fd}. We denote the
objective by $f(x)$, where $f$ is concave.

We can thus write our problem in the more compact form:
\begin{align}
\max_{x,y}f(x) \label{eq-obj} \\
\mbox{subject to } g(y) + Ax + b & = 0 \label{eq-ncc}\\
h(x) & \ge 0 \label{eq-cc}.
\end{align}
We write its augmented Lagrangian (\cite{bertsekas99,powell1978algorithms,hestenes1969multiplier}):
\begin{align}
  {\cal L}_\rho(x,y;\lambda) =& f(x) + \lambda'(g(y)+Ax+b) \nonumber \\
  &- \frac \rho 2 \|g(y)+Ax+b\|^2,\label{eq:OPF-augmLagr}
\end{align}
where $\rho$ is the weight of the quadratic penalty term added to the classic Lagrangian function, and $\lambda$ is the vector of Lagrange multipliers associated with the equality constraints~\eqref{eq-ncc}.

Our centralized iterative algorithm for solving the OPF is based on
the method of multipliers (\cite[\S 4.2]{bertsekas99}). This method
was first introduced for solving iteratively non-linear equality
constrained problems. It is shown to converge under more general
conditions than dual ascent
\cite{boyd2011distributed}. Algorithm~\ref{alg:Centralized_OPF}
summarizes the proposed centralized algorithm, and
Theorem~\ref{thm:central_cv} characterizes its convergence.


\begin{algorithm}[b!]
  \caption{Centralized algorithm for the OPF
    (\ref{eq-obj})-(\ref{eq-cc})}
\label{alg:Centralized_OPF}
\begin{itemize}
\item{Set $k{=}0$ and initialize control variables $x$ and $y$:\\
    $\bar{S}_g^0 = 0,\bElp^0 = \bElm^0 = 1, V_b^0 = 1, \varphi_b^0=0,
    i_{{\ell}^+}^0 = i_{{\ell}^-}^0 = 0$ (per-unit), Lagrange
    multipliers $\lambda^0 = 0$, increasing gain sequence
    $(\rho^k)_k$, $\rho^k\to\infty$.}
\end{itemize}
\begin{algorithmic}[1]
  \REPEAT
  \STATE  Maximize the augmented Lagrangian for fixed $\lambda = \lambda^k$:
  \label{alg:step-iter1}
  \begin{align}
    \label{eq:method_multipl_central}
    (x^{k+1}, y^{k+1})= \argmax_{x,y: h(x)\ge 0} L_{\rho^k}(x, y;\lambda^{k}).
  \end{align}
  \STATE  Update the Lagrange multipliers:
  \begin{align}
    \label{eq:method_multipl_lagr}
    \lambda^{k+1} = \Pi_{[-\bar\lambda, \bar\lambda]}\left\{\lambda^k
      + \rho^k\left[g(y^{k+1})+Ax^{k+1}+b\right]\right\}
  \end{align}
  \STATE $k \leftarrow k{+}1$
  \UNTIL{the maximum number of iterations has been reached} \OR {the change in the Lagrange multipliers between two consecutive iterations is less than a tolerance $\delta>0$}
\end{algorithmic}
\end{algorithm}
The main advantage of the method of multipliers is that there exists a
finite value $\bar\rho$ such that the
problem~\eqref{eq:method_multipl_central} is locally convex for all
$\rho^k>\bar\rho$. Note also that the algorithm bounds the value of
$\lambda$ at each iteration. The next vector of multiplier estimates
$\lambda$ is obtained after a projection on the set $[-\bar \lambda,
\bar \lambda]$ defined as $[-\bar \lambda_1, \bar
\lambda_1]\times[-\bar \lambda_2, \bar \lambda_2]\times\dots$; the
constant vector $\bar\lambda$ is chosen such that the sought optimal
vector of Lagrange multipliers $\lambda^*$ lies in
$[-\bar\lambda,\bar\lambda]$ (see~\cite[\S
2.2.2]{bertsekas1996constrained}).
\begin{theorem}
  \label{thm:central_cv}
  For smooth objective function $f\in{\cal C}^2$ and suitably chosen
  $\bar\lambda$ such that the optimal vector of Lagrange multipliers
  $\lambda^*$ satisfies $\lambda^*\in[-\bar\lambda,\bar\lambda]$,
  Algorithm~\ref{alg:Centralized_OPF} converges to a local minimum of
  the nonlinear progam~\eqref{eq-obj}-\eqref{eq-cc}.
\end{theorem}
\begin{proof}
  By \cite[Proposition 1.23]{bertsekas1996constrained}, our problem
  satisfies assumption (S) from \cite[\S
  2.2]{bertsekas1996constrained}, since the equality constraint is a
  ${\cal C}^2$ function of $y$, and the objective function is chosen
  to be ${\cal C}^2$. Proposition 2.7 from the same reference
  guarantees the desired convergence, if the iterates $(x^k, y^k,
  \lambda^k)$ reach the set $D$ from Proposition 2.4 of
  \cite{bertsekas1996constrained}, {\it i.e.}, if there exists a $\bar
  k$ such that $(x^{\bar k}, y^{\bar k}, \lambda^{\bar k}) \in D$ (for
  all the following indices $k>\bar k$, the iterates stay in $D$, and
  convergence ensues). The existence of such a $\bar k$ follows from
  the choice of the divergent increasing sequence of gains $(\rho^k)$
  and from the boundedness of the sequence $(\lambda^k)$.
\end{proof}

Due to the quadratic terms in the expression of the augmented
Lagrangian~\eqref{eq:OPF-augmLagr}, the optimization problem in
(\ref{eq:method_multipl_central}) does not decouple across the network
and, therefore, cannot be solved in a distributed manner. In the
following section, we reformulate this problem in an equivalent way
that leads to a distributed algorithm for its resolution.

\subsection{Distributed Solution of the OPF Problem}
We adopt a primal decomposition method~\cite{palomar2006tutorial} that
gives an iterative algorithm for the minimization of the problem in
\textit{Step~\ref{alg:step-iter1}} of
Algorithm~\ref{alg:Centralized_OPF}. In \eqref{eq:OPF-augmLagr} the
line voltages $y = (\bElp, \bElm)$ are ``coupling'' variables. If
these variables are fixed to a specific value, then problem
(\ref{eq:method_multipl_central}) decouples in smaller, easier
(convex) problems, that can be solved by local agents.

Specifically, to solve (\ref{eq:method_multipl_central}) iteratively
for fixed values of the Lagrange multiplier estimates $\hat\lambda$
and fixed gain $\hat\rho$ we take the following approach: At the
$n$-th iteration, the value of the coupling variables $y^n =
(\bElp^{n}, \bElm^{n})$ is assumed fixed. The $x$ variables, {\it
  i.e.}, the power set-points of the controllable devices
($\bar{S}_g$), the bus voltages ($\bar{V}_{b}$), and the slack
variables $i_{{\ell}^+}, i_{{\ell}^-}$, are computed by solving the
following constrained convex optimization problem:

\begin{align}
  \label{eq:method_multipl_iterative_x}
  x^{n+1} = \argmax_{x:h(x)\ge 0} L_{\hat\rho}(x, y^n, \hat\lambda).
\end{align}

 Next, the coupling variables $y$ are updated as follows:
\begin{align}
  \label{eq:method_multipl_iterative_y}
  y^{n+1} = y^n + \alpha^n(\nabla_yL_{\hat\rho})(x^{n+1}, y^n, \hat\lambda),
\end{align}
where $\alpha^{n}$ is a positive step-size sequence of the gradient
descent. The choice of the step-size is related to the topology of the
network and the parameters of the lines (i.e., the network admittance
matrix). For example, a large constant step-size might not allow the
algorithm to converge, whereas a small constant step-size could cause
slow convergence\footnote{In order to properly tune this parameter, a
  dedicated off-line study can be performed before deployment of the
  proposed algorithm.}.

The algorithm stops when the norm of the update in the $y$ variables
is less than some small positive tolerance $\varepsilon$, {\it i.e.},
when $\|\nabla_yL_{\hat\rho}(x^{n+1}, y^n, \hat\lambda)\| \le
\varepsilon$.

\begin{theorem}
  \label{thm:inner_dist_cv}
  The algorithm
  \eqref{eq:method_multipl_iterative_x}-\eqref{eq:method_multipl_iterative_y}
  with tolerance $\varepsilon$ in the stopping criterion converges to
  a vicinity ${\cal B}((x^*, y^*),\delta)$ of a local optimum $(x^*,
  y^*)$ of problem (\ref{eq:method_multipl_central}). If (\ref{eq:method_multipl_central}) is strongly locally convex in $y$ in a vicinity of $(x^*, y^*)$, then $\delta =
  \Theta(\varepsilon^2)$.
\end{theorem}
\begin{proof}
  (Sketch) Denote $v(y) = \max_{x:h(x)\ge 0} L_{\hat\rho}(x, y,
  \hat\lambda)$ and $x^*(y)$ the value of $x$ that achieves this
  maximum ~\eqref{eq:method_multipl_iterative_x}. Theorem 2.1
  of~\cite{Geoffrion72} says that the optimum $(x^*(y^*), y^*)$ of
  $\max_y v(y)$ coincides with the one
  of~\eqref{eq:method_multipl_central}. Moreover, a $\delta$-optimal
  solution $(x^*(y_\delta), y_\delta)$ of $\max_y v(y)$ (that is,
  $v(y_\delta) \ge v(y^*) - \delta$) is also $\delta$-optimal
  for~\eqref{eq:method_multipl_central}.

  We now show that it holds that $\nabla_yv(y) = (\nabla_y
  L_{\hat\rho})(x^*(y),y,\hat\lambda)$, or equivalently, $\frac {D
    x^*(y)} {Dy} (\nabla_xL_{\hat\rho})(x^*(\bar y), \bar
  y,\hat\lambda) = 0$. If we can show this, then the algorithm
  \eqref{eq:method_multipl_iterative_x}-\eqref{eq:method_multipl_iterative_y}
  is equivalent to a gradient ascent in $y$ on $v(y)$. It is easy to show that the function $v(y)$ is ``smooth'' (${\cal C}^2$). By the
  strong local convexity around $(x^*, y^*)$ of the augmented Lagrangian, \cite[Exercise 1.2.10]{bertsekas99} allows us to conclude that $\delta =
  \Theta(\varepsilon^2)$.

  Note that problem~\eqref{eq:method_multipl_iterative_x} is
  convex. Consider the optimal multipliers $\mu^*$ corresponding to
  the constraints $h(x)\ge 0$. They satisfy the KKT conditions:
  \begin{align*}
    (\nabla_{x}L_{\hat\rho})(x^*(y),y,\hat\lambda) &= \sum_i
    \mu^*_i(y)\nabla_xh_i(x^*(y))\\
    \mu^*_i(y)h_i(x^*(y)) &= 0; \quad \mu^*_i \ge 0.
  \end{align*}
  Define the following functions: $\psi_i(y) := h_i(x^*(y))$. Since
  $x^*(y)$ is always feasible, it means that $\psi_i(y) \ge
  0$. Consider the set of indices ${\cal I}_0(y) := \{ i : h_i(x^*(y))
  = 0\}$. Take some $i\in {\cal I}_0(y)$. In this case the function
  $\psi_i(y)$ has an extremal point in $y$, which implies that
  $\nabla_y\psi_i(y) = 0$, or again that $\frac {D x^*(y)}
  {Dy}\nabla_xh_i(x^*(y)) = 0$. For all $i\not \in {\cal I}_0(y)$, by
  KKT we have $\mu^*_i(y) = 0$. By the above arguments,
  \begin{align*}
    &\frac {D x^*(y)} {Dy} (\nabla_xL_{\hat\rho})(x^*(y),
    y,\hat\lambda) \\ &= \sum_i \mu^*_i(y)\frac {D x^*(y)}
    {Dy}\nabla_xh_i(x^*(y)) = 0.
  \end{align*}
\end{proof}

Thanks to its separability property, problem
(\ref{eq:method_multipl_iterative_x}) can be solved in a distributed
manner. Bus agents can be responsible for updating the power
set-points of the controllable devices ($\bar{S}_g$) that are
connected to them, as well as their voltages ($\bar{V}_{b}$) in
parallel, and lines can be responsible for updating the slack
variables ($i_{{\ell}^+},i_{{\ell}^-}$). Specifically, the `power
set-points ($\bar{S}^{n+1}_g$) of devices in bus $b$ are obtained by
solving the following convex problem:
\begin{align*}
  (\bar S_g^{n+1}) =& \argmax_{\bar S_g\in{\cal H}_g}\sum_{g\in b}W_g(\bar
  S_g) \\
  &- \frac{\hat\rho}{2} \Big|\sum_{g \in b} \bar{S}_g + \bar{S}(b) + \sum_{\mathclap{\beta(\ell^+)=b}} \bar{S}^n_{\ell^+} +\sum_{\mathclap{\beta(\ell^-)=b}} \bar{S}^n_{\ell^-} - \frac{\hat{\lambda}_b}{\hat\rho} \Big|^2,
\end{align*}
where $\hat\lambda_b$ is the given multiplier corresponding to the
constraint~\eqref{eq:OPF-Sbal} of bus $b$. The other problems (for the
other $x$ variables) have simpler expressions that we do not reproduce
for brevity sake.

Similarly, (\ref{eq:method_multipl_iterative_y}) can be decomposed
across the different network lines: line-agents can update the
voltages at their two ends in parallel. In terms of required
information, each bus agent needs to know only the voltage values of
the lines that are incident to it, the constraints of the devices, and
the state of the loads that are connected to it. Finally, in order to
compute the partial derivatives of
(\ref{eq:method_multipl_iterative_y}) with respect to its voltages,
each line requires solely the information of the power balance and the
voltage values of its two adjacent buses. The actual implementation of
the distributed synchronous OPF algorithm is summarized below in
Algorithm~\ref{alg:Distr_synch}.

\begin{algorithm}[b!]
  \caption{Distributed algorithm for the OPF
    (\ref{eq-obj})-(\ref{eq-cc})}
\label{alg:Distr_synch}
\begin{itemize}
\item{Set $k{=}0$ and initialize control variables $x$ and $y$:\\
    $\bar{S}_g^0 = 0,\bElp^0 = \bElm^0 = 1, V_b^0 = 1, \varphi_b^0=0,
    i_{{\ell}^+}^0 = i_{{\ell}^-}^0 = 0$ (per-unit), Lagrange
    multipliers $\lambda^0 = 0$, increasing diverging gain sequence
    $(\rho^k)_k$, $\rho^k\to\infty$, decreasing tolerance sequence
    $(\varepsilon^k\ge 0)_k$, $\varepsilon^k\to0$.}
\end{itemize}
\begin{algorithmic}[1]
  \REPEAT

  \STATE $n\leftarrow0; \ \tilde x^0 \leftarrow x^k; \ \tilde y^0
  \leftarrow y^k$

  \REPEAT
  \label{alg2:step-iter1}

  \STATE $\tilde x^{n+1} = \argmax_{x:h(x)\ge 0} L_{\rho^k}(x, \tilde y^n, \lambda^k)$

  \STATE $\tilde y^{n+1} = \tilde y^n + \alpha^n(\nabla_yL_{\rho^k})(\tilde x^{n+1}, \tilde y^n, \lambda^k)$

  \STATE $n \leftarrow n{+}1$

  \UNTIL{$\|\nabla_yL_{\rho^k}(\tilde x^{n+1}, \tilde y^n, \lambda^k)\| \le
    \varepsilon^k$}
  \STATE $x^{k+1}\leftarrow \tilde x^{n+1}; \ y^{k+1}\leftarrow \tilde y^{n+1}$
  \STATE $\lambda^{k+1} = \Pi_{[-\bar\lambda,
    \bar\lambda]}\left\{\lambda^k +
    \rho^k\left[g(y^{k+1})+Ax^{k+1}+b\right]\right\}$
  \STATE $k \leftarrow k{+}1$

  \UNTIL{the maximum number of iterations has been reached} \OR {the change in the Lagrange multipliers between two consecutive iterations is less than a tolerance $\delta>0$}
\end{algorithmic}
\end{algorithm}

\begin{theorem}
  \label{thm:dist_cv}
  For smooth objective function $f\in{\cal C}^2$ and suitably chosen
  $\bar\lambda$ such that the optimal vector of Lagrange multipliers
  $\lambda^*$ satisfies $\lambda^*\in[-\bar\lambda,\bar\lambda]$,
  Algorithm~\ref{alg:Distr_synch} converges to a local minimum of
  the nonlinear progam~\eqref{eq-obj}.
\end{theorem}
\begin{proof}
  The proof is similar to the one of Theorem~\ref{thm:central_cv}. It
  uses Proposition 2.16 of \cite{bertsekas1996constrained} for
  convergence, which only requires at each iteration a
  $\delta^k$-optimal solution for \eqref{eq:method_multipl_central}
  with $\delta^k\to 0$. By Theorem~\ref{thm:inner_dist_cv} we can
  conclude.
\end{proof}

In a realistic setting, in order to take full advantage of the
distributed formulation of the OPF algorithm, as described above and
to avoid the overhead cost of coordination between agents, the updates
should be performed in an asynchronous fashion. Contrary to ADMM-based
algorithms, which require a synchronized implementation of the
updates, the proposed algorithm can be implemented in an asynchronous
manner. In this direction, we assume that each of the bus and line
agents has its own two local poisson clocks with different rates. The
clock with the lower rate ($C1$) triggers the multiplier update
(\ref{eq:method_multipl_lagr}) and the clock with the higher rate
($C2$) triggers the events described in steps
(\ref{eq:method_multipl_iterative_x})-(\ref{eq:method_multipl_iterative_y}).

In detail, all the control variables and the Lagrange multipliers are
first initialized. Then, every time the $C2$ clock of a bus ticks,
this bus performs local update operations by using the most recent
stored values for the voltages of its incident lines and for the
associated Lagrange multipliers. Once the bus updates its power and
voltage values, it informs the incident lines of the
changes. Similarly, when the $C2$ clock of a line ticks, the line
agent updates the variables ($i_{\ell^+},i_{\ell^-}$) by taking into
account the most recent values of the line current flows and
associated Lagrange multipliers. In addition to this update, the
updates of the voltages of its two end-points are triggered. In order
to compute the new values, the line uses the most recent stored values
for the adjacent buses' powers and voltages, and once the updates are
completed the line communicates this information to its neighboring
buses. Now, when the $C1$ clock of a bus or a line ticks, then the
corresponding agent updates the Lagrange multipliers
(\ref{eq:method_multipl_lagr}). It is worth noting, that we no
longer have a serial implementation of the various updates like the
ones presented in Algorithm~\ref{alg:Distr_synch}. On the contrary,
the different rates of the clocks are chosen in such a way to ensure
that, on average, a sufficient number of the updates occurs before an
update of the corresponding Lagrange multiplier takes place.

In this section, we investigate the performances and convergence properties of the centralized Algorithm~\ref{alg:Centralized_OPF} in several different scenarios. In particular, we consider the cases presented in Part I of the paper, where the BFM convexification leads to an incorrect solution of the OPF problem and ADMM fails to converge to a solution. Additionally, we investigate the performances of the proposed centralized algorithm under different initial conditions of the electrical-network state. In order to do so, we consider the same 4-bus test network that was used in Part I of the paper. We assume a first test case where the controllable device connected to bus $4$ is a generator, whereas controllable loads are connected to buses $2$ and $3$. The network characteristics, the base values, the capability limits of the controllable resources, and the voltage and ampacity limits are given in Part I (Fig. 3 and Table 7). In what follows, the objective function accounts for the minimization of the network losses, as well as for the utility of the generating units, namely:
\begin{align}
\label{eq:OPF-BFM_objII}
\min_{\bar{S}_g,\bar{S}_\ell,V_b,\varphi_b,|\bar{I}_\ell|} -\sum_{g \in \mathcal{G}} Re(\bar{S}_{g}) + \sum_{\ell \in \mathcal{L}} Re(\bar{Y}_\ell)|\bar{I}_\ell|^2
\end{align}

\subsection{Effect of the Line Length, Network Rated Value and Network State on the Convergence of Algorithm~\ref{alg:Centralized_OPF}}
\begin{figure}
\begin{centering}
\includegraphics[width=1\linewidth,clip=true]{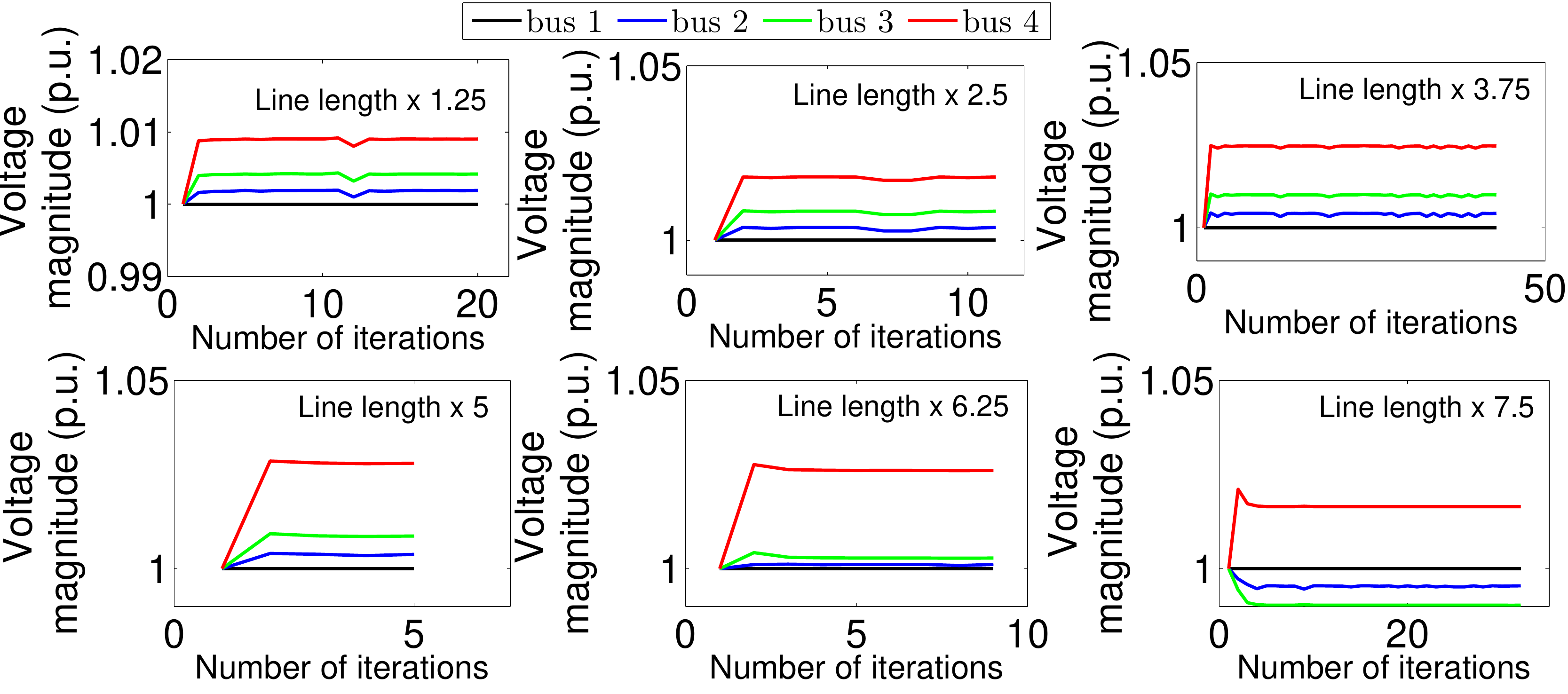}
\vspace{-25pt}
\caption{Evolution of the magnitude of network voltages for various line lengths.}
\label{fig:voltages_vs_line_length}
\end{centering}
\end{figure}

\begin{figure}[t!]
\begin{centering}
\includegraphics[width=1\linewidth,clip=true,  trim = 0 0 10 0]{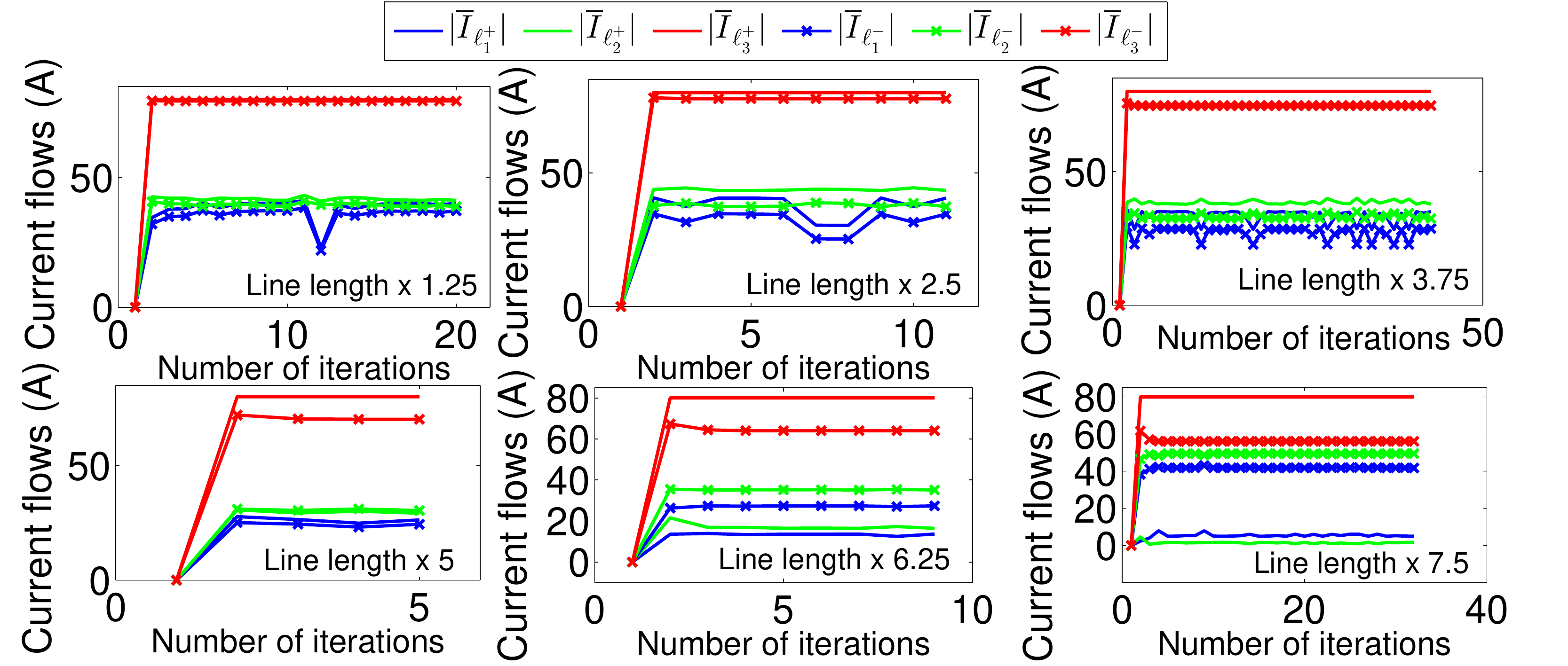}
\vspace{-25pt}
\caption{Evolution of the line current flows for various line lengths.}
\label{fig:currents_vs_line_length}
\end{centering}
\end{figure}

\begin{figure}
\begin{centering}
\includegraphics[width=1\linewidth,clip=true]{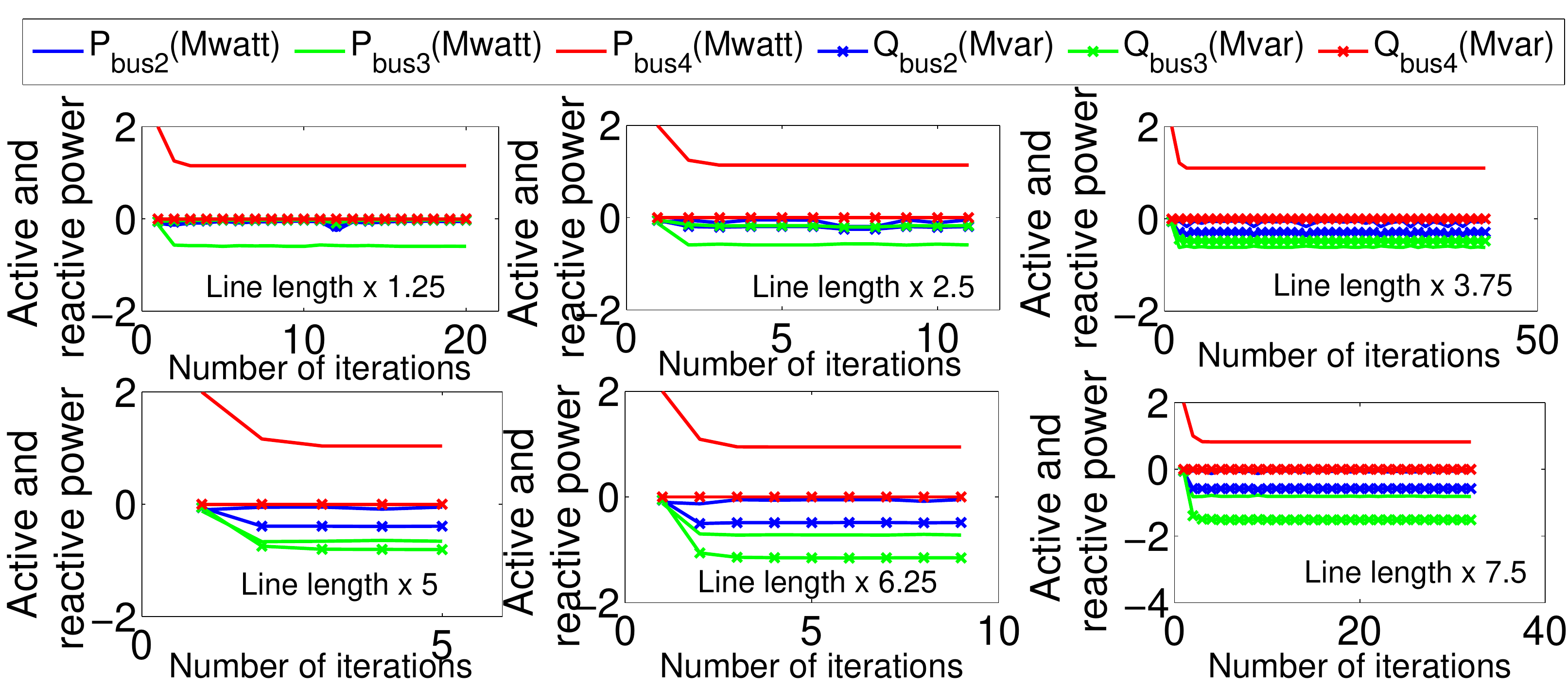}
\vspace{-25pt}
\caption{Evolution of the active and reactive power of the controllable devices for various line lengths.}
\label{fig:powers_vs_line_length}
\end{centering}
\end{figure}

\begin{figure}[h!]
\begin{centering}
\includegraphics[width=1\linewidth,clip=true]{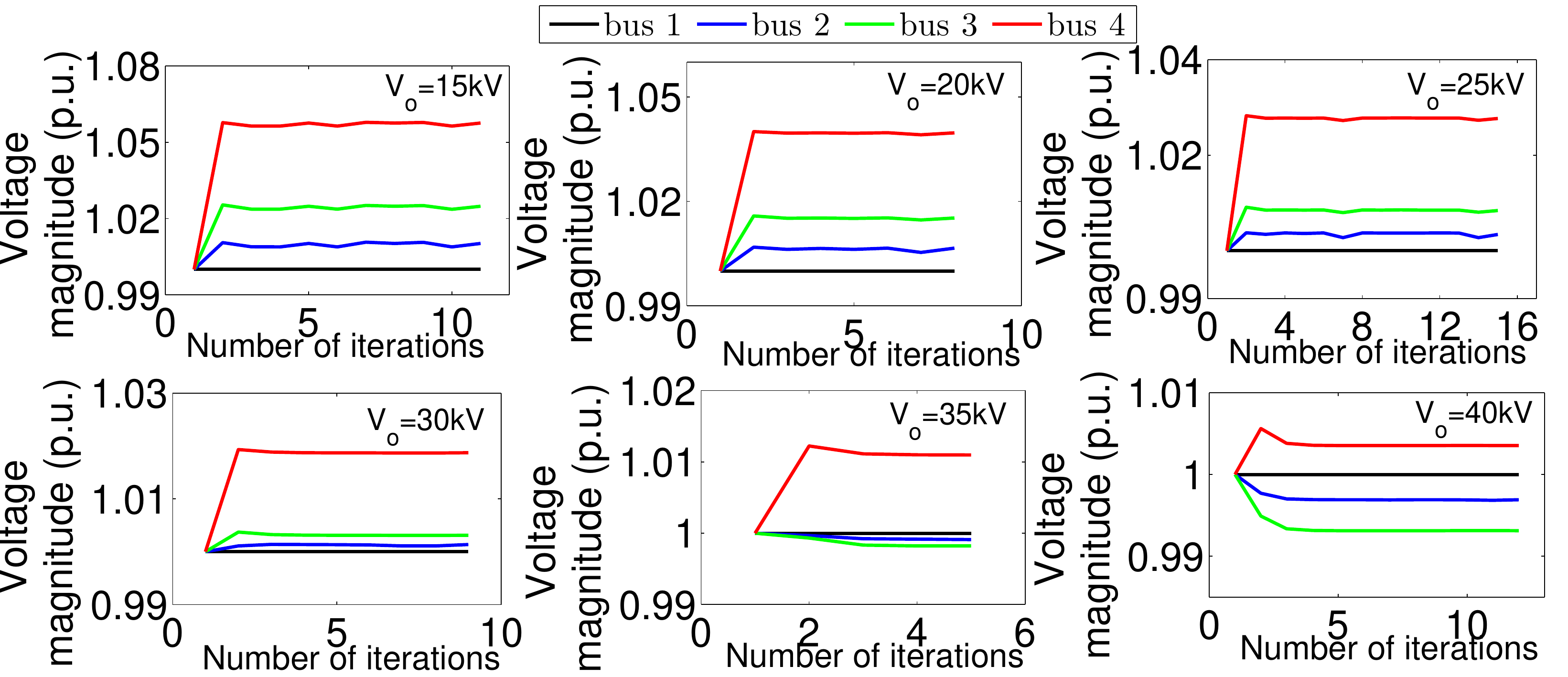}
\vspace{-25pt}
\caption{Evolution of the magnitude of network voltages for various values of the network rated voltage.}
\label{fig:voltages_vs_vlevel}
\end{centering}
\end{figure}

\begin{figure}
\begin{centering}
\includegraphics[width=1\linewidth,clip=true]{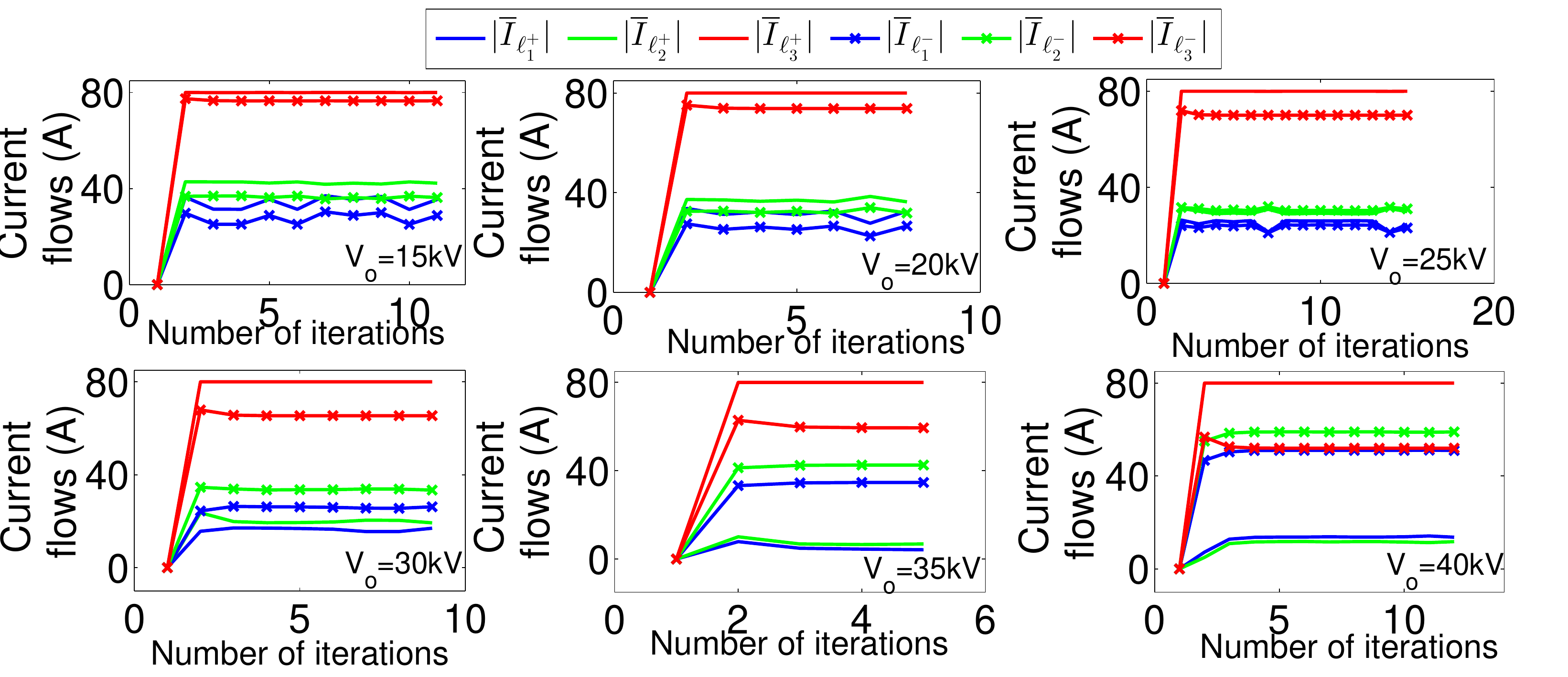}
\vspace{-25pt}
\caption{Evolution of the line current flows for various values of the network rated voltage.}
\label{fig:currents_vs_vlevel}
\end{centering}
\end{figure}

\begin{figure}
\begin{centering}
\includegraphics[width=1\linewidth,clip=true]{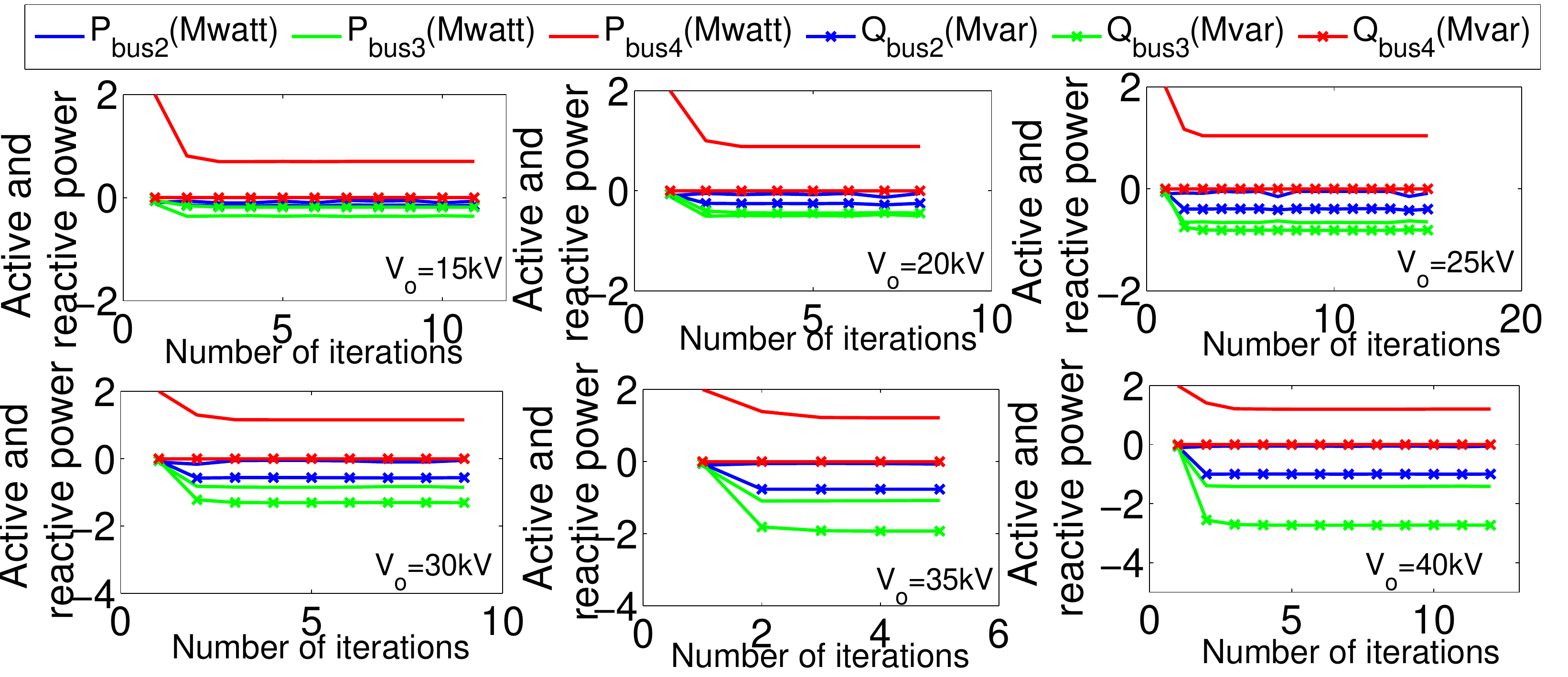}
\vspace{-25pt}
\caption{Evolution of the magnitude of active and reactive power of the controllable devices for various values of the network rated voltage.}
\label{fig:powers_vs_vlevel}
\end{centering}
\end{figure}

\begin{figure}
\begin{centering}
\includegraphics[width=1\linewidth,clip=true,  trim = 50 0 50 100]{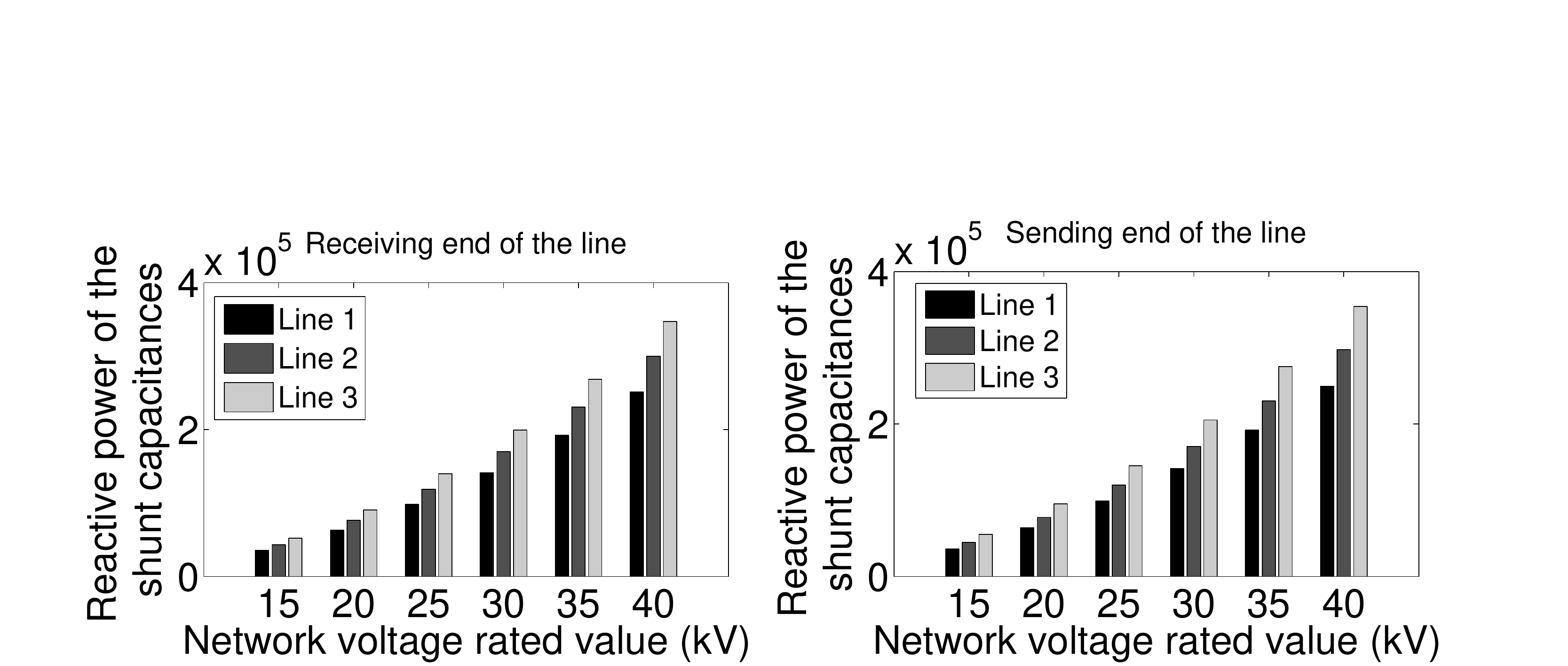}
\vspace{-25pt}
\caption{Reactive power produced by the shunt elements of the lines for various values of the network voltage rated value.}
\label{fig:Qc_vs_vlevel}
\end{centering}
\end{figure}

\begin{figure}
\begin{centering}
\includegraphics[width=1\linewidth,clip=true,  trim = 50 0 50 150]{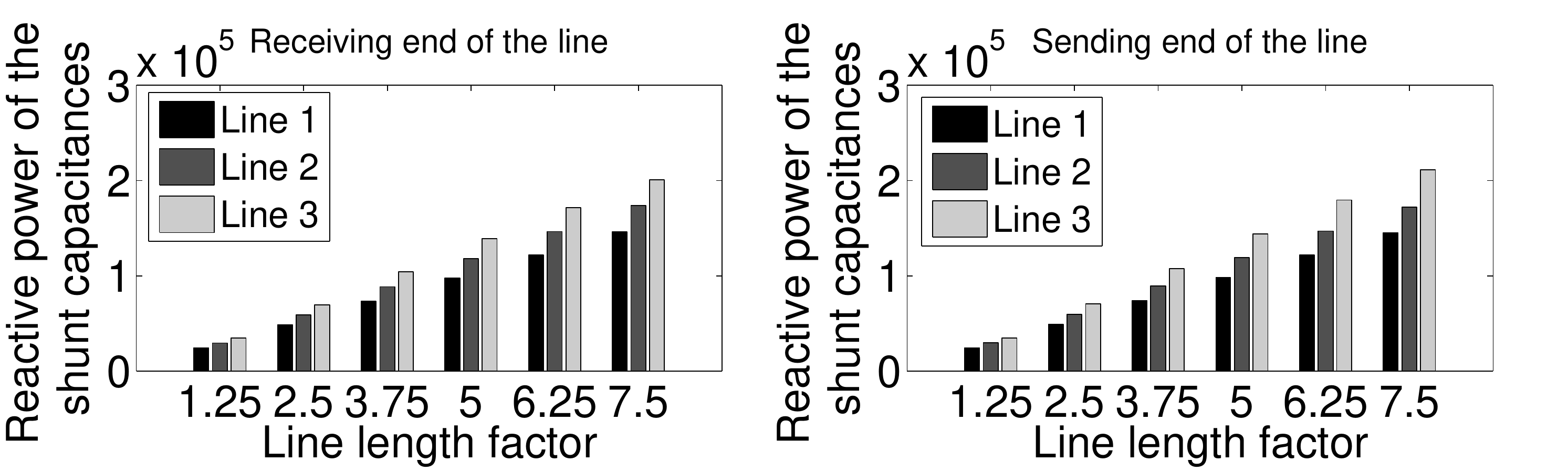}
\vspace{-25pt}
\caption{Reactive power produced by the shunt elements of the lines for various line lengths.}
\label{fig:Qc_vs_linelen}
\end{centering}
\end{figure}

In order to compare the performances of the proposed algorithm with the OPF algorithm proposed in~\cite{farivar2013branch,6507352}, we solve the OPF problem for various line lengths and network voltage rated values as in Part I of the paper. In particular, we assume that the line lengths are uniformly multiplied by a factor in the range $[1.25-7.5]$ (while keeping the network voltage rated value to its nominal value) and the network voltage rated value varies in the range $[15-40]kV$ (while keeping the line lengths to their nominal values). The evolution of the bus voltages, the line-current flows, as well as the active and reactive powers, are shown in Figures~\ref{fig:voltages_vs_line_length}-\ref{fig:powers_vs_vlevel}. It is worth noting that in all the cases the proposed algorithm converges in a few iterations. Furthermore, we observe from Fig.~\ref{fig:currents_vs_line_length} and Fig.~\ref{fig:currents_vs_vlevel} that the line-current flows satisfy the line ampacity limit, once the algorithm has converged, in all cases. In particular, in Fig.~\ref{fig:currents_vs_line_length} it is worth observing that as the line length increases the receiving and sending-end current flows of the same line become significantly different. The behavior of the current flows as the voltage rated value increases is similar (Fig.~\ref{fig:currents_vs_vlevel}). This effect is due to the increasing contribution of the current flow toward the shunt elements of the lines. In fact, we show, in Figures~\ref{fig:Qc_vs_vlevel} and~\ref{fig:Qc_vs_linelen}, the amount of reactive power produced by the shunt elements of the lines for the various values of the line lengths and the network voltage rated values. We observe that as the line length increases or the rated value of the voltage increases the reactive power produced by the shunt elements of the line increases as well.


\begin{figure}
\begin{centering}
\includegraphics[width=1\linewidth,clip=true, trim = 0 0 0 90]{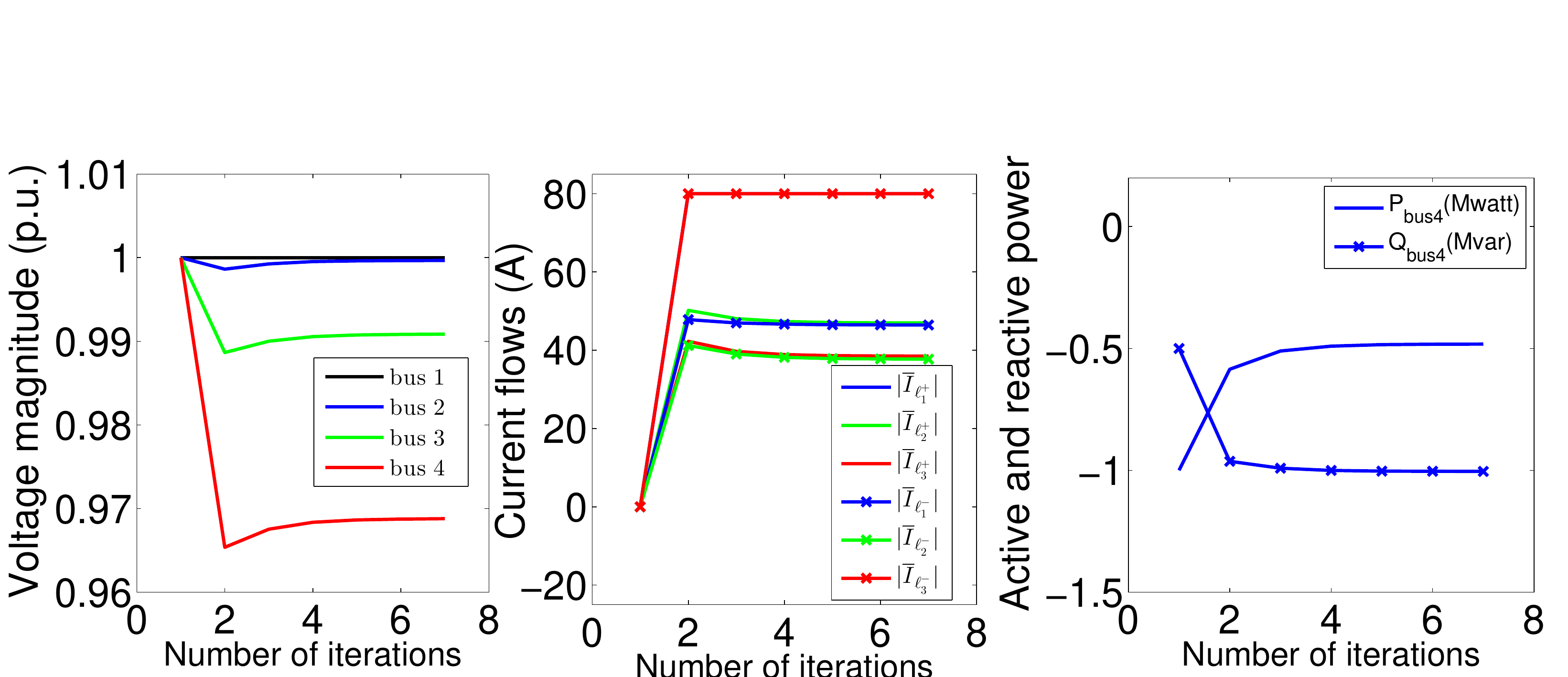}
\vspace{-25pt}
\caption{Evolution of the magnitude of network voltages, current flows, as well as active and reactive power of the controllable load at bus 4 for the case of low generation and high load in the network.}
\label{fig:diff_op}
\end{centering}
\end{figure}

We investigate, in addition to the effect of the line lengths and the network voltage rated value, the performance of the proposed algorithm under a different network operating point. To this end, we consider a second test-case where the controllable device connected to bus $4$ is a controllable load and generators are connected to buses $2$ and $3$. In this respect, we consider an extra term in the objective function, which represents the utility associated with the controllable load and is given by $(P_L-P_o)^2$, where $P_o$ represents a constant amount of load that has to be served. The capability limits of the controllable resources are shown in Table~\ref{bfm_emtp_OPF_param_op_state}. The convergence of the voltages, current flows, as well as active and reactive powers are shown in Fig.~\ref{fig:diff_op}. For the sake of brevity, we only show the evolution of the active and reactive power of the controllable load of bus 4, as the controllable generators are small and reach their maximum value upon convergence.
\begin{table}
\begin{center}
\caption{Parameters of the test network used for the investigation of the performance of the proposed OPF algorithm under a different operating point}
\label{bfm_emtp_OPF_param_op_state}
\begin{tabular}{|c|c|}
  \hline
  \textbf{Parameter} &  \textbf{value} \\ \hline
  $[P_{g_{min}},P_{g_{max}}]$(bus 2) (MW)& $[0,0.01]$\\ \hline
  $[P_{g_{min}},P_{g_{max}}]$(bus 3) (MW)& $[0,0.012]$\\ \hline
  $(P_{c_{min}},Q_{c_{min}})$(MW,Mvar) (bus 4)& $0.3,0.15$\\ \hline
  $P_{o}$(MW) (bus 4)& $1$\\ \hline
\end{tabular}
\end{center}
\end{table}

\subsection{Performance Evaluation of the Proposed Algorithm in the Presence of Shunt Capacitors in the Network}
In what follows, we consider the same network adopted in the previous section and a case where each network bus, apart from the slack, has a load and a generator connected to it. The demand in the network is assumed to be non-controllable, whereas the generators are assumed to be distributed solar panels with typical PV-type capability constraints given by (\ref{eq:OPF-fd}). For this scenario, the capability limits and the values of loads and generation are shown in Table~\ref{bfm_emtp_OPF_param_admmII}. In addition to the loads and generation, we consider that a shunt capacitor is connected to bus 2. In order to model this shunt capacitor, we consider that it is part of the first line. In particular, we consider that the shunt capacitance on the sending end of the $\pi$-model of the line that connects buses 1 and 2 is modified accordingly, to account for the shunt capacitor. For this particular test case, it is worth noting that ADMM exhibits oscillations and fails to converge to a solution (see Part I, Fig.9-12).

\begin{table}
\begin{center}
\caption{Parameters of the test network used for the evaluation of Algorithm 1 in the presence of shunt capacitors in the network}
\label{bfm_emtp_OPF_param_admmII}
\begin{tabular}{|c|c|}
  \hline
  \textbf{Parameter} &  \textbf{Value} \\ \hline
  Generators' power, $|\bar{S}_{i_{g_{max}}}|, i=2,3,4$ (MVA)& $0.40,0.39,0.46$\\ \hline
  Generators' power factor, $cos\phi_{i_{g}}, i=2,3,4$ & $0.9$ \\ \hline
  Loads' active power, $P_{i_{c}}, i=2,3,4$ (MW)& $2.76,2.16,2.46$\\ \hline
  Loads' reactive power, $Q_{i_{c}}, i=2,3,4$ (MW)& $1.38,1.08,1.23$\\ \hline
  Shunt capacitor (bus 2)(uF) & $859$ \\ \hline
  Penalty term gain, $\rho$ & $10^4$ \\ \hline
  Tolerance and maximum number of iterations & $10^{-4}, 10^4$\\ \hline
  $[V_{min},V_{max}]$ (p.u)& $[0.9,1.1]$\\ \hline
\end{tabular}
\end{center}
\end{table}

\begin{figure}
\begin{centering}
\includegraphics[width=1\linewidth,clip=true, trim=0 0 0 180]{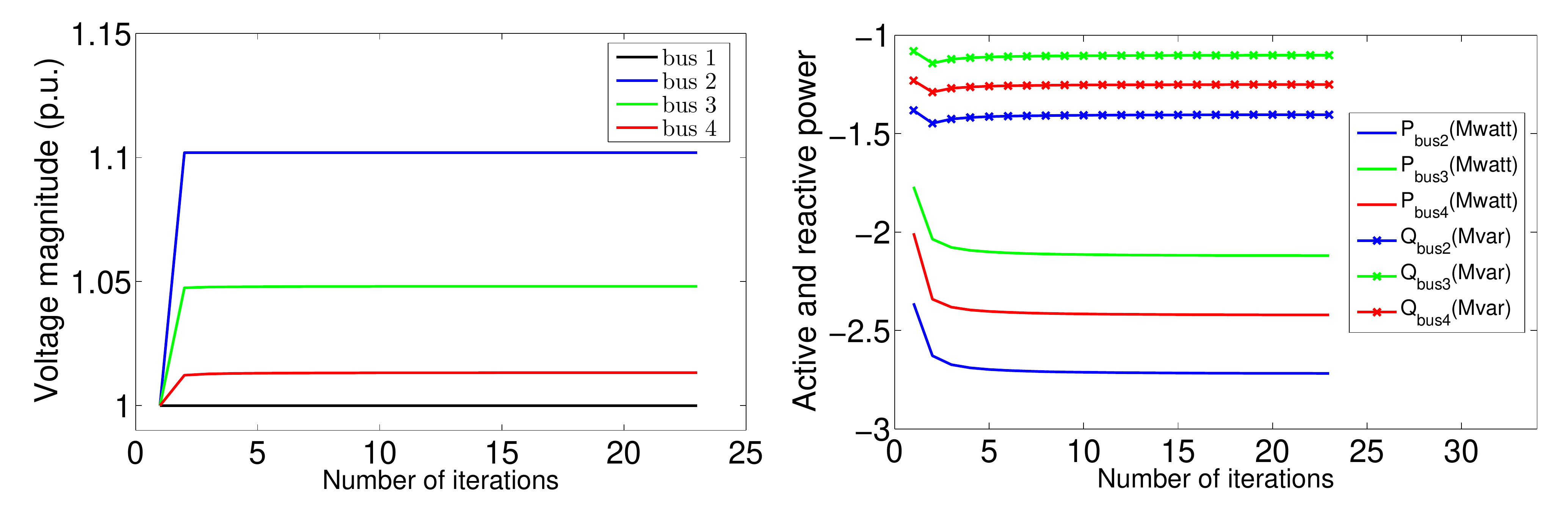}
\vspace{-25pt}
\caption{Evolution of the active and reactive power, as well as the voltages of the buses when a shunt capacitor is connected to bus 2.}
\label{fig:power_admm}
\end{centering}
\end{figure}

The results for this specific test-case, for the voltage magnitudes and the active and reactive power of the buses, are shown in Fig.~\ref{fig:power_admm}. It is worth observing that the proposed algorithm converges to a solution within a few tens of iterations; which is contrary to the ADMM-based solution of the OPF problem.

\subsection{Coordinated Control of DERs Power Set-points and OLTC}
{\color{black}{In this section we investigate the performances of the centralized OPF algorithm in the case of coordinated control of the DERs' nodal power injections and the OLTCs positions. We assume that at the primary substation there is an OLTC that can vary the secondary voltage within a $\pm6\%$ range in $\pm36$ steps. In order to do so, we assume as described in Section 4.2 of Part I of the paper that the first network branch connecting network nodes 1 and 2 has a longitudinal admittance that is a function of the OLTC ratio $\alpha$ and the transformer nominal admittance, namely $\alpha \bar{Y}_t$. The shunt elements of the receiving and sending ends of this specific branch are also a function of the new control variable, i.e., $\alpha (\alpha-1) \bar{Y}_t$  and $(1-\alpha) \bar{Y}_t$ respectively (based on the OLTC model in~\cite{stevenson1994power}). We model this control variable $\alpha \in \mathbb{R}$ as pseudo-continuous and round it to the nearest integer when the algorithm has converged to a solution. We also include a cost function in the objective function of the problem related to the OLTC operation. In particular, the changes of OLTC need to be penalized as these devices are typically used by the DNO rarely due to their increased cost and their limited lifetime. This is why we include a quadratic penalty function at each time-step in order to minimize the OLTC changes from their previous position. In the application example that follows, we consider the initial OLTC position set to -18, the rest of the grid parameters are as in Table~\ref{bfm_emtp_OPF_param_admmII} except for the voltage limits that are set in this case to $\pm 5\%$ of the network rated value. The results for this test case are shown in Fig.~\ref{fig:oltc_v} and~\ref{fig:oltc_p} below. In particular, Fig.~\ref{fig:oltc_v}  shows the evolution of the voltage magnitudes in all network buses and in solid black line the evolution of the control variable $\alpha$, which upon solution of the problem is rounded to position -4. Fig.~\ref{fig:oltc_p} shows the evolution of the rest of the control variables, which are essentially the nodal active and reactive power injections. Overall we can observe that for this test case the OPF algorithm converges in 31 iterations and the OLTC ratio converges to the value which is closest to its initial setting and at the same time allows the voltage magnitudes at all buses to lie within the acceptable limits for safe operation.

\begin{figure}[t!]
\begin{centering}
\includegraphics[width=1\linewidth,clip=true, trim= 0 120 0 120]{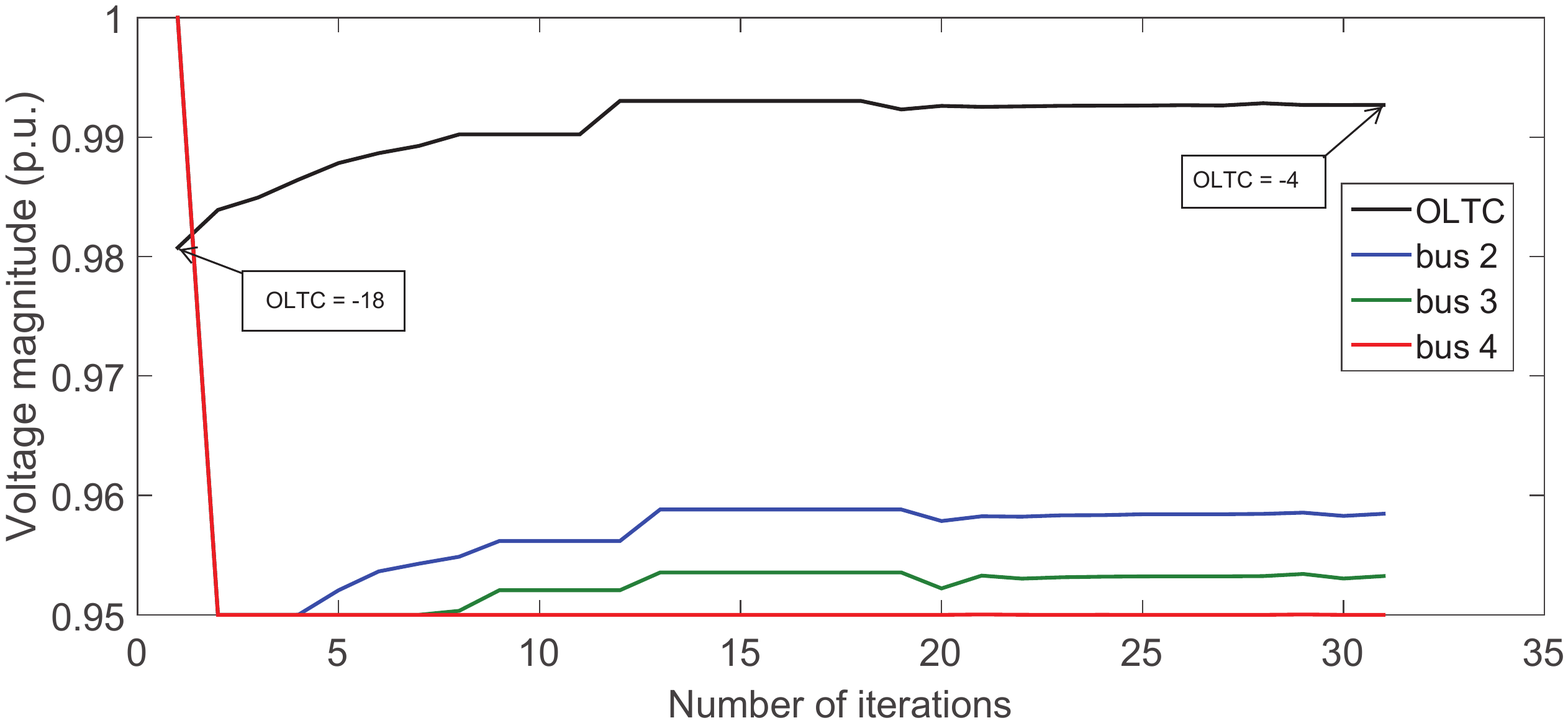}
\vspace{-25pt}
\caption{Evolution of the magnitude of network voltages when OLTC is included as a control variable.}
\label{fig:oltc_v}
\end{centering}
\end{figure}
\begin{figure}[t!]
\begin{centering}
\includegraphics[width=1\linewidth,clip=true, trim= 0 0 0 0]{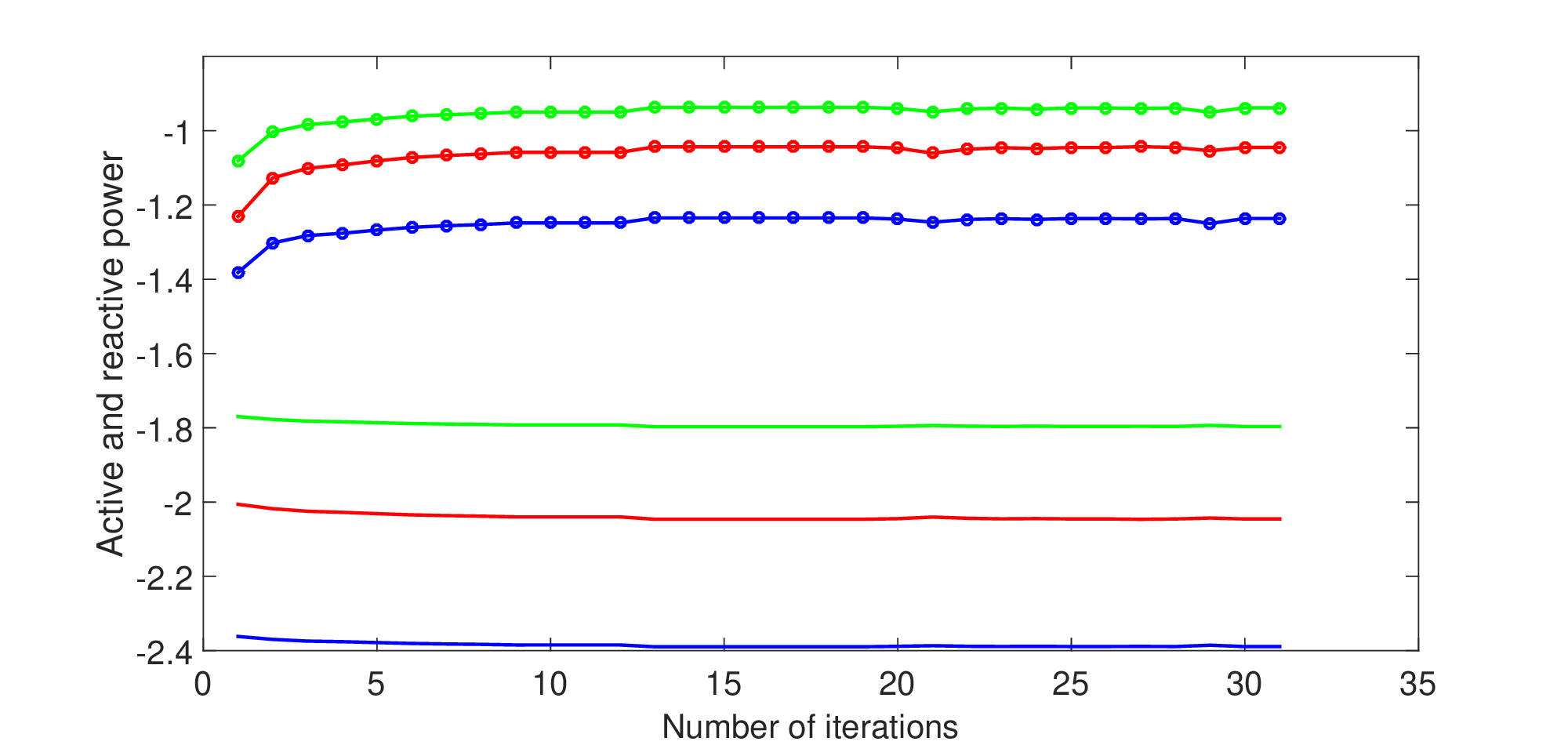}
\vspace{-25pt}
\caption{Evolution of the magnitude of the nodal active and reactive power of the network buses when OLTC is included as a control variable.}
\label{fig:oltc_p}
\end{centering}
\end{figure}}}
\subsection{Performance Evaluation of the Proposed Algorithm under Different Initial Conditions of the Network State}
\begin{figure}
\begin{centering}
\includegraphics[width=1\linewidth,clip=true, trim= 0 0 0 250]{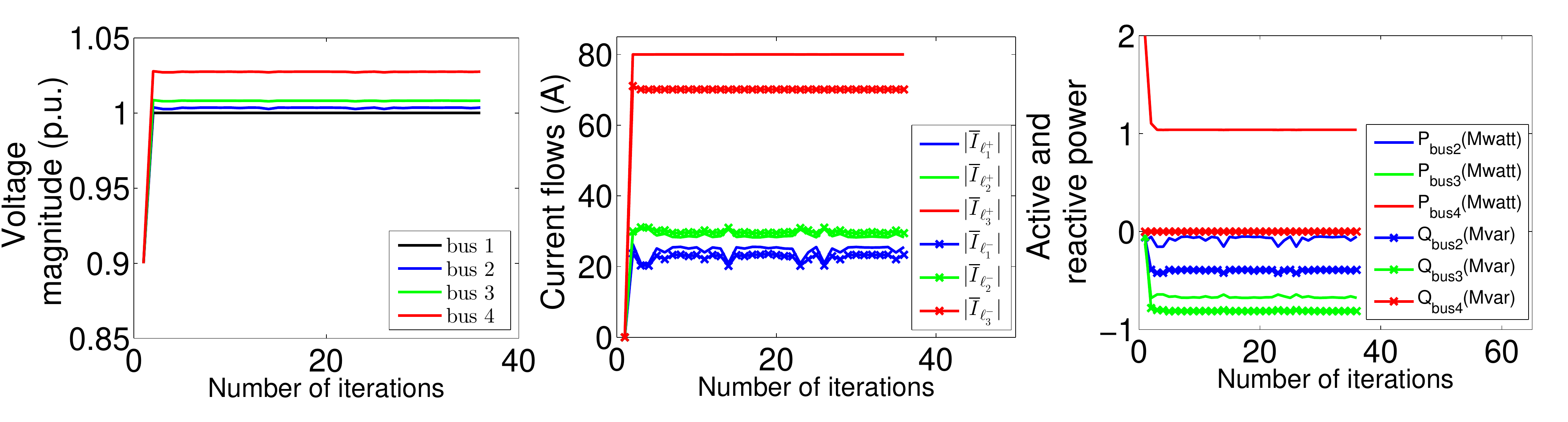}
\vspace{-25pt}
\caption{Evolution of the magnitude of network voltages, line current flows and active and reactive power of the controllable devices when the initial voltage magnitudes are set to 0.9 and the voltage angles to $-\pi/6$.}
\label{fig:volt_sp}
\end{centering}
\end{figure}

\begin{figure}[t!]
\begin{centering}
\includegraphics[width=1\linewidth,clip=true, trim= 0 0 0 250]{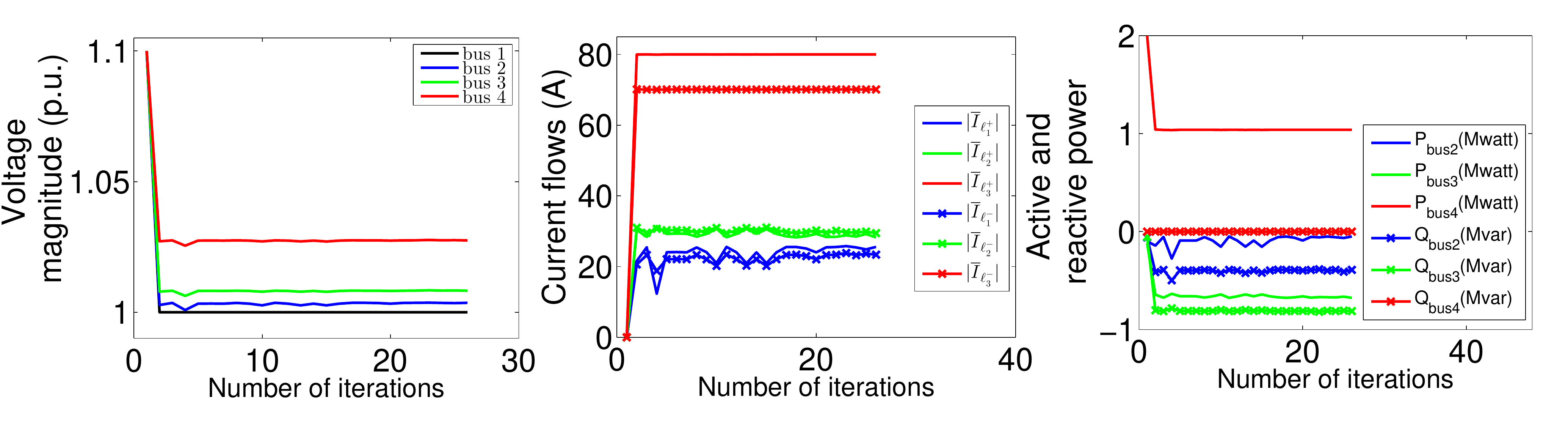}
\vspace{-25pt}
\caption{Evolution of the magnitude of network voltages, line current flows and active and reactive power of the controllable devices when the initial voltage magnitudes are set to 1.1 and the voltage angles to $\pi/6$.}
\label{fig:pow_sp}
\end{centering}
\end{figure}

Finally, we investigate the performances of the proposed algorithm under different initial conditions of the network state variables. In order to do so, we initialize the magnitude of the control variables $\bElp^0, \bElm^0, V_b^0$ in Algorithm~\ref{alg:Centralized_OPF} in the range $[0.9,1.1]$ and their angle in the range $[-\pi/6,\pi/6]$, totaling 121 different cases. For each combination, we solve the centralized OPF problem for the same network adopted in Part I (Fig. 3). In all the cases the algorithm converges to the same solution within a few tens of iterations. In Table~\ref{table:starting_point}, the mean value of the number of iterations, as well as the 95-th percentile are shown. For the sake of brevity, we show in Fig.~\ref{fig:volt_sp}-\ref{fig:pow_sp} the convergence results for the voltage, as well as for the current flows and the active and reactive power profiles for the two extreme cases, specifically when the voltage magnitude is set to 0.9 (1.1) and the voltage angle is set to $-\pi/6$ ($\pi/6$).

\begin{table}[H]
\caption{Number of iterations for the solution of the OPF problem (Algorithm 1)}
\label{table:starting_point}
\begin{center}
\begin{tabular}{ c |c | c |  }
\cline{2-3}
         &\multicolumn{1}{ c|}{ \textbf{Mean number of iterations}}&\multicolumn{1}{ c|} {\textbf{95-th Percentile}} \\\hline
    \multicolumn{1}{|c|}{\textbf{Algorithm 1}  }  & $18.21 $ & $46.45$ \\\hline
\end{tabular}
\end{center}
\label{table:comparison}
\end{table}


\section{Performance Evaluation of the Proposed Distributed Asynchronous OPF Algorithm}
For the sake of completeness, in this section, we assess the performance of the proposed algorithm with respect to a realistic grid represented by a modified IEEE 13-node test feeder (\cite{916993}). The modifications are (i) balanced lines, (ii) inclusion of secondary substations where voltage independent PQ-injections are placed, and (iii) lines ten times longer. We use this benchmark to assess the behavior of the proposed distributed asynchronous OPF algorithm. Also, we compare the solution and convergence of the distributed version of the algorithm to the centralized one.

We consider a test case where each network bus, apart from the slack bus, has a load and a generator connected to it. The demand in the network is assumed to be non-controllable, whereas the generators are assumed to be distributed solar panels with typical PV-type capability constraints. For this test case, the capability limits and the values of loads and generation are shown in Table~\ref{table:async-PV-loads}.
\begin{table}
\begin{center}
\caption{Capability limits and values of loads and generation for the evaluation of Algorithm 2}
\label{table:async-PV-loads}
\begin{tabular}{|c|c|c|c|c|c|}
  \hline
  \textbf{Bus} & ${S}_{{g_{max}}}$& $P_c(MW)$/ & \textbf{Bus}& ${S}_{{g_{max}}}$& $P_c$(MW)/\\
     &(MVA) & $Q_c$(Mvar) &   &(MVA) &$Q_c$(Mvar)\\ \hline
2	&	0.0437	&	0.0025	/	0.0011	&	8	&	0.0347	&	0.0031	/	0.0014	\\ \hline
3	&	0.0480	&	0.0029	/	0.0012	&	9	&	0.0403	&	0.0031	/	0.0013	\\ \hline
4	&   0.0506	&	0.0032   /	0.0013	&	10	&	0.0373	&	0.0031	/	0.0013	\\ \hline
5	&	0.0367	&	0.0029	/	0.0012	&	11	&	0.0482	&	0.0024	/	0.0010	\\ \hline
6	&	0.0443	&	0.0029	/	0.0012	&	12	&	0.0399	&	0.0030	/	0.0013	\\ \hline
7   &	0.0426	&	0.0025	/	0.0010	&	13  &	0.0436	&	0.0029	/	0.0012	\\ \hline
\end{tabular}
\end{center}
\end{table}
\begin{figure}[b!]
\begin{centering}
\includegraphics[width=1\linewidth, clip=true, trim=0 0 0 0]{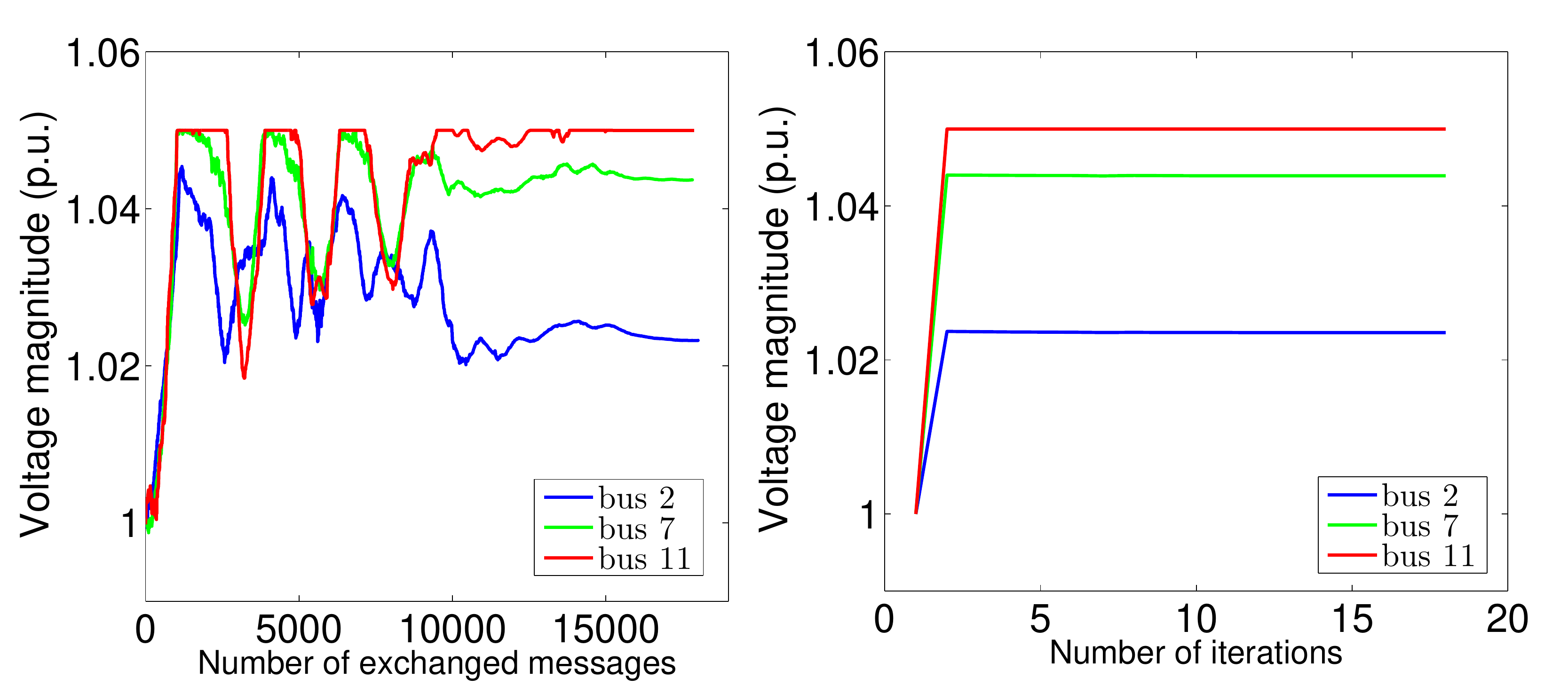}
\vspace{-25pt}
\caption{Evolution of the voltage magnitude for the distributed asynchronous algorithm as a function of the number of messages exchanged (left) and for Algorithm 1 as a function of the number of iterations (right).}
\label{fig:V_async}
\end{centering}
\end{figure}

\begin{figure}[t!]
\begin{centering}
\includegraphics[width=1\linewidth, clip=true, trim=0 0 0 0]{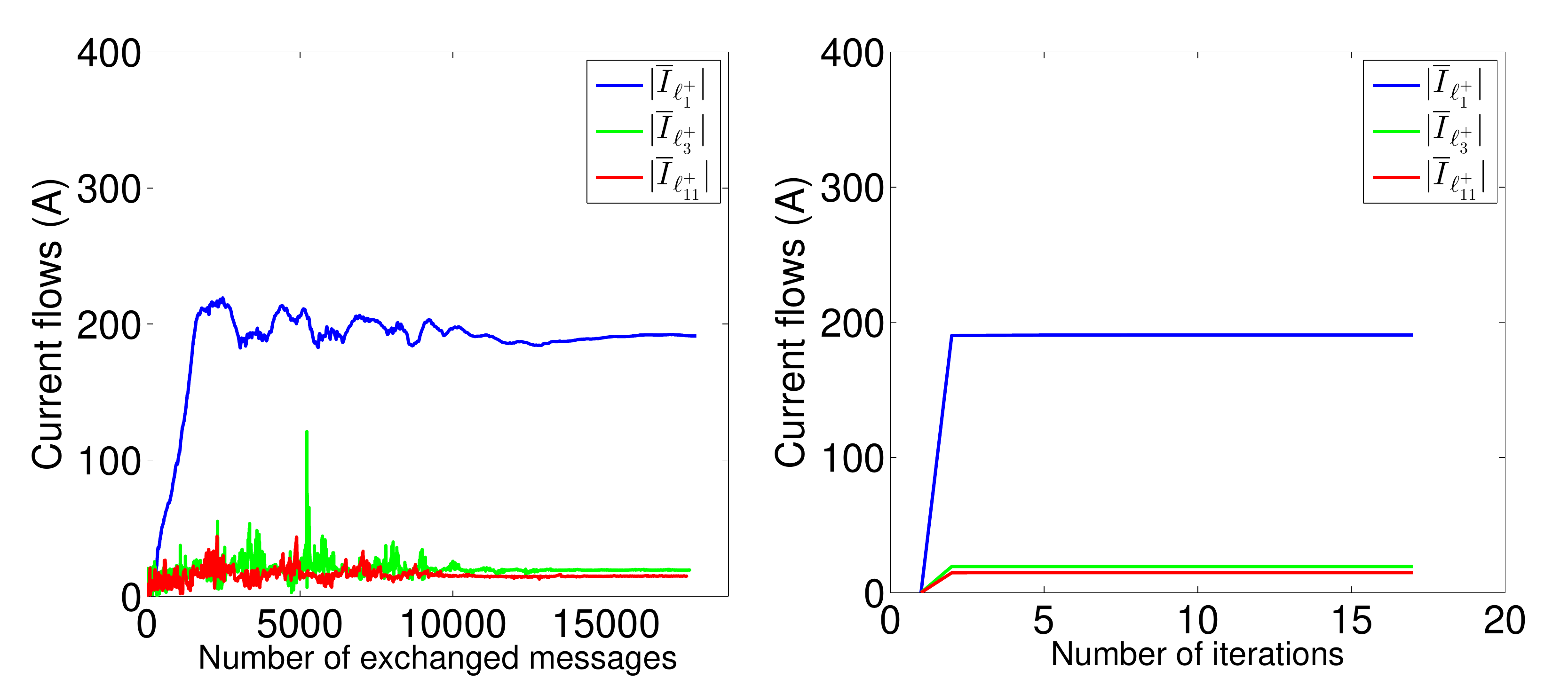}
\vspace{-25pt}
\caption{Evolution of the current flows for the distributed asynchronous algorithm as a function of the number of messages exchanged (left) and for Algorithm 1 as a function of the number of iterations (right).}
\label{fig:I_async}
\end{centering}
\end{figure}

\begin{figure}[h!]
\begin{centering}
\includegraphics[width=1\linewidth, clip=true, trim=0 0 0 0]{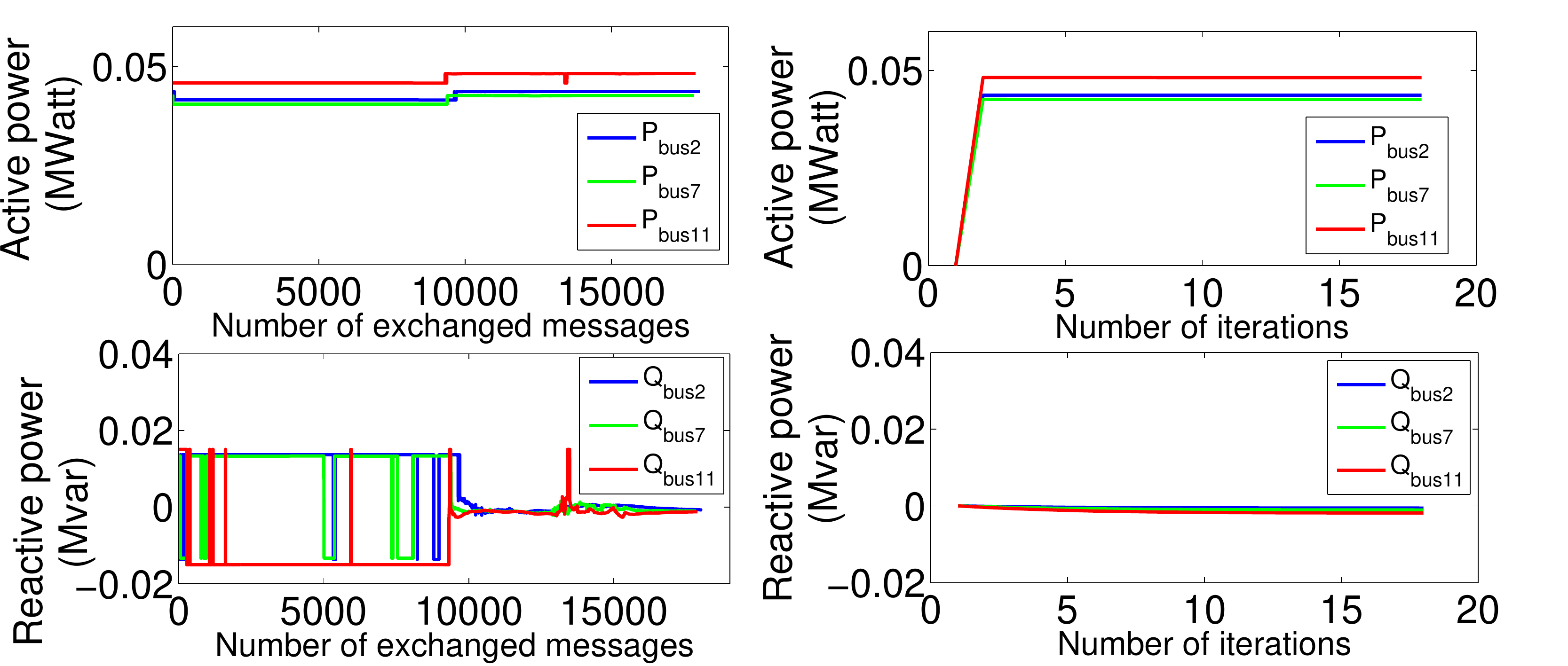}
\vspace{-25pt}
\caption{Evolution of the active and reactive power for the distributed asynchronous algorithm as a function of the number of messages exchanged (left) and for Algorithm 1 as a function of the number of iterations (right).}
\label{fig:powasync}
\end{centering}
\end{figure}

We solve the OPF problem in (\ref{eq:OPF-global2})-(\ref{eq:OPF-fd}) using Algorithm~\ref{alg:Centralized_OPF}, as well as the asynchronous implementation of Algorithm~\ref{alg:Distr_synch}. The results are shown in Fig.~\ref{fig:V_async}-~\ref{fig:powasync}. For the sake of brevity, we plot only the evolution of the magnitudes of the minimum voltage, the maximum voltage and the median value of the voltage. We plot also the evolution of the minimum, maximum and mean values of the current flows on the receiving-end of the line and the evolution of the active and reactive powers. It is worth observing that Algorithm 1 converges to the optimal solution within a few iterations and also that the distributed asynchronous implementation of Algorithm~\ref{alg:Centralized_OPF} converges to the same solution as its centralized counterpart.


\section{Conclusion}
To overcome the limitations identified in Part I, we have proposed algorithms for the solution of the AC non-convex OPF problem in radial networks that are proven to converge to a local minimum. These algorithms use an augmented Lagrangian approach and rely on the method of multipliers for the OPF solution. The two algorithms solve the centralized and decentralized (asynchronous) formulation of the targeted OPF. We have shown the robustness of the centralized version with respect to the following elements: (i) various line lengths, (ii) various network-rated voltage values and (ii) different network operating points (cases where the BFM convexification leads to an incorrect solution),  (iii) the presence of shunt capacitors in the grid (where ADMM failed to converge to a solution) and (iv) different initial conditions of the electrical network state. Finally, we have verified the equivalence of the two proposed algorithms for the case of the IEEE 13-node test distribution feeder where realistic operating conditions have been considered.